\documentclass[fleqn,usenatbib]{mnras}

\usepackage{newtxtext,newtxmath}
\usepackage{mathptmx}
\usepackage{txfonts}

\usepackage[T1]{fontenc}

\usepackage{threeparttablex}
\usepackage{longtable}
\usepackage{graphicx}	% Including figure files
\usepackage{amsmath}	% Advanced maths commands
\usepackage{amssymb}	% Extra maths symbols
\usepackage{caption}

\title[Broad-lined luminous type II SNe]{Broad-emission-line dominated hydrogen-rich luminous supernovae}

\author[P. J. Pessi et al.]{
P.~J. Pessi,$^{1,2,3}$\thanks{E-mail: priscila.pessi@astro.su.se}
J.~P. Anderson,$^{2,4}$ 
G. Folatelli,$^{3,5,6}$ 
L. Dessart,$^{7}$ 
S. González-Gaitán,$^{8}$ 
A. M\"{o}ller,$^{9}$
\newauthor
C.~P. Guti\'{e}rrez,$^{10,11}$ 
S. Mattila,$^{10,11}$ 
T.~M. Reynolds,$^{11,12}$ 
P. Charalampopoulos,$^{11}$
A.~V. Filippenko,$^{13}$
\newauthor
L. Galbany, $^{14,15}$
A. Gal-Yam, $^{16}$
M. Gromadzki,$^{17}$
D. Hiramatsu, $^{18,19}$
D.~A. Howell, $^{20,21}$
C. Inserra,$^{22}$
\newauthor
E. Kankare,$^{11}$
R. Lunnan,$^{1}$
L. Martinez, $^{5,23}$ 
C. McCully, $^{18}$
N. Meza, $^{24}$
T.~E. M\"uller-Bravo, $^{14,15}$ 
\newauthor
M. Nicholl, $^{25,26}$
C. Pellegrino, $^{20,21}$
G. Pignata,$^{4,27}$
J. Sollerman, $^{1}$
B.~E. Tucker, $^{28,29,30}$
X. Wang, $^{31,32}$
\newauthor
and 
D. R. Young$^{26}$
\\
$^{1}$The Oskar Klein Centre, Department of Astronomy, Stockholm University, AlbaNova 106 91, Stockholm, Sweden\\
$^{2}$European Southern Observatory, Alonso de C\'ordova 3107, Casilla 19, Santiago, Chile\\
$^{3}$Facultad de Ciencias Astron\'{o}micas y Geof\'{i}sicas (FCAG), Universidad Nacional de La Plata (UNLP), Paseo del bosque S/N, 1900, Argentina\\
$^{4}$  Millennium Institute of Astrophysics MAS, Nuncio Monsenor Sotero Sanz 100, Off. 104, Providencia, Santiago, Chile\\
$^{5}$Instituto de Astrofísica de La Plata (IALP), CCT-CONICET-UNLP, Paseo del Bosque s/n, B1900FWA La Plata, Argentina\\
$^{6}$Kavli Institute for the Physics and Mathematics of the Universe (WPI), The University of Tokyo, 5-1-5 Kashiwanoha, Kashiwa,Chiba 277-8583, Japan\\
$^{7}$Institut d'Astrophysique de Paris, CNRS-Sorbonne Universit\'e, 98 bis boulevard Arago, F-75014 Paris, France\\
$^{8}$CENTRA, Instituto Superior Técnico, Universidade de Lisboa, Av. Rovisco Pais 1, 1049-001 Lisboa, Portugal\\
$^{9}$Centre for Astrophysics $\&$ Supercomputing, Swinburne University
of Technology, Victoria 3122, Australia\\
$^{10}$Finnish Centre for Astronomy with ESO (FINCA), FI-20014 University of Turku, Finland\\
$^{11}$Tuorla Observatory, Department of Physics and Astronomy, FI-20014 University of Turku, Finland\\
$^{12}$Niels Bohr Institute, University of Copenhagen, Jagtvej 128, 2200 Copenhagen, Denmark\\
$^{13}$Department of Astronomy, University of California, Berkeley, CA 94720-3411, USA\\
$^{14}$Institute of Space Sciences (ICE, CSIC), Campus UAB, Carrer de Can Magrans, s/n, E-08193 Barcelona, Spain\\
$^{15}$Institut d’Estudis Espacials de Catalunya (IEEC), E-08034 Barcelona, Spain\\
$^{16}$Department of Particle Physics and Astrophysics, Weizmann Institute of Science, 76100 Rehovot, Israel\\
$^{17}$Astronomical Observatory, University of Warsaw, Al. Ujazdowskie 4, 00-478 Warszawa, Poland\\
$^{18}$Center for Astrophysics \textbar{} Harvard \& Smithsonian, 60 Garden Street, Cambridge, MA 02138-1516, USA\\
$^{19}$The NSF AI Institute for Artificial Intelligence and Fundamental Interactions, USA\\
$^{20}$Las Cumbres Observatory, 6740 Cortona Drive, Suite 102, Goleta, CA 93117-5575, USA\\
$^{21}$Department of Physics, University of California, Santa Barbara, CA 93106-9530, USA\\
$^{22}$Cardiff Hub for Astrophysics Research and Technology, School of Physics \& Astronomy, Cardiff University, Queens Buildings, The Parade, Cardiff, CF24 3AA, UK\\
$^{23}$Universidad Nacional de Río Negro. Sede Andina, Mitre 630 (8400) Bariloche, Argentina\\
$^{24}$Department of Physics and Astronomy, University of California, Davis, 1 Shields Avenue, Davis, CA 95616-5270, USA\\
$^{25}$Birmingham Institute for Gravitational Wave Astronomy and School of Physics and Astronomy, University of Birmingham, Birmingham B15 2TT, UK \\
$^{26}$Astrophysics Research Centre, School of Mathematics and Physics, Queen's University Belfast, Belfast BT7 1NN, UK \\
$^{27}$Instituto de Astrof\'{i}sica, Departamento de F\'{i}sica, Universidad Andres Bello, Avda. Rep\'{u}blica 252, 8320000 Santiago, Chile\\
$^{28}$Mt Stromlo Observatory, The Research School of Astronomy and Astrophysics, Australian National University, ACT 2611, Australia\\
$^{29}$National Centre for the Public Awareness of Science, Australian National University, ACT 2601, Australia\\
$^{30}$The ARC Centre of Excellence for All-Sky Astrophysics in 3 Dimension (ASTRO 3D), Australia\\
$^{31}$Physics Department, Tsinghua University, Beijing, 100084, China\\
$^{32}$Beijing Planetarium, Beijing Academy of Sciences and Technology, Beijing, 100044, China\\
}

\date{Accepted 2023 June 13. Received 2023 June 12; in original form 2023 February 15}

\pubyear{2023}

\begin{document}
\label{firstpage}
\pagerange{\pageref{firstpage}--\pageref{lastpage}}
\maketitle
\clearpage

\begin{abstract}

Hydrogen-rich Type II supernovae (SNe~II) are the most frequently observed class of core-collapse SNe (CCSNe). However, most studies that analyse large samples of SNe~II lack events with absolute peak magnitudes brighter than $-$18.5~mag at rest-frame optical wavelengths. Thanks to modern surveys, the detected number of such luminous SNe~II (LSNe~II) is growing. There exist several mechanisms that could produce luminous SNe~II. The most popular propose either the presence of a central engine (a magnetar gradually spinning down or a black hole accreting fallback material) or the interaction of supernova ejecta with circumstellar material (CSM) that turns kinetic energy into radiation energy. In this work, we study the light curves and spectral series of a small sample of six LSNe~II that show peculiarities in their H$\alpha$ profile, to attempt to understand the underlying powering mechanism. We favour an interaction scenario with CSM that is not dense enough to be optically thick to electron scattering on large scales --- thus, no narrow emission lines are observed. This conclusion is based on the observed light curve (higher luminosity, fast decline, blue colours) and spectral features (lack of persistent narrow lines, broad H$\alpha$ emission, lack of H$\alpha$ absorption, weak or nonexistent metal lines) together with comparison to other luminous events available in the literature. We add to the growing evidence that transients powered by ejecta-CSM interaction do not necessarily display persistent narrow emission lines.

%It should be a single paragraph not more than 250 words (200 words for Letters).
%No references should appear in the abstract.
\end{abstract}


%%%%%%%%%%%%%%%%%%%%%%%%%%%%%%%%%%%%%%%%%%%%%%%%%%

%%%%%%%%%%%%%%%%% BODY OF PAPER %%%%%%%%%%%%%%%%%%

\section{Introduction}
\label{sec:intro}

Type II supernovae (SNe~II) arise from the core-collapse-induced explosion of massive stars (zero age main sequence mass $\gtrsim 8$--10~M$_{\sun}$). This supernova (SN) type is characterised by the presence of prominent hydrogen features throughout their entire spectral evolution\footnote{There exists a class of transitional events that show prominent hydrogen spectral lines at early times that disappear soon after light-curve peak \citep{Filippenko1997,2017hsn..book..195G}. This class is known as Type IIb and will not be considered on this work.} \citep{1941PASP...53..224M}. For such features to be observed, the progenitors of SNe~II must have retained most of their hydrogen envelopes before explosion. There are several subclassifications within the Type II family based on spectral or photometric properties. Events whose spectral evolution shows persistent, relatively narrow emission lines are classified as Type IIn \citep{1990MNRAS.244..269S}. SNe with slowly rising light curves, resembling that of SN~1987A, are classified as 87A-like (\citealt{2012A&A...537A.141P}, \citealt{2016A&A...588A...5T}, and references therein). Events that show a peak absolute magnitude brighter than $\sim -$20~mag in the $V$ band are classified as Type II superluminous SNe (SLSNe~II; see \citealt{2019ARA&A..57..305G} for a review). SNe displaying a ``plateau'' in their light curves were historically classified as Type IIP, while those displaying fast linearly declining (in magnitudes) light curves were historically classified as Type IIL \citep{1979A&A....72..287B}. Recent works have found a continuum of observed properties in the light curves of the SN~IIP/IIL subtypes, arguing against the division and for simply considering these as SNe~II \citep{2014ApJ...786...67A, 2015ApJ...799..208S, 2016AJ....151...33G, 2016ApJ...820...33R, 2016MNRAS.459.3939V, 2019MNRAS.490.2799D}. Throughout this work, we will refer to the mentioned historically studied SNe~IIP/IIL as ``regular'' SNe~II or simply SNe~II.

Although there are several systematic studies of samples of regular SNe~II that consider increasing numbers of events through the years, most of these works do not include objects with rest-frame light-curve peaks brighter than $\sim -$18.5~mag in the $V$ band \citep[e.g.][]{2014ApJ...786...67A,2016MNRAS.459.3939V}. Yet, there exist a growing number of such events. We will refer to these as luminous SNe~II (LSNe~II). Such objects were already noticed by \cite{1994A&A...282..731P}, who studied a sample of 51 SNe~II and observed the existence of a gap between regular SNe~II and brighter ($\lesssim -$18.5 mag in the $B$ band) events. According to \cite{1994A&A...282..731P}, these more-luminous SNe~II display both fast and slow (linear and plateau) light-curve declines and were theoretically predicted by \cite{1991ApJ...374..266S} after studying three different models: carbon deflagration of a near-Chandrasekhar C-O core, electron-capture-induced collapse of a Chandrasekhar mass O-Ne-Mg core, and Fe core collapse of a massive star. They concluded that carbon deflagration can be ruled out for SNe~II and that fast-declining events (SNe~IIL) can be explained by the electron-capture-induced collapse of an O-Ne-Mg core formed after the star underwent helium enhancement of its envelope by core penetration and dredge-up. \citet{1991ApJ...374..266S} mentioned that if the proposed model is correct, mass loss must have reduced the initial envelope mass of the star. Although electron capture SNe models typically suggest low explosion energy and ejecta velocity for these events \citep[e.g.][]{2017hsn..book..483N}, this mechanism has recently been invoked for LSNe~II \citep[][see Section~\ref{sec:possiblescena}]{2022MNRAS.509.2013Z}.

There are many theoretical works that compare observations to hydrodynamical models of hydrogen-rich SN explosions. One of the most recent is that of \cite{2022A&A...660A..42M}. These authors compare a large number of observations of regular SNe~II with a grid of modeled light curves and velocities, concluding that the explosion energy is the main driver of much of the observed light-curve diversity, with higher explosion energies producing more-luminous events \citep[see][for another example]{2009ApJ...703.2205K}. However, they do not have LSNe~II in their analysis. While it would be tempting to assume that LSNe~II arise from progenitors that present similar characteristics to those of regular SNe~II but which exploded with larger energies, it should be noted that similar events can be reproduced considering different combinations of progenitor radius, mass, and energy \citep[e.g.,][]{1985SvAL...11..145L}. The SN explosion energy promptly transforms into kinetic energy that drives the ejecta and radiation energy. Some of the radiation energy is lost at early times due to photon trapping in the optically thick ejecta. The duration of the SNe somewhat depends on the trapping timescale. The luminosity of the SNe can be calculated from the radiation energy diffusion rate. If a large amount of $^{56}$Ni is produced during the explosion, its radioactive decay can become a significant contributor to the SN powering mechanism and boost the observed luminosity. Besides a large production of $^{56}$Ni, a variety of alternative powering-mechanism scenarios have been proposed to explain the features observed in recent LSNe. One possibility is the presence of a central engine in which part of the energy that powers the light curve arises from the thermalisation of the energy produced by the gradual spindown of a central magnetar or from the accretion of fallback material into a central black hole. Another possibility is the interaction of the SN ejecta with circumstellar material (CSM), in which the kinetic energy of the outflow is thermalised by the interaction shock and then radiated \citep[see][for a review of alternative powering sources]{2017hsn..book..939K}.

There is extensive evidence that SNe~II undergo CSM interaction shortly after explosion, which explains the diversity observed in their early-time light curves \citep[e.g.,][]{2015MNRAS.451.2212G,2018NatAs...2..808F,2020ApJ...891L..32M}. Depending on the characteristics of this interaction, the spectral features and overall luminosity of an event could be affected \citep[e.g.,][]{2019A&A...631A...8H}. If the surrounding CSM is dense enough, the spectral series will show persistent narrow lines. However, the absence of such persistent narrow lines does not necessarily rule out a CSM interaction scenario \citep[e.g.,][]{2011ApJ...729L...6C,2012ApJ...747..118M,2018MNRAS.477...74A,2019A&A...631A...8H}. In particular, \cite{2012ApJ...747..118M} argue that the diversity in the density slope of a wind produced by nonsteady mass loss can account for spectral differences observed in LSNe~II. If CSM is present, this could reprocess the radiation from the SN and release it on a diffusion timescale, which would result in broad-boxy emission features. If this CSM is not dense enough to be optically thick to electron scattering, narrow lines will not be visible \citep[][and references therein]{2022A&A...660L...9D}.

Two of the best-observed hydrogen-rich LSNe are SN~1979C \citep{1981ApJ...244..780B,1981PASP...93...36D} and SN~1998S \citep{2000A&AS..144..219L,2000MNRAS.318.1093F}. The former has long been considered a prototype of fast-declining SNe~II, although it is more luminous than most regular SNe~II \citep[see, for example,][]{2014AJ....147..118R}. The latter has been considered as a prototype of SNe~IIn although it loses its narrow emission lines within 10 days \citep{2000ApJ...536..239L,2015ApJ...806..213S,2016MNRAS.458.2094D}. For this reason, \cite{2017hsn..book..403S} suggests that SN~1998S is part of a transitional group of SNe~IIn where narrow lines could be missed if sufficiently early observations do not exist. \citet{2017A&A...605A..83D} question whether a SN~II should be classified as SN~IIn if narrow lines can only be seen for a few days. Such lines could be missed if follow-up observations are not started early enough with respect to the explosion. There is, in fact, a large fraction of SNe~II that do show narrow lines only in their early-time spectra \citep{2017NatPh..13..510Y,2021ApJ...912...46B}. It has been proposed that both SN~1979C and SN~1998S interact with CSM. In the case of SN~1979C, \cite{1984A&A...132....1F} analysed ultraviolet (UV) observations and concluded that the observed spectral lines were formed in a constant-velocity shell close to the photosphere. Later, \cite{1993A&A...273..106B} proposed that the peak brightness results from reradiation of UV light into optical wavelengths produced by the presence of a dense stellar wind. 
Late-time radio observations of the SN remnant of SN~1979C have aided to uncover the CSM structure around it \citep[e.g.,][]{2000ApJ...532.1124M,2008ApJ...682.1065B}; these data provide strong support for the CSM powering mechanism interpretation. In the case of SN~1998S, it is widely accepted that the narrow lines indicate CSM interaction \citep[e.g.,][and references therein]{2016MNRAS.458.2094D,2017hsn..book..403S}, even if they are seen for only a short period of time; the density, morphology, and distribution of the CSM should be different for events that display different narrow-line features. 

The exact magnitude at which an event is considered to be an SLSN instead of an LSN is arbitrary, and it is not clear whether a continuum exits between them \citep[e.g.,][]{2016ApJ...819...35A, 2018MNRAS.475.1046I,2019MNRAS.487.2215A}. A well-known example of hydrogen-rich SLSN is SN~2008es \citep{2001MNRAS.325..907F,2009ApJ...690.1313G,2009ApJ...690.1303M}. This event is often considered an archetype of the class. It shows spectral evolution similar to that of some LSNe~II \citep[see][for an example]{2020MNRAS.493.1761R}, and its overall characteristics resemble those observed in SN~1979C. The extreme luminosity of SN~2008es ($M_{V} = -$22.3 mag) has been explained invoking CSM interaction. \cite{2019MNRAS.488.3783B} argued in favour of a CSM interaction powering mechanism and disfavoured a magnetar scenario based on the analysis of the late-time bolometric light curve. Other SLSNe have also been proposed to be powered by some degree of CSM interaction. One remarkable exception is OGLE-2014-SN-073 \citep{2017NatAs...1..713T}; its luminosity has been better explained by the presence of a magnetar. However, the morphology of the OGLE-2014-SN-073 light curve is more similar to that of 87A-like SNe~II. These morphologies can be reproduced including a magnetar, although some CSM might be necessary at early phases in some cases \citep{2018A&A...613A...5D,2018A&A...619A.145O}. Nevertheless, 87A-like SNe~II are outside the scope of this work.

Given that there are only a few studies of LSNe~II in the literature and that the powering mechanisms necessary to produce these events are still under debate (as exemplified by the cases of SN~1979C and SN~1998S), we started a follow-up campaign to obtain photometric and spectroscopic data of LSNe~II to attempt to constrain their powering mechanism and progenitor properties. We obtained a sample of 35 LSNe~II (see Section~\ref{sec:observations}) that display a large variety of spectroscopic and photometric features. A diversity in observed features usually hints toward the need for more than one physical interpretation of the explosion scenarios. Therefore, in this work we concentrate on a subsample of events that show common properties in their H$\alpha$ profiles (specifically, a lack of absorption and evidence for multiple emission components -- see Section~\ref{sec:sampleprops}). Given that the H$\alpha$ profile is the most class-defining feature of SNe~II and given its importance for interpreting the explosion and spectral line formation conditions (e.g., \citealt{2014ApJ...786L..15G}), we assume that similarities in this feature imply similarities in the underlying powering mechanisms. The considered subsample includes six LSNe~II for which we present optical light curves and spectral series. The paper is structured as follows. Section~\ref{sec:observations} describes the observations and data reduction. In Section~\ref{sec:sampleprops} we characterise the sample, which we analyse in Section~\ref{sec:analy}. A discussion of the observed features is given in Section~\ref{sec:discussion}. We summarise our conclusions in Section~\ref{sec:conclusion}.

\section{Observations and data reduction}
\label{sec:observations}

The majority of the events in this study were observed through the extended Public ESO Spectroscopic Survey for Transient Objects (ePESSTO, an extension of the project described by \citealt{2015A&A...579A..40S}). The discovery and classification of each event was done by different surveys; the respective discovery and classification reports for the presented subsample are cited in each LSN~II subsection. To obtain our sample, we regularly inspected the ePESSTO Marshall \citep{2015A&A...579A..40S}, searching for SNe classified as Type II (by ePESSTO or others) that had an initial estimate of absolute magnitude brighter than $-$18.5~mag at optical wavelengths and apparent magnitudes brighter than the ePESSTO follow-up limit of $\sim$ 20~mag. The ePESSTO Marshall is ideal for these kind of tasks since it provides the user with a detailed overview of each transient event by cross-correlating all the associated metadata available from various sources  (e.g., SN photometric measurements from different surveys, probable host-galaxy associations and the resulting distance estimates, etc). We obtained a sample of 35 SNe observed between 2017 and 2019. The bright nature of the peak absolute magnitude of each event was corroborated by our own analysis (see Section~\ref{sec:sampleprops}). The aim of our project is to produce a systematic characterisation of LSNe~II to understand the powering mechanism behind their higher luminosities. We noticed that one of our follow-up targets, SN~2018bsz, was initially misclassified and is a Type I SLSN instead of a Type II event \citep{2018ATel11674....1A}. Therefore, we removed it from the sample. The analysis of SN~2018bsz was presented by \cite{2018A&A...620A..67A} and \cite{2021arXiv210907942C}. The remaining 34 events, to which we will refer as the ``full sample,'' display a large diversity of light-curve morphologies and spectral-evolution features.

To better explore the involved powering mechanisms, we focus the present study on six LSNe~II that stand out from the rest of the objects in the full sample because of their spectral properties (see Section~\ref{sec:sampleprops}). A thourough analysis of the full sample will be presented in a future work. Our follow-up campaign made use of different observing facilities. In Section \ref{subsec:optspec} we describe the facilities involved in the spectral observations and the applied spectral reduction techniques. Section \ref{subsec:optphot} presents the same for photometric observations.

\subsection{Optical spectroscopy}
\label{subsec:optspec}

We have a total of 71 spectra for our six LSNe~II. The median phase of the first observed spectrum for the sample is $\sim 14$ days after explosion (see Section~\ref{sec:sampleprops} for details of how explosion epochs were derived), and the median phase of the last observed spectrum is $\sim 82$ days after explosion. The spectral log is given in Table~\ref{tab:lum-specs}. Most of the spectra were obtained with the ESO Faint Object Spectrograph and Camera (EFOSC2) mounted on the 3.6~m New Technology Telescope (NTT) as part of ePESSTO, using mostly the grism Gr$\#$13 (3685--9315~\AA) but also the grisms Gr$\#$11 (3380--7520~\AA) and Gr$\#$16 (6015--10,320~\AA)\footnote{Specific properties for each of the EFOSC2 grisms can be found in the dedicated ESO instrument webpage, \url{https://www.eso.org/sci/facilities/lasilla/instruments/efosc/inst/Efosc2Grisms.html}}. Data reduction was performed using the ePESSTO dedicated pipeline\footnote{\url{https://github.com/svalenti/pessto}} as described by \citet{2015A&A...579A..40S}, following standard procedures. Some spectra were obtained through Las Cumbres Observatory (LCO; \citealt{2013PASP..125.1031B}) as part of both ePESSTO and the ``Global Supernova Project'' (GSP). The LCO facilities were particularly useful for bright events having a declination such that they were not observable by the NTT. In these cases the monitoring was done exclusively by GSP. The reduction of LCO spectra was performed by the above-mentioned project using a PyRAF-based dedicated pipeline\footnote{\url{https://github.com/LCOGT}}. We restrict the analysis of these spectra to the 4800--9000~\AA\ region to avoid noisy edges. When available, public spectra from other sources were also included.

Although we present all obtained spectra, in order to be able to make a meaningful comparative analysis we only study spectra with signal-to-noise ratio (S/N) $\geq 5.5$. The S/N was measured using the {\sc IRAF}\footnote{{\sc IRAF} is distributed by the National Optical Astronomy Observatory, which is operated by the Association of Universities for Research in Astronomy, Inc., under cooperative agreement with the U.S. National Science Foundation.} \citep{1986SPIE..627..733T,1993ASPC...52..173T} routine  \textit{splot} at the continuum near the H$\alpha$ emission profile.

\subsection{Photometry}
\label{subsec:optphot}

We present optical photometry in a number of different bands for our six LSNe~II. The median first $V$-band photometric point for the sample was observed $\sim 15.5$ days after explosion and the median last $V$-band photometric point was observed $\sim 90$ days after explosion. Consequently, maximum $V$-band brightness was not observed for any of the presented LSNe~II and the first observed photometric point was considered as the SN peak. Moreover, most of them do not have observations of the radioactive tail that typically sets in around 100 days after explosion in regular SNe~II. Imaging in the $griBV$ optical bands was obtained with the LCO 1.0~m telescope network as part of both ePESSTO and the ``Las Cumbres Observatory SN Key Project.'' Bias and flatfield correction was performed automatically with the {\sc BANZAI} pipeline \citep{2022ascl.soft07031M}. Differential aperture photometry using a fixed radius was extracted through a self-developed code that implements the routines available in the {\sc Python} package {\sc photutils} \citep{larry_bradley_2020_4044744}. The images were calibrated using the the ATLAS All-Sky Stellar Reference Catalog \citep[Refcat2;][]{2018ApJ...867..105T}. Bands $gri$ are presented in the catalog's system, but $BV$ bands are presented in the Johnson system and were obtained using the transformations given by \cite{2012ApJ...750...99T}. Template subtractions were not achieved owing to the lack of template images, although the effects of this should only be important at late phases that are generally unimportant for our analysis. LCO photometry is listed in Table~\ref{tab:lum-photlcogt}. When available, we present $o$-band (corresponding to roughly the $r + i$ range) ATLAS survey photometry \citep{2018PASP..130f4505T} obtained from the ATLAS forced-photometry server\footnote{\url{https://fallingstar-data.com/forcedphot/}}. The ATLAS plotted and tabulated values are the error-weighted mean values from the four measurements from each night. For one of the studied SNe, we present $gri$ SkyMapper photometry. Although filters from different surveys do not necessarily have the same efficiency curves, the changes are negligible for our analysis purposes and thus no corrections are applied. SkyMapper photometry was extracted from images from the Transient Survey \citep{2017PASA...34...30S,2019IAUS..339....3M} taken using the set of SDSS-like $griz$ filters available in the telescope. Images were reduced using the difference-imaging pipeline described by \cite{2017PASA...34...30S} and calibrated using APASS DR7. The ATLAS and SkyMapper photometry is given in Table~\ref{tab:lum-photatlas} and Table~\ref{tab:lum-photskymap}, respectively.

\section{Sample description}
\label{sec:sampleprops}

\begin{figure}
	\includegraphics[width=\columnwidth]{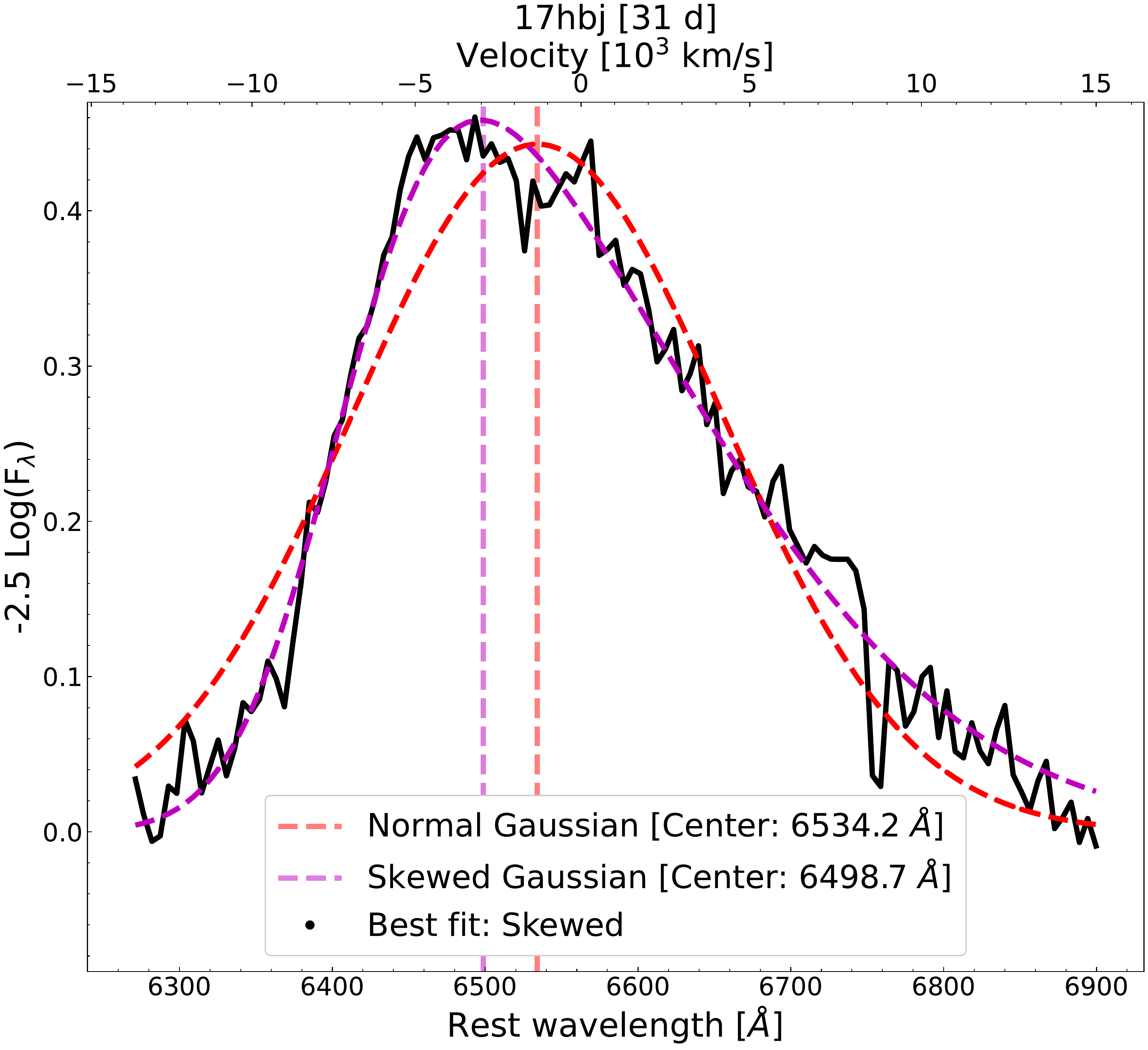}
    \caption{Example of Gaussian fit to the H$\alpha$ emission line in the spectrum of SN~2017hbj taken at $\sim 31$ days. The red dashed line shows a normal Gaussian fit, while the purple dashed line is a skewed Gaussian fit. In both cases the vertical line represents the centre of the fit. The best fit was selected based on the AIC criterion to be the skewed Gaussian. This is consistent with the blueshift observed in the peak of the H$\alpha$ feature.}
    \label{fig:geg}
\end{figure}

\begin{figure*}
	\includegraphics[width=\textwidth]{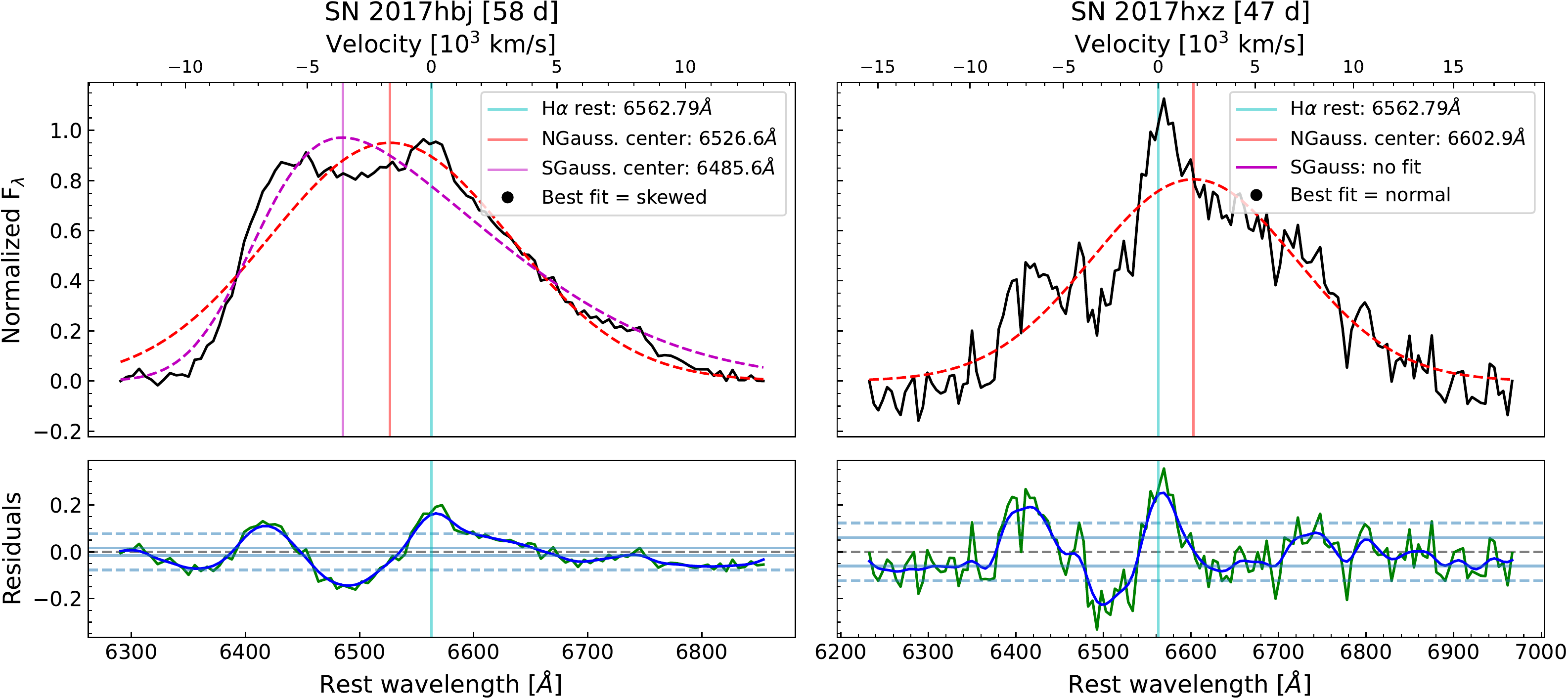}
    \caption{Example of residual excess blueward of the H$\alpha$ rest wavelength. Top panels show the line fitting; the fitting is not expected to accurately reproduce the observed feature as it is only used to obtain evidence of multiple components from the observed residuals. The red lines represent the normal Gaussian (NGauss.) fit while magenta lines display the skewed Gaussian (SGauss.) fit. Note that in the top right panel an SGauss. model was performed but no fit was found. Red and magenta vertical lines indicate the centre of the former and latter fit, respectively. Cyan vertical lines mark the rest-wavelength position of H$\alpha$. Bottom panels display the residuals with respect to the best fit. The residuals are shown as green lines, and the gaussian convolution of the residuals is shown as blue lines. Light-blue horizontal solid lines illustrate the root-mean-squared uncertainty of the residuals while light blue horizontal dashed lines show the standard deviation ($\sigma$). If the convolution of the residuals exceeds 1$\sigma$ we consider it to be an excess. An excess can be seen in both bottom panels.}
    \label{fig:eg_excess}
\end{figure*}

As already mentioned, to define our sample we first selected SNe~II brighter than $\sim -$18.5 mag in $V$ from the ePESSTO Marshall (34 SNe) discovered between April 2017 and April 2019. We then inspected the available spectra. Out of the 34 SNe in the full sample, 18 show at least one spectral phase in which the absorption component of the H$\alpha$ P-cygni profile is detectable. There are 10 SNe with no visually detected absorption in their H$\alpha$ feature. The quality of the spectral follow up  (either number of observed spectral phases or S/N) of the remaining events is not good enough to confirm a detection of the absorption component. After fitting a Gaussian profile to the H$\alpha$ emission profiles of the SNe with no detected H$\alpha$ absorption, we note that out of these events, six show a blue excess. At this point, we are not able to asses whether this blue excess should be expected in every event with no H$\alpha$ absorption because of low number statistics. The excess was identified using a quantitative method. The method consists of fitting a single Gaussian profile to the H$\alpha$ feature of each spectrum using the {\sc Python} package {\sc lmfit}. Some emission profiles are better fit by a single skewed Gaussian model instead of a single normal Gaussian (see Fig.~\ref{fig:geg}); this might be a consequence of the typical blueshift observed in SN~II emission peaks \citep{2014MNRAS.441..671A}. Thus, every emission feature was fit with both a skewed and a normal Gaussian model. The best fit was selected based on the Akaike information criterion (AIC; \citealt{1100705}). The AIC estimator is used to compare models rewarding goodness of fit while penalising an increment on the number of estimated parameters. After selecting the best fit to the H$\alpha$ emission, we analysed the associated residuals. These residuals were convolved with a Gaussian of standard deviation equal to the resolution element of the EFOSC2's gr$\#$13 grism. If the resulting convolution exceeds the standard deviation of the residuals, we consider the excess to be produced by the presence of an additional (or multiple additional) feature in the emission profile (see Fig.~\ref{fig:eg_excess}). This method allow us to find the epoch at which the excess appears. Note that our goal is not to accurately reproduce the shape of the studied feature but to evaluate the residuals of a single component Gaussian fit to assess the possible evidence of (at least) an extra component. 

We focus our analysis on the six spectroscopically distinct LSNe~II: SN~2017cfo, SN~2017gpp, SN~2017hbj, SN~2017hxz, SN~2018aql, and SN~2018eph. Their general properties are listed in Table~\ref{tab:info}. The SN redshifts ($z$) were obtained from underlying host \ion{H}{II} regions present in the SN spectrum, except in the case of SN~2018aql for which there is no other available information than that obtained by spectral matching performed using the Supernova Identification (SNID; \citealt{2007ApJ...666.1024B}) software. SNID was also used to find events among the template sample that could resemble those presented here. We note that although we found some matches (see below), the phases are not accurate as our SNe are often quite distinct (spectroscopically) from the template events in SNID. Absolute magnitudes were calculated by correcting for Milky Way extinction obtained from the NASA Extragalactic Database (NED)\footnote{The NASA/IPAC Extragalactic Database (NED) is operated by the Jet Propulsion Laboratory, California Institute of Technology, under contract with the National Aeronautics and Space Administration.} assuming a reddening law with $\mathrm{Rv} = 3.1$. The distance moduli (DM) were calculated using the estimated $z$ and the Cosmology Calculator III\footnote{\url{http://faraday.uwyo.edu/~chip/misc/Cosmo2/cosmo.cgi .}} provided by NED, adopting NED's cosmological parameters (H$_{0} = 73$ km s$^{-1}$ Mpc$^{-1}$, $\Omega _{\mathrm{Matter}} = 0.27$, $\Omega _{\mathrm{Lambda}} = 0.73$). Note that we do not correct for host-galaxy extinction. This is because (a) there are no reliable methods for doing so, especially for unusual SNe~II as presented here; (b) we do not see signs of strong \ion{Na}{I}~D interestellar absorption, and (c) these objects are generally blue (see Section~\ref{sec:colorsBV}). Additional correction for host extinction would simply make the sample even brighter and bluer than regular SNe~II, thus strengthening our results below. 

In every case the explosion epoch is considered to be the midpoint between the last nondetection and the first detection, adopting half of this range as the uncertainty. All mentioned phases are considered in rest-frame days after the respective explosion epoch unless stated otherwise. Given that none of the SNe in our sample show a $V$-band maximum, the presented peak magnitude, considered to be that of the first observed photometric point, is a lower limit. Gaussian process (GP) interpolations, performed using the {\sc Python} package {\sc GPy}\footnote{\url{https://gpy.readthedocs.io/en/deploy/}.}, were used to estimate the $V$-band decline rates as $100 \times (m_{50\mathrm{d}} - m_{\mathrm{peak}}) / (t_{50\mathrm{d}} - t_{\mathrm{peak}})$, where $m_{50\mathrm{d}}$ and $m_{\mathrm{peak}}$ are the $V$ magnitudes at 50 days past explosion and at peak (respectively), and $t_{50\mathrm{d}}$ and $t_{\mathrm{peak}}$ are the phases in days since explosion of the corresponding magnitudes.

All of the selected LSNe~II show initially fast declining light curves (decline rate faster than 1.4~mag/(100~d), a limit that has been used in the literature to separate slow- and fast-declining SNe~II; see \citealt{2019ApJ...887....4D}, and references therein).
\cite{2014ApJ...786L..15G} note that luminous and fast declining SNe~II show a weak H$\alpha$ absorption component. The H$\alpha$ profile of our subsample of LSNe~II at no point in the spectral evolution shows (noticeable) signs of the typical absorption component seen in regular SNe~II (by design of our selection criteria). Instead, the H$\alpha$ profile exhibits only an emission component that broadens with time. The broadening is such that at a given time in the spectral evolution a single Gaussian, typically used to fit and characterise SN spectral lines, becomes insufficient to fit the profile. At this time we assume that the profile shows (at least) one additional component in the emission that is seen as a blue excess in the H$\alpha$ feature (see fourth column of Table~\ref{tab:info} for the phase of the first excess). In addition, they exhibit an emission feature at $\sim 5800$~\AA\ that is identified as \ion{He}{I} (see Section \ref{sec:heid}). Below we present a short description of each object in the sample. We then describe general characteristics and make comparisons with regular SNe~II (focusing on measurements presented for the ``Carnegie Supernova Project'' sample, \citealt{2006PASP..118....2H, 2019PASP..131a4001P}, see Section~\ref{sec:analy}). In Section~\ref{sec:discussion} we include comparisons with other LSNe~II from the literature to better understand the observed features. 

\begin{table*}
  \caption{General information for each LSN~II} 
  \begin{threeparttable}
  \setlength{\tabcolsep}{4pt}
    \begin{tabular}{lcccccccclc}
%\tablenum{3}
\hline
Object        & Exp. date     & 1st spec & $t_{1}$ H$\alpha$ excess & 1st $V$       & $M_{V}$\tnote{$^{a}$} & Decline rate\tnote{$^{b}$} & $z$     & $A_{V}^{\mathrm{MW}}$  &  host~galaxy &  DM\tnote{$^{c}$}  \\
              & [MJD]         & [d]      & [d]                      & [d]           & [mag]                 & [mag/(100~d)]               &        & [mag]  &       &   [mag]   \\
\hline                                                                                 
SN 2017cfo & 57822.2(5.2)  & 14.6     & 46.2                    & 16.7          & -19.0(0.2)           &  4.2                      & 0.042  & 0.066   &  SDSSJ103812.75+280704.0     &  36.2 \\      
SN 2017gpp & 57995.0(1.0)  & 11.5     & 84.2                    & $\cdots$      & $\cdots$             &  $\cdots$                 & 0.058  & 0.045   &  2MASXJ22074707-4412416      &  37.0 \\     
SN 2017hbj & 58023.5(5.5)  & 12.6     & 31.4                    & 8.3           & -18.3(0.2)           &  3.9                      & 0.018  & 0.095   &  ESO084-G021	              &  34.4 \\  
SN 2017hxz & 58048.0(5.0)  & 19.8     & 35.5                    & 21.6          & -19.4(0.2)           &  8.7                      & 0.076  & 0.128   &  GALEXASCJ033410.88-135616.7 &  37.6  \\  
SN 2018aql & 58193.0(13.0) & 22.3     & 51.7                    & 26.5          & -19.1(0.2)           &  2.9                      & 0.074  & 0.052   &  SDSSJ165705.00+392253.8     &  37.5  \\     
SN 2018eph & 58331.2(3.1)  & 4.1      & 64.6                    & 10.7          & -18.8(0.2)           &  1.6                      & 0.029  & 0.066   &  2MASXiJ0455502-614521       &  35.4  \\
\hline
\end{tabular}
\end{threeparttable}
\label{tab:info}
  \begin{tablenotes}
    \small
    \item $^{a}$ $V$-band magnitude of the first observed photometric point (adopted as peak magnitude).  
    \item $^{b}$ Decline rate calculated between first observed photometric point and 50 days.
    \item $^{c}$ DM obtained from the calculated $z$. The associated error owing to host-galaxy peculiar velocities is conservatively assumed to be 0.2~mag in every case (following NED's guide for use of cosmology calculator \url{https://ned.ipac.caltech.edu/help/objresult_help.html#DerivedValues}). 
  \end{tablenotes}
\end{table*}

\subsection{SN~2017cfo}

\begin{figure}
	\includegraphics[width=\columnwidth]{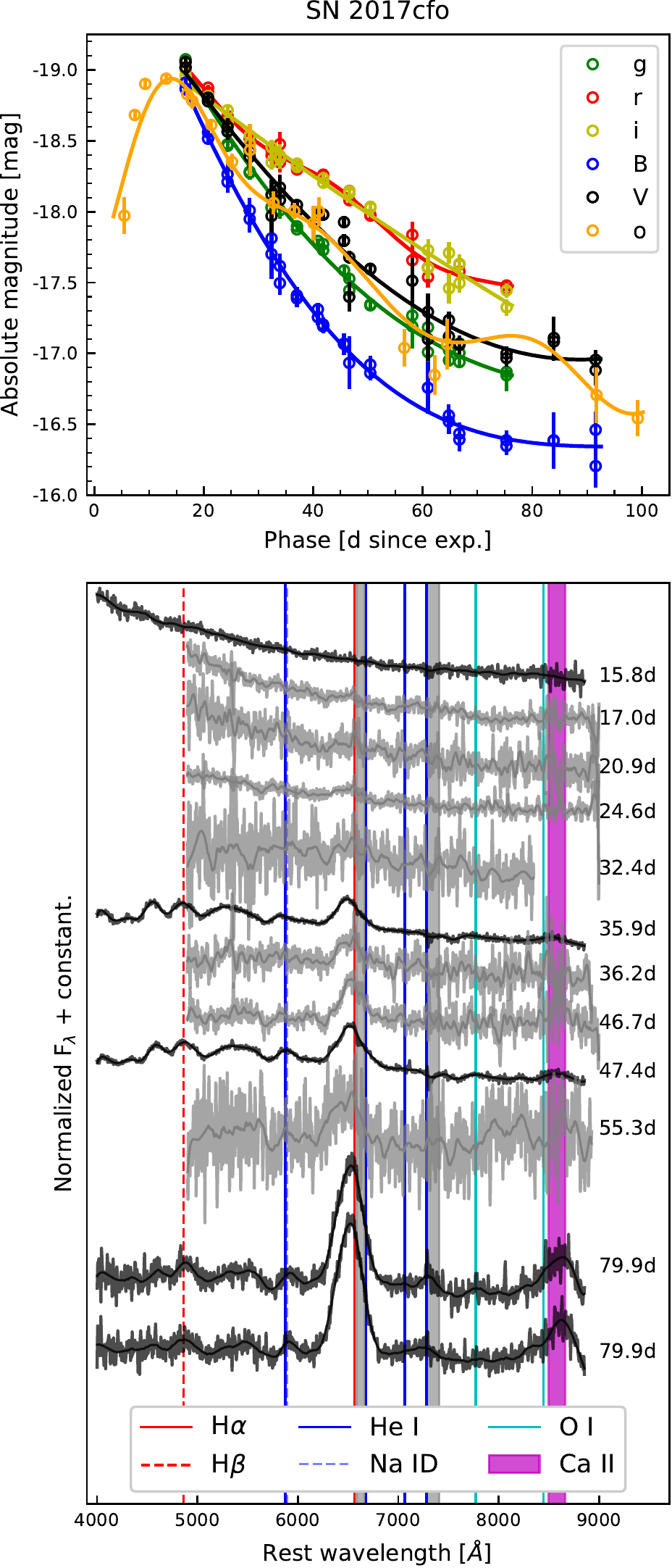}
    \caption{SN~2017cfo. {\it Top panel:} optical light curves. The photometric points are presented as dots. GP fits are presented as solid lines. {\it Bottom panel:} spectral evolution. Thin solid lines show Savitzky-Golay smoothing \citep{1964AnaCh..36.1627S}. Phase in rest-frame days after explosion is annotated to the right of each spectrum. NTT spectra are plotted in black while LCO spectra are plotted in grey. Grey vertical regions indicate the locations of strong telluric lines \protect\citep{2015A&A...576A..77S}.}
    \label{fig:evspec17cfo}
\end{figure}

\cite{2017TNSTR.320....1T} reported the discovery of SN~2017cfo on 17 March 2017, although ATLAS observations exist from 15 March 2017. \cite{2017ATel10212....1F} classified this event as a possible SN while \cite{2017ATel10225....1P} provided the Type II classification. At early times, no good visual spectral matches are found using SNID. However, at later phases ($> 45$ days) we find good agreement with SN~1979C and SN~1998S.

The spectral time series is presented in the bottom panel of Fig.~\ref{fig:evspec17cfo}. The first spectrum was obtained at $\sim 15$ days by ePESSTO; it is rather blue and featureless. Hints of H$\alpha$ can be seen at 17 days, although no good spectral match is found when using SNID, possibly because of the low S/N in the spectrum. At 17 days a feature appears to be visible at $\sim 5800$~\AA, and this line is seen more clearly at $\sim 36$ days, remaining present throughout the rest of the observed evolution. We identify this line as He~{\sc I} (see Section~\ref{sec:heid} for further discussion). At $\sim 36$ and 47 days, a feature is seen near 7770~\AA\ and can be identified as O~{\sc I}. At $\sim 36$ days the Ca~{\sc II} near-infrared triplet (NIR3) starts to be visible and becomes stronger with time. From $\sim 36$ days onward, H$\beta$ can be seen in all spectra that have sufficient wavelength coverage. 

The available $ogriBV$ photometry for SN~2017cfo is presented in the top panel of Fig.~\ref{fig:evspec17cfo}\footnote{The Legacy Survey Data Release 7 (\url{https://www.legacysurvey.org/dr7/description/}) includes photometry for SN~2017cfo but we do not consider it because it is sparse and would not significantly impact our results}. We note that the peak of the light curve is only observed in the $o$ band. The first observed $V$-band photometric point was obtained at $\sim 17$ days and shows an absolute magnitude of $-19.0 \pm 0.1$~mag. SN~2017cfo declines fast after peak, showing a decline rate of 4.2~mag/(100~d) between the first observed photometric point and 50 days. The light curves decline almost monotonically for $\sim 75$ days, after which the $BV$ light curves show a flattening, which could be produced by contamination from host-galaxy light.

\subsection{SN~2017gpp}

\begin{figure}
	\includegraphics[width=\columnwidth]{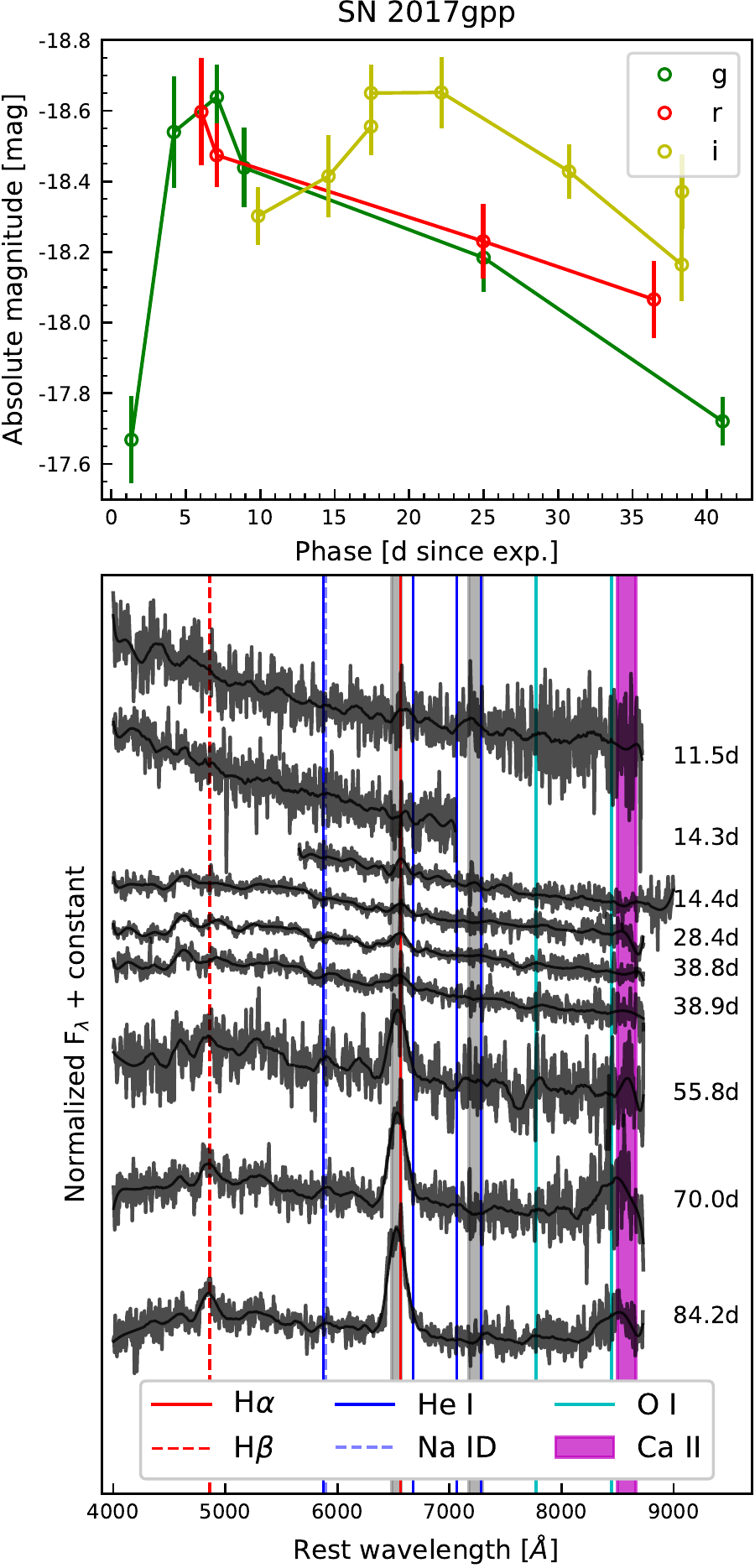}
    \caption{{\small SN~2017gpp. {\it Top panel:} optical light curves. The photometric points are presented as dots connected by solid lines; GP is not used here to avoid smoothing out the second peak of the $i$-band light curve. {\it Bottom panel:} spectral evolution. Thin solid lines show Savitzky-Golay smoothing \citep{1964AnaCh..36.1627S}. Phase in rest-frame days after explosion is annotated to the right of each spectrum. NTT spectra are plotted in black. Grey vertical regions indicate the locations of strong telluric lines \protect\citep{2015A&A...576A..77S}.}}
    \label{fig:evspec17gpp}
\end{figure}

SN~2017gpp was discovered on 31 August 2017 by \cite{2017TNSTR.974....1M}, who report a last nondetection on 29 August 2017. It was classified 11 days later by \cite{2017ATel10732....1G} as ``other.'' Based on the hydrogen spectral features, we classify this event as an SN~II. There are narrow lines on top of the H$\alpha$ emission profile during the full spectral evolution; they are consistent with poorly subtracted emission from host-galaxy \ion{H}{II} regions. 

The spectral time series is presented in the bottom panel of Fig.~\ref{fig:evspec17gpp}. The first spectrum was obtained at $\sim 12$ days by ePESSTO; it is rather blue but has low S/N. Two additional spectra were taken at $\sim 14$ days; the one with higher S/N shows a clear H$\alpha$ feature, although running SNID for this spectrum results in no match. The H$\alpha$ feature evolves slowly and is only prominent after $\sim 56$ days. At $\sim 70$ days there seems to be fairly weak H$\alpha$ absorption, although it could be an artifact of the noise. Hints of H$\beta$ can be seen in all the spectra, although the S/N is not ideal. We note the presence of a telluric region right on top of the H$\alpha$ emission, but this does not appear to affect the evolution or the ``boxiness'' of the feature. At $\sim 39$ days there seems to be a hint of He~{\sc I} (see Section~\ref{sec:heid} for further discussion), but again the S/N is not ideal for identification. The Ca~{\sc II} NIR3 can be seen in the last two available spectra, for which SNID shows SN~1998S as a good match, although the H$\alpha$ emission does not fit well visually. 

The available $gri$ photometry for SN~2017gpp is presented in the top panel of Fig.~\ref{fig:evspec17gpp}. Unfortunately there are no $V$ observations available for this object. Nevertheless, the peak absolute magnitude in $g$ is $-18.6 \pm 0.2$ mag, and the $g$ band is close enough to $V$ to assume that SN~2017gpp is also more luminous than $\sim -$18.5 mag in $V$. We note an odd behaviour of the $i$-band light curve. While both the $g$ and $r$ light curves evolve very fast, the $i$ light curve seems to take $\sim 10$ days longer to reach the maximum. The origin of this later and wider light curve peak is unknown. It is impossible to assess if a previous peak exists in the $i$ band since we do not have earlier observations. We note that \cite{2015MNRAS.451.2212G} find that some regular SNe~II also exhibit a late $i$-band maximum.

\subsection{SN~2017hbj}

\begin{figure}
	\includegraphics[width=\columnwidth]{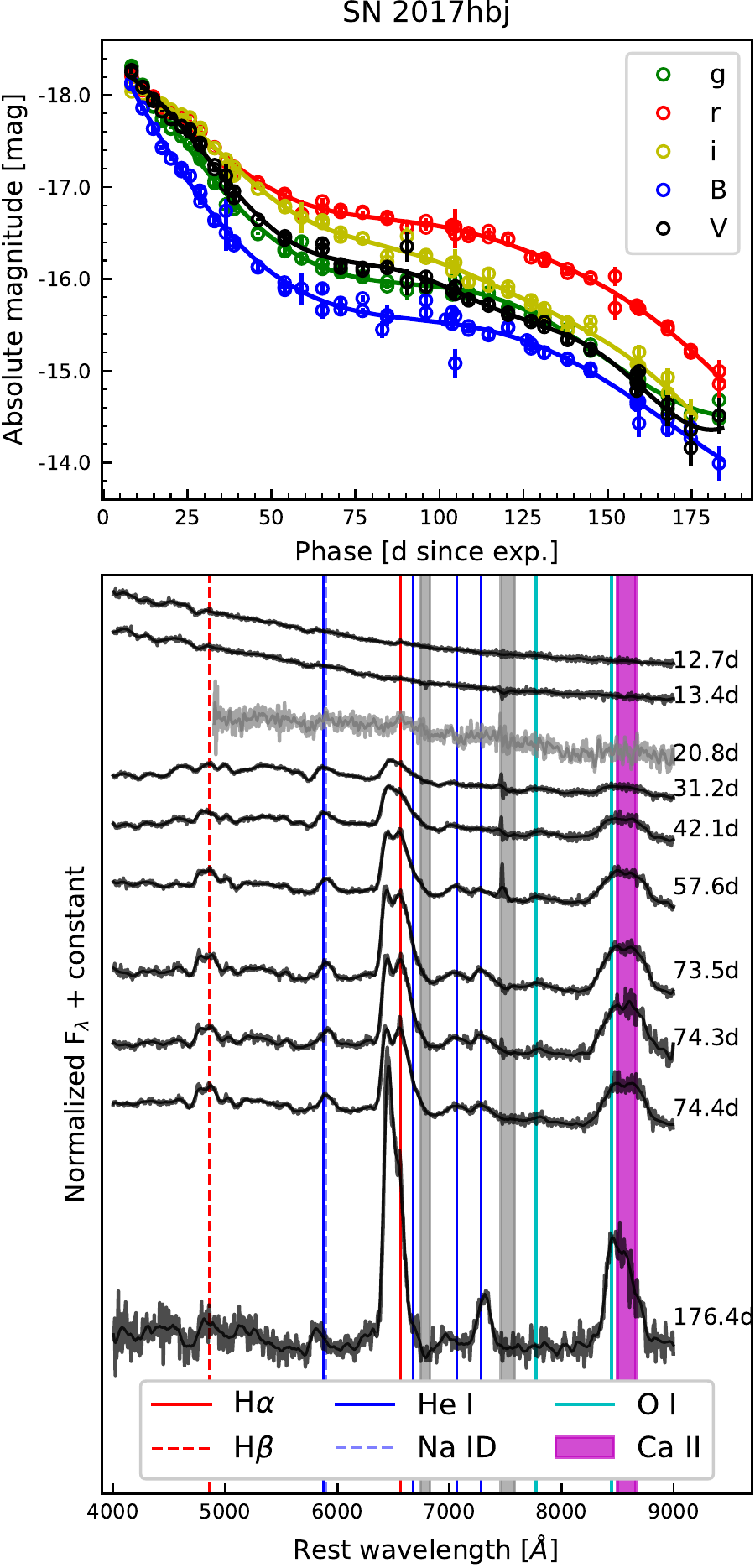}
    \caption{SN~2017hbj. {\it Top panel:} optical light curves. The photometric points are presented as dots. GP fits are presented as solid lines. {\it Bottom panel:} spectral evolution. Thin solid lines show Savitzky-Golay smoothing \citep{1964AnaCh..36.1627S}. Phase in rest-frame days after explosion is annotated to the right of each spectrum. NTT spectra are plotted in black while LCO spectra are plotted in grey. Grey vertical regions indicate the locations of strong telluric lines \protect\citep{2015A&A...576A..77S}.}
    \label{fig:evspec17hbj}
\end{figure}

SN~2017hbj was discovered on 3 October 2017 by \cite{2017TNSTR1067....1P}. The first spectrum was obtained by \cite{2017ATel10836....1K} on 9 October 2017 and was used to classify the SN as Type II. Our study shows that the first available spectrum presents an acceptable visual match in SNID with the Type II SN~2014G. For later phases SNID shows relatively good visual matches with SN~1979C and SN~1998S, although the H$\alpha$ feature is not well matched.

The spectral time series is presented in the bottom panel of Fig.~\ref{fig:evspec17hbj}. The available early-time spectra are rather blue, presenting a small number of very weak features. The H$\alpha$ and H$\beta$ emission profiles become strong at $\sim 31$ days. At $\sim 58$ days, both these profiles develop a trough that creates a red and blue peak, with the red peak being stronger at early times and the blue peak being much stronger at late times. At $\sim 31$ days the He~{\sc I} $\lambda$5876 and Ca~{\sc II} NIR3 features are clearly visible, and remain observable for the rest of the spectral evolution, with the Ca~{\sc II} NIR3 feature becoming more prominent with time. 

The available $griBV$ photometry is presented in the top panel of Fig.~\ref{fig:evspec17hbj}. The ePESSTO Marshall's first-order absolute peak magnitude was estimated to be $-$18.4~mag; given the closeness of this value to our selection limit, we decided to include this SN in our sample.  Note that the peak of the light curve is not observed in any band. The first $V$-band photometric point was obtained at $\sim 8$ days and has an absolute magnitude of $-18.3 \pm 0.1$~mag. While this is dimmer than our selection criteria, we choose to keep this event in our analysis given that the SN shows similar properties to the other five in the sample. SN~2017hbj declines at a rate of 3.9~mag/(100~d) between the first observed photometric point and 50 days. The light curves in all photometric bands decline monotonically up to $\sim 65$ days, at which time there is a subtle flattening that produces a small slowdown of the decline rate, followed by a second change of decline rate at $\sim 125$ days.

\subsection{SN~2017hxz}

\begin{figure}
	\includegraphics[width=\columnwidth]{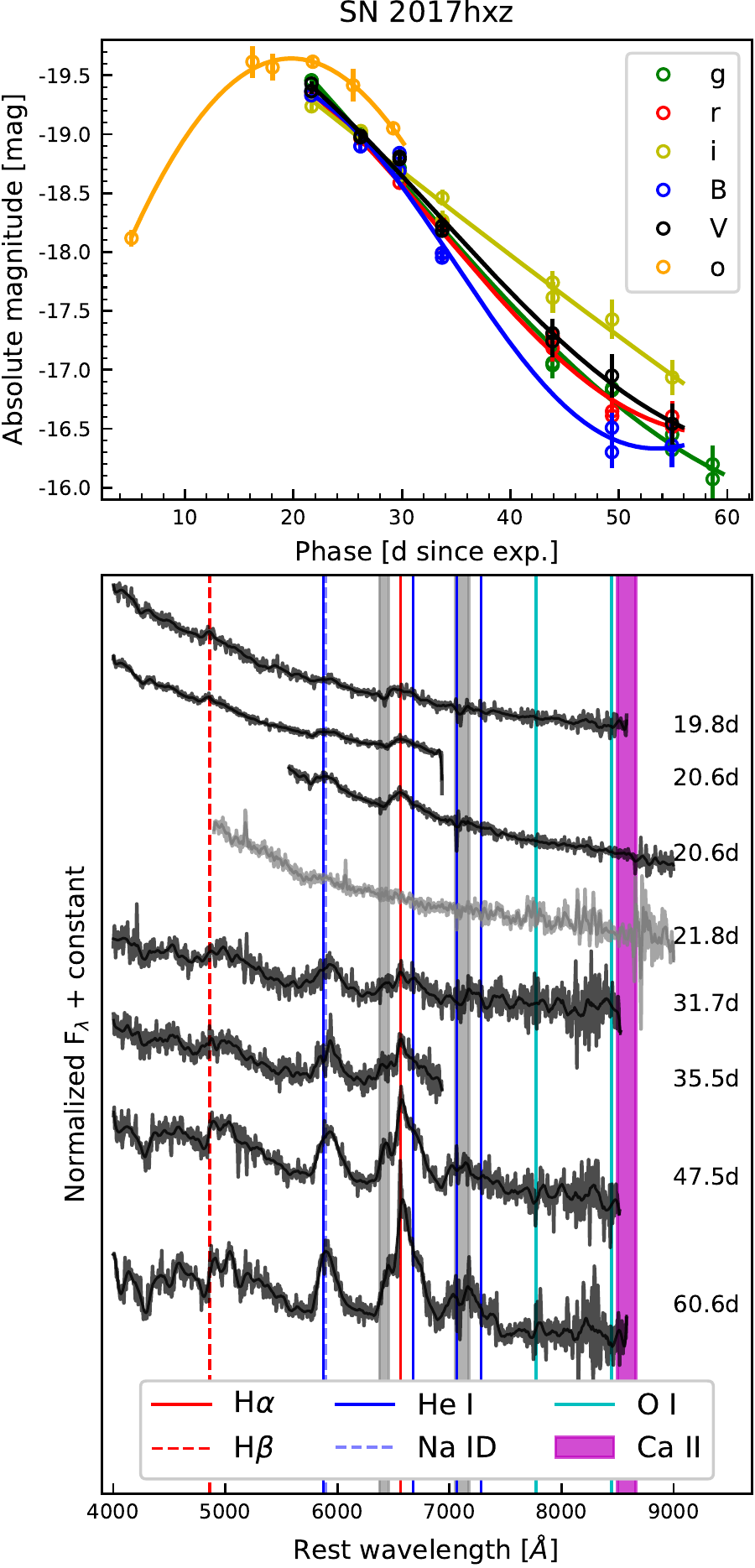}
    \caption{SN~2017hxz. {\it Top panel:} optical light curves. The photometric points are presented as dots. GP fits are presented as solid lines. {\it Bottom panel:} spectral evolution. Thin solid lines show Savitzky-Golay smoothing \citep{1964AnaCh..36.1627S}. Phase in rest-frame days after explosion is annotated to the right of each spectrum. NTT spectra are plotted in black while LCO spectra are plotted in grey. Grey vertical regions indicate the locations of strong telluric lines \protect\citep{2015A&A...576A..77S}.}
    \label{fig:evspec17hxz}
\end{figure}

SN~2017hxz was discovered on 10 November 2017 by \cite{2017ATel10960....1B}. However, there exists a previous detection on 27 October 2017 by ATLAS, which obtained deeper observations than those reported in the discovery alert. The classification as an SN~II was reported by \cite{2017ATel10961....1C}. Our study shows that at early times ($\lesssim 10$ days), SNID produces a decent visual spectral match with SN~2012aw, although at late times no good visual match is found.

The spectral time series is presented in the bottom panel of Fig.~\ref{fig:evspec17hxz}. The first spectrum was obtained at $\sim 20$ days. The spectrum is still rather blue at this epoch. At $\sim$ 32 days, a narrower emission appears on top of the H$\alpha$ profile. The next available spectrum is the first one to show an H$\alpha$ blue excess. This and all the following spectra exhibit a sharp trough on top of the H$\alpha$ profile that divides it into a blue and red side. Both sides evolve with time, although the red much more than the blue. Note that the last (and possibly the second to last) spectrum seems to have an additional trough on top of the red side that could indicate further components. The first available spectrum shows He~{\sc I} $\lambda$5876 (see Section~\ref{sec:heid} for further discussion). This feature evolves with time and becomes comparable in strength to H$\alpha$ at late epochs. Not many other metallic lines are visible during the spectral evolution.  

The available $ogriBV$ photometry is presented in the top panel of Fig.~\ref{fig:evspec17hxz}. Note that the peak of the light curve is only observed in the $o$ band. The first observed $V$-band photometric point was obtained at $\sim$ 21.6 days, $\sim$ 1.3 days after the $o$-band maximum, and shows an absolute magnitude of $-19.4 \pm 0.1$ mag. SN~2017hxz is a fast decliner; the light curves in all the available photometric bands decline almost monotonically, and the $V$ light curve has a decline rate of 8.7~mag/(100~d) between the first observed photometric point and 50 days, making this SN the fastest decliner of the sample. For further discussion of SN~2017hxz see Section~\ref{sec:17hxzstandsout}.

\subsection{SN~2018aql}

\begin{figure}
	\includegraphics[width=\columnwidth]{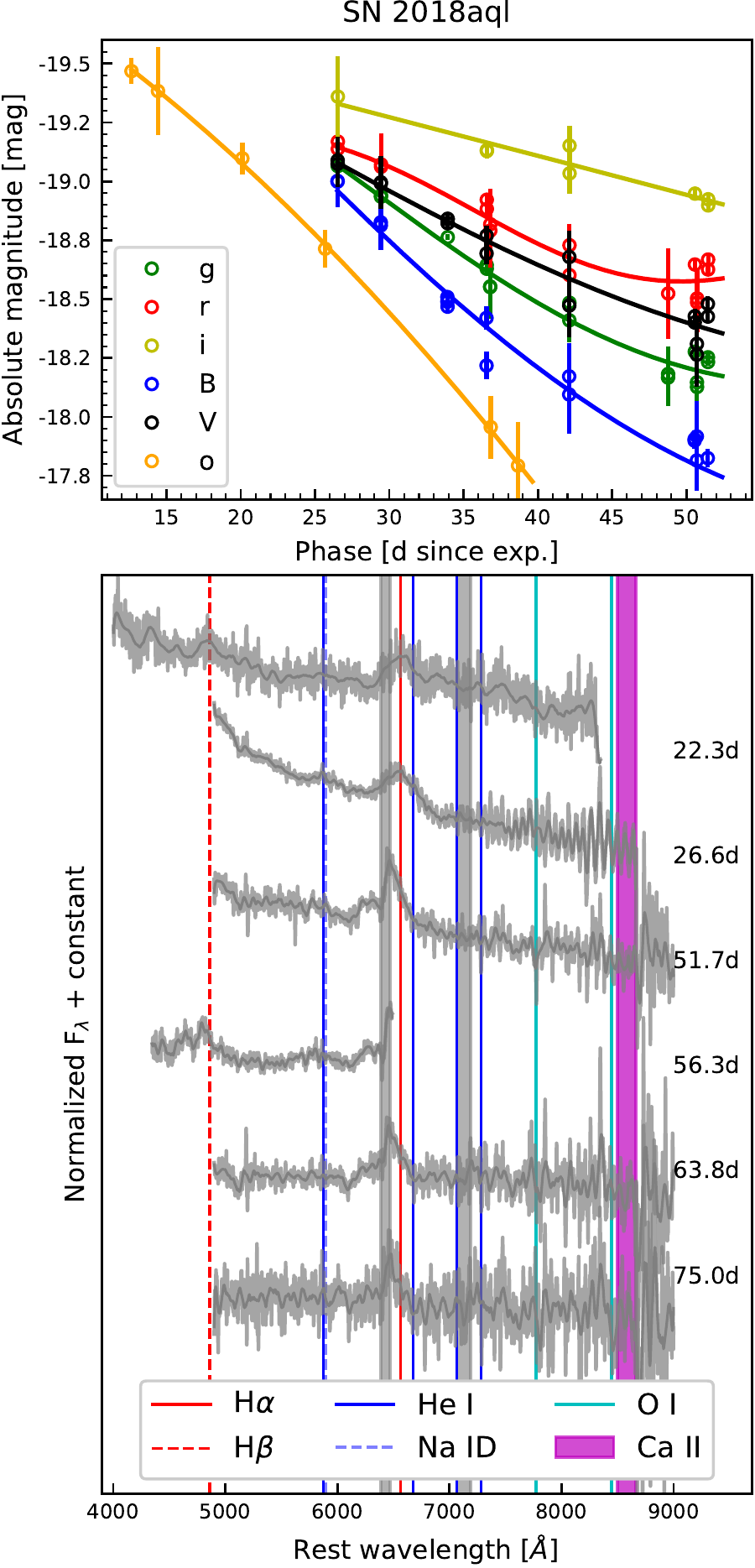}
    \caption{SN~2018aql. {\it Top panel:} optical light curves. The photometric points are presented as dots. GP fits are presented as solid lines. {\it Bottom panel:} spectral evolution. Thin solid lines show Savitzky-Golay smoothing \citep{1964AnaCh..36.1627S}. Phase in rest-frame days after explosion is annotated to the right of each spectrum. LCO spectra are plotted in grey. Grey vertical regions indicate the locations of strong telluric lines \protect\citep{2015A&A...576A..77S}.}
    \label{fig:evspec18aql}
\end{figure}

The discovery of SN~2018aql was reported on 6 April 2018 by \cite{2018TNSTR.455....1X}. However, ATLAS provides better limiting magnitude constraints on the last nondetection (3 March) and the first detection (29 March). The classification as an SN~II was reported by \cite{2018ATel11525....1Z}. Our study shows that the first two spectra match well with several known SNe~II in SNID. From $\sim$ 52 days onward, no good visual matches can be found.

The spectral time series is presented in the bottom panel of Fig.~\ref{fig:evspec18aql}. The first available spectrum is the one used for classification, obtained by \cite{2018ATel11525....1Z} $\sim$ 22 days after explosion. The spectrum shows H$\alpha$ and H$\beta$, as well as a subtle hint of He~{\sc I} (see Section~\ref{sec:heid} for further discussion). The declination of this SN is out of NTT's observability range; hence, the whole spectral time series was obtained through LCO. At $\sim$ 52 days, the top of the H$\alpha$ emission presents a sharp trough similar to the one observed in SN~2017hxz at earlier epochs (see above), although SN~2018aql does not show prominent He~{\sc I} features at late times. We note that the trough is present under a telluric region. However, the difference in strength of the H$\alpha$ emission at each side of the trough and the similarity to SN~2017hxz suggest that the observed trough is real, not related to the telluric correction. There are no distinguishable metallic lines throughout the spectral evolution, although the S/N is low, the red part of each spectrum being particularly noisy. 

The available  $ogriBV$ photometry is presented in the top panel of Fig.~\ref{fig:evspec18aql}. The first $V$-band photometric point was observed at $\sim$ 26 days and has an absolute magnitude of $-19.1 \pm 0.2$~mag. Given the position of SN~2018aql in the host galaxy, we can assume the photometry is contaminated by the host, which is reflected in the large photometric error bars. Nevertheless, the brightness decline seems to be consistent throughout all the observed bands. In particular, the $V$ band declines at a rate of 2.9~mag/(100~d) between the first observed photometric point and 50 days, consistent with fast-declining SNe. 

\subsection{SN~2018eph}

\begin{figure}
	\includegraphics[width=\columnwidth]{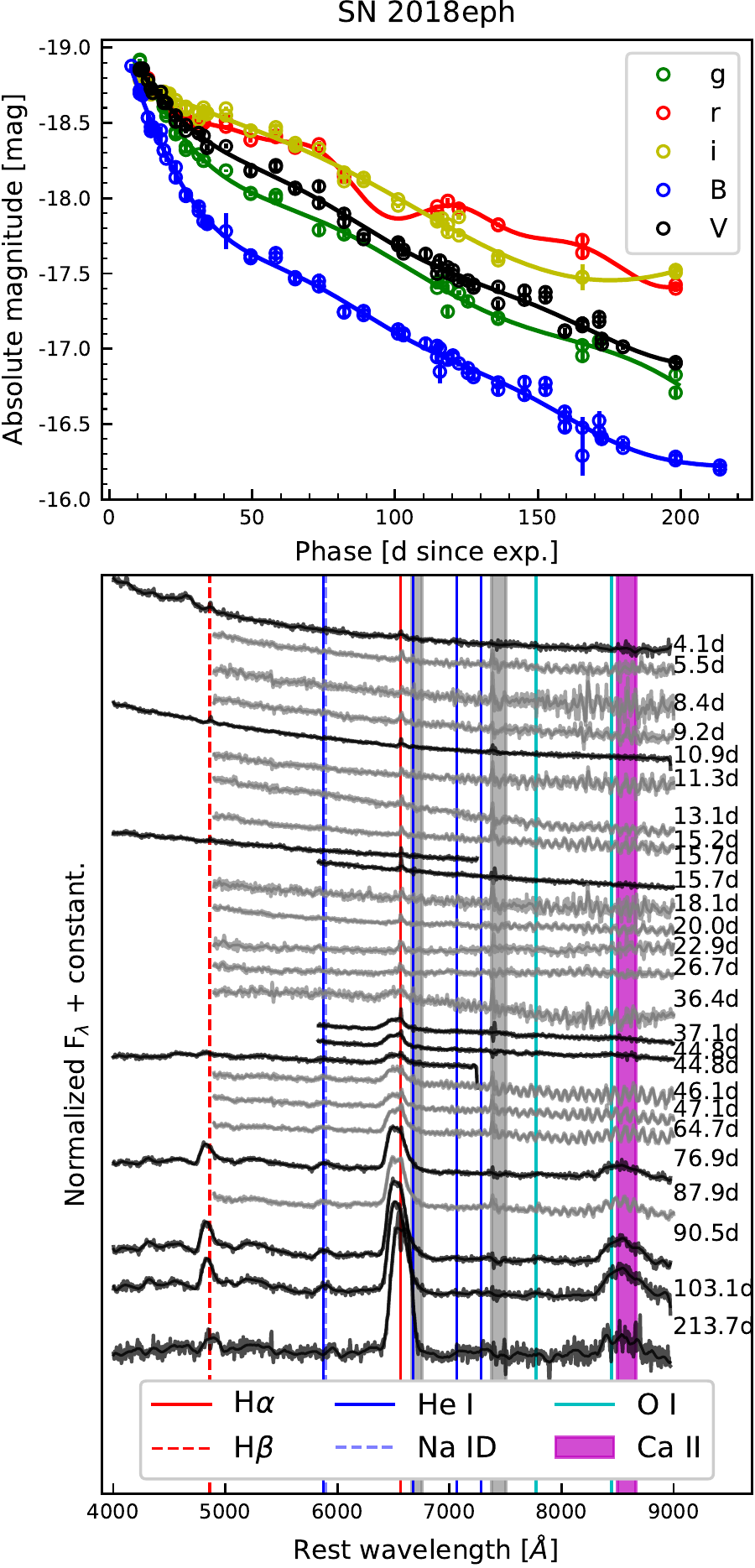}
    \caption{SN~2018eph. {\it Top panel:} optical light curves. The photometric points are presented as dots. GP fits are presented as solid lines. {\it Bottom panel:} spectral evolution. Thin solid lines show Savitzky-Golay smoothing \citep{1964AnaCh..36.1627S}. Phase in rest-frame days after explosion is annotated to the right of each spectrum. NTT spectra are plotted in black while LCO spectra are plotted in grey. Grey vertical regions indicate the locations of strong telluric lines \protect\citep{2015A&A...576A..77S}.}
    \label{fig:evspec18eph}
\end{figure}

SN~2018eph was discovered on 4 August 2018 by \cite{2018ATel11933....1B} who also report a last nondetection on 29 July 2018. It was classified as an SN~II the next day by \cite{2018ATel11916....1O}. The early-time spectra of SN~2018eph are rather blue and featureless. The spectrum obtained at 4 days show flash spectroscopy features at the bluer end. The first two spectra show a relatively good visual match to SN~1998S using SNID. No other good visual match is found from these phases up to $\sim$ 37 days when SNID provides a relatively good visual match with SN~2004fc. At late epochs, SNID gives good visual matches with SN~1979C and SN~1998S.

The spectral time series is presented in the bottom panel of Fig.~\ref{fig:evspec18eph}. SN~2018eph has the best spectral follow-up observations of our sample, with the first spectrum obtained at $\sim$ 4 days and the last at $\sim$ 214 days; no other SN in the present sample has a spectrum obtained as early or as late. Yet, there is barely any spectral evolution until $\sim$ 30 days. At $\sim$ 37 days, H$\alpha$ becomes strong and hints of He~{\sc I} $\lambda$5876 and Ca~{\sc II} NIR3 are detectable. At $\sim$ 65 days, the emission profile becomes boxy. H$\alpha$, H$\beta$, He~{\sc I}, and Ca~{\sc II} NIR3 become stronger with time. Not many metallic lines are visible during the spectral evolution. A narrow H$\alpha$ emission line can be seen throughout the entire spectral evolution. This feature is consistent with host-galaxy contamination.

The available $griBV$ photometry\footnote{SN~2018eph has also been observed by the Transiting Exoplanet Survey Satellite (TESS). Analysis of the TESS data is presented by \cite{2021MNRAS.500.5639V} and is not included in this work.} is presented in the top panel of Fig.~\ref{fig:evspec18eph}. The ePESSTO Marshall's first-order absolute peak magnitude was estimated to be $-$19.3 mag. The peak of the light curve is not observed in any band. The first $V$-band photometric point was obtained at 10.7 days and has an absolute magnitude of $-18.8 \pm 0.1$~mag. SN~2018eph declines at a rate of 1.6~mag/(100~d) between the first observed photometric point and 50 days, which positions it near the lower end of the fast-declining SNe (considering a limit of 1.4~mag/(100~d) as mentioned above). The light curves in all photometric bands decline monotonically with time.

\section{Analysis}
\label{sec:analy}

After characterising our sample above, we now analyse different aspects of the dataset below to further elucidate the nature of these SNe.

\subsection{$B-V$ colours}
\label{sec:colorsBV}

\begin{figure}
	\includegraphics[width=\columnwidth]{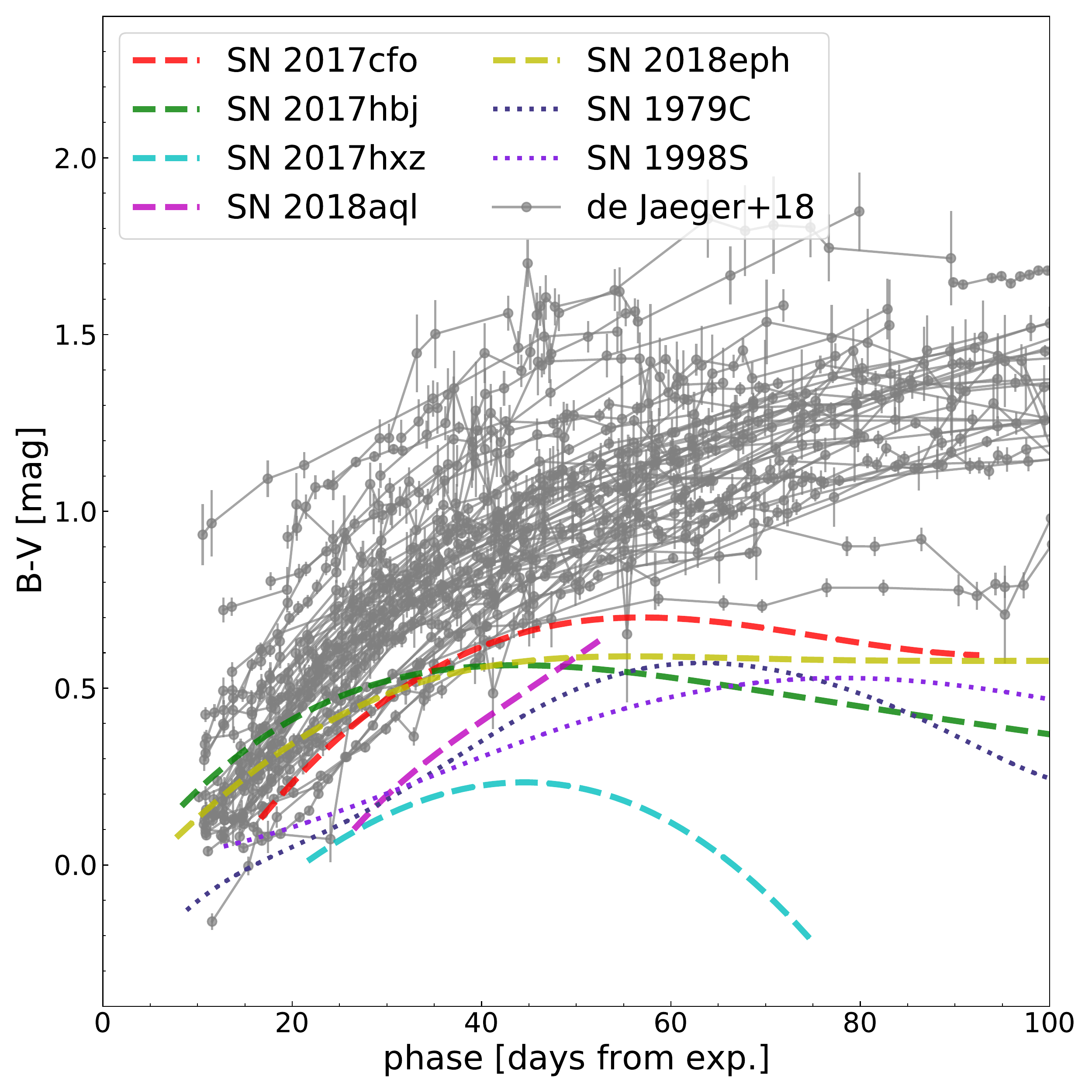}
    \caption{$B-V$ colours of the LSNe~II included in this sample are presented with coloured dashed lines. SN~1979C and SN~1998S are included in the comparison with coloured dotted lines, since these are often the closest spectral matches to our sample as given by SNID. Following the work of \protect\cite{2020MNRAS.493.1761R}, we adopt the parameters of \protect\cite{1981PASP...93...36D,1982A&A...116...43B} and \protect\cite{2000ApJS..128..431F} for SN~1979C and the parameters of \protect\cite{2000MNRAS.318.1093F} for SN~1998S. $B-V$ colours of the sample of SNe~II studied by \protect\cite{2018MNRAS.476.4592D} are presented in grey for comparison.}
    \label{fig:colors}
\end{figure}

When possible, we computed $B-V$ colours making use of the Gaussian process interpolation for each LSN~II in our sample. The obtained $B-V$ colours were then compared to those of the regular SN~II sample studied by \cite{2018MNRAS.476.4592D}, presented in their Figure 13. The resulting comparison is shown in Fig.~\ref{fig:colors}. It can be seen that, at early times, the LSNe~II in our sample are among the bluest end of regular SNe~II and overall stay bluer than regular SNe~II as time evolves. 
Unlike the comparison SNe~II, the LSNe~II colour curves seem to reach a phase where the evolution stalls and the colours remain constant or even start becoming blue again (except maybe for SN~2018aql, although the photometry does not cover phases later than $\sim$ 55 days). It can also be seen that SN~1979C and SN~1998S show similar behaviour. Note that the photometry presented in this work was not host-subtracted nor corrected for intrinsic host-galaxy extinction. We consider the former as a caveat, although the sample of \cite{2018MNRAS.476.4592D} was also not corrected for intrinsic host-galaxy extinction. In fact, the authors conclude that colours might be dominated by differences in the photospheric temperature. Further analysis will be performed in the full sample to evaluate if this is also the case for LSNe~II.

\begin{figure}
	\includegraphics[width=\columnwidth]{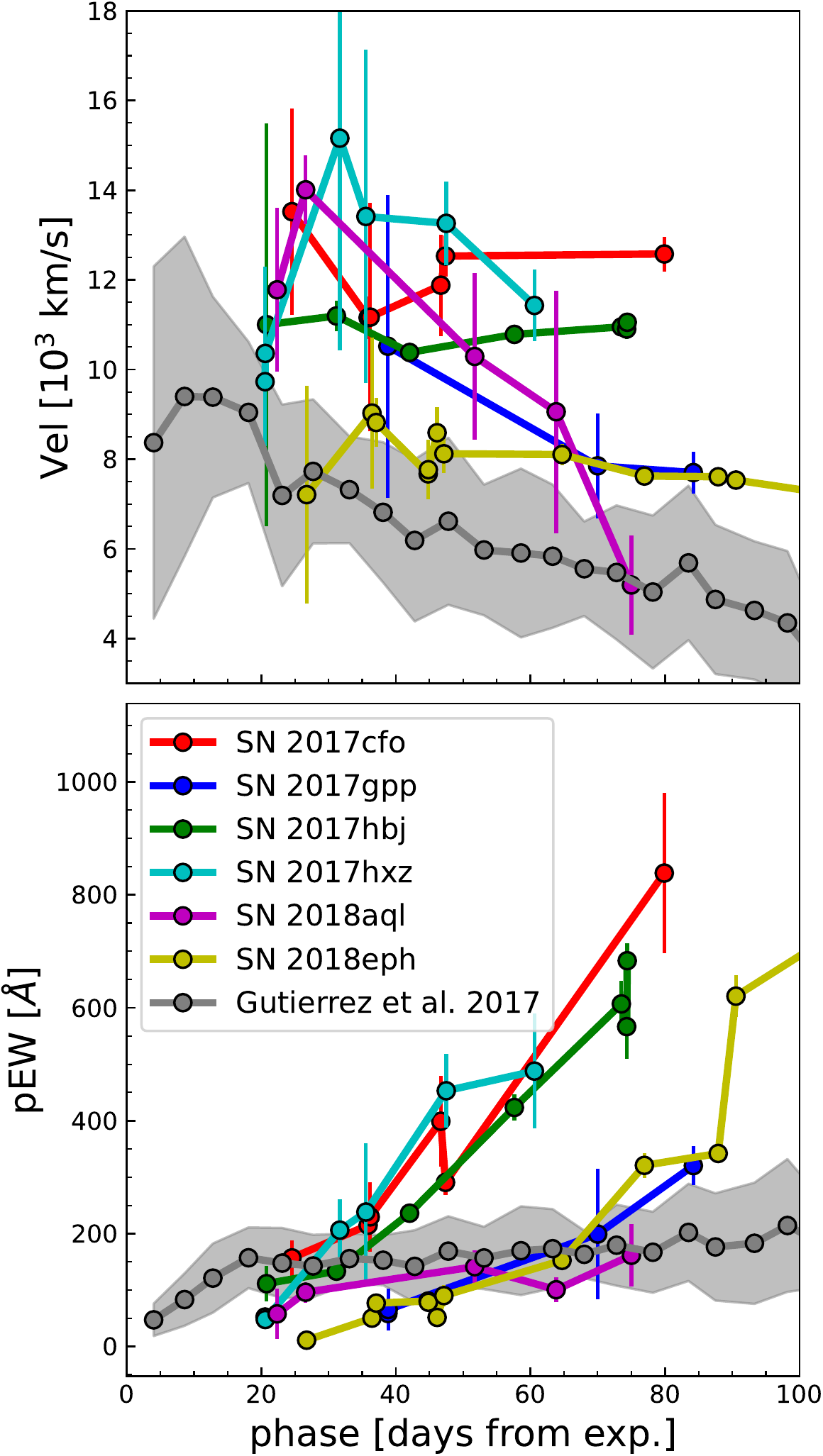}
    \caption{{\it Top panel:} H$\alpha$ FWHM velocities. {\it Bottom panel:} H$\alpha$ pseudo-equivalent widths. The values for the sample of LSNe~II are presented in colour. Mean values for regular SNe~II from the sample of \protect\cite{2017ApJ...850...89G} are presented in grey.}
    \label{fig:velpew}
\end{figure}

\subsection{H$\alpha$ velocities and pseudo-equivalent widths.}

Considering the peculiar shape of the H$\alpha$ emission features of the studied LSN~II sample, we followed the work of \cite{2014ApJ...786L..15G} and make use of the full width at half-maximum intensity (FWHM) of the H$\alpha$ emission profile to estimate velocities. Gaussian fitting was performed using models available in the {\sc lmfit} library to each emission profile. To obtain the velocity value and its associated uncertainty, we first selected the left and right edges (e$_{\mathrm{left}}$ and e$_{\mathrm{right}}$, respectively) of the feature by eye and then took a window of 5~\AA\ on each side of each edge ([e$_{\mathrm{left}}-5$, e$_{\mathrm{left}}+5$] and [e$_{\mathrm{right}}-5$, e$_{\mathrm{right}}+5$], respectively). We used these windows to define different traces of the continuum considering all possible combinations of the wavelengths contained in the right and left windows with a step of 1~\AA. Multiple Gaussian fits were obtained considering each of the resulting traces of the continuum. The mean FWHM of the fits was used to calculate the velocities. The respective standard deviation was considered to be the associated uncertainty. The results can be seen in the top panel of Fig.~\ref{fig:velpew}. We also show the H$\alpha$ FWHM velocities of the sample of SNe~II studied by \cite{2017ApJ...850...89G} for comparison. Overall, the velocities of the studied LSNe~II are larger than those of regular SNe~II at all available epochs.

A popular parameter to study the strength of spectral lines is the pseudo-equivalent width (pEW). The true SN spectral continuum level is not easy to identify owing to feature superposition, so the EW is measured considering a pseudo-continuum. We measure the pEW of the H$\alpha$ emission profile in order to characterise its strength at each observed epoch, utilising a straight line that connects the edges of the profile as pseudo-continuum. Again, multiple measurements were performed considering different traces of the continuum as explained above, and the respective standard deviation is considered to be the associated uncertainty. The bottom panel of Fig.~\ref{fig:velpew} shows the obtained pEW values along with the pEW values of the sample of SNe~II studied by \cite{2017ApJ...850...89G} for comparison. At early times, the pEW of LSNe~II are smaller than those of regular SNe~II, but become much larger at later times.

\subsection{Evolution of spectral metallic features}

The spectra of all LSNe~II in our sample remain blue and almost featureless until $\sim 30$--40 days, after which they develop only a small number of metal features. This is particularly obvious in the spectral series of SN~2018eph for which we have the best spectral coverage, starting at $\sim$ 4 days and finishing at $\sim$ 214 days. It can be seen in Fig.~\ref{fig:evspec18eph} that even at these late phases the spectra are very much dominated by H$\alpha$ with a lack of other features. The spectral evolution of our LSNe~II contrasts with the evolution observed in regular SNe~II that develop prominent metal features after $\sim$ 15 days \citep{2017ApJ...850...89G}. The main features of each of our LSNe~II at $\sim$ 15, 30, and 70 days are shown in comparison with other events in Figs.~\ref{fig:speccomp10d}, \ref{fig:speccomp30d}, and \ref{fig:speccomp70d}, respectively. 

\subsection{Summary of observed properties of six LSNe~II}
\label{sec:summary}

In this work we present the characteristics of a sample of six LSNe~II that stand out of a larger sample because of their overall characteristics. In summary, our LSNe~II were selected to show
\begin{itemize}
    \item light curves brighter than $\sim -$18.5~mag in the $V$ band,
    \item peculiar H$\alpha$ feature with no absorption component, and
    \item a blue excess in the H$\alpha$ emission profile.
\end{itemize}
After analysis we see that they also show
\begin{itemize}
    \item fast-declining light curves \citep[considering a decline rate of 1.4~mag/(100~d) as the separation between slow- and fast-declining events; e.g.,][and references therein]{2019ApJ...887....4D},
    \item bluer $B-V$ colours than regular SNe~II (see Fig.~\ref{fig:colors}),
    \item blue and (practically) featureless early-time spectra,
    \item large, persistent H$\alpha$ FWHM velocities (see top panel of Fig.~\ref{fig:velpew}),
    \item low pEW of the H$\alpha$ emission at early times that becomes very large at late times (see bottom panel of Fig.~\ref{fig:velpew}),
    \item a (somewhat) strong persistent emission at $\sim 5800$~\AA\ that we identify as \ion{He}{I} (see Section~\ref{sec:heid}), and
    \item a lack of typical metal lines observed in regular SNe~II \citep[see][]{2017ApJ...850...89G}.
\end{itemize}

 In addition, the presented LSNe~II share characteristics with SN~1979C and SN~1998S as well as with other LSNe~II previously studied in the literature (see Section~\ref{sec:comptolit}). In the following section we discuss the observed features and their implications.

\section{Discussion}
\label{sec:discussion}

In the previous sections we have outlined the observed properties of a selected sample of six LSNe~II. Here we present a discussion of those properties and attempt to link them to their underlying progenitor and explosion physics. Special attention is paid to SN~2017hxz, the fastest-declining event in the sample. Then, we compare our sample of LSNe~II to others available in the literature. Finally, we discuss the implications of the observed properties and their origin.

\subsection{Bright and fast-declining light curves with blue colours}
\label{sec:brightandfast}

%Light curves
The follow-up observations of the $V$-band light curves of our LSNe~II started after light-curve peak. Thus, we consider the first available photometric point to be the peak of the light curve, although the actual peak is probably brighter than the reported values. Taking this into consideration, the $V$ peak absolute brightness of the sample ranges from $-$18.3 to $-$19.4~mag. The light curves of all the LSNe~II in the sample are fast decliners \citep[and would have been historically classified as SNe~IIL;][]{1979A&A....72..287B}, with the slowest declining object displaying a $V$ decline rate of 1.6~mag/(100~d). Previous studies have found that more-luminous SNe~II also decline faster \citep[e.g.,][]{1994A&A...282..731P, 2014ApJ...786...67A, 2022A&A...660A..42M}, although most of them do not include LSNe~II. Some of the $V$ light curves show subtle bumps between $\sim$ 25 and 70 days, depending on the LSN. These bumps can be seen in Fig.~\ref{fig:lccomp}, although they become smoothed out by the GP interpolations. The most noticeable bump is seen in SN~2017cfo, which shows a break at $\sim$ 35 days. The $V$ light curve of SN~1979C is also somewhat bumpy (see Fig.~\ref{fig:lccomp}). Although these bumps could be related to noise in the light curves, it is interesting to note that \cite{2022arXiv220606497F} mention the presence of similar early-time bumps in the $gri$ light curves of the Type IIn SN~2019zrk, which also shows much more prominent bumps at late times. In the context of this study, although an SN~IIn, SN~2019zrk is interesting because its spectral evolution shows a dramatic broadening of the H$\alpha$ emission feature at late times, similar to what is observed in our LSNe~II. \cite{2022arXiv220606497F} claim that the late-time bumps seen in the optical light curves of SN~2019zrk are indicative of interaction with different CSM shells, which would suggest separate events of mass ejection. The authors do not provide any interpretation for the early-time bumps, although it could be argued that they have a similar origin. Unfortunately, we do not have sufficiently late-time photometric observations to further study possible light-curve bumps. In addition, the LCO aperture photometry presented here is not host-galaxy subtracted and the absolute magnitude of each SN might vary once PSF photometry of host-subtracted images is computed. However, we consider that the shape of the light curves should not differ much, especially at early phases when the SNe are brighter.  

%Colors
The $B-V$ colours of our sample are overall bluer than those of the large sample of SNe~II studied by \cite{2018MNRAS.476.4592D}. The most luminous LSN~II in our sample, SN~2017hxz, is also the one showing the bluest colours (at all epochs). At the same time, the least luminous in our sample, SN~2017hbj, has the reddest colours at early epochs. This is in agreement with the result of \cite{2018MNRAS.476.4592D} that redder (bluer) SNe~II have fainter (brighter) absolute magnitude at peak. They propose that this result originates from intrinsic colours rather than from dust effects, and might relate to the presence or absence of CSM close to the progenitor. Nevertheless, \cite{2018MNRAS.476.4592D} find an anticorrelation between the strength of the H$\alpha$ absorption and the slope of the colour curve. This is in contrast with the observed behaviour of the LSNe~II in our sample which show no (or almost no) H$\alpha$ absorption (by selection criteria), and become redder very slowly at early times. In Fig.~\ref{fig:colors} we see that SN~1979C and SN~1998S show a similar $B-V$ colour evolution as the sample of LSNe~II; these two SNe are included here because of the consistent SNID matches to their spectra.

\subsection{Peculiar spectral features}

%spectral sequence
Unfortunately, there are no early-time or nebular observations of our LSNe~II, except in the case of SN~2018eph. All of the spectral series are blue and almost featureless from the first observed spectrum until $\sim$ 30 days and do not develop the prominent metallic features observed in regular SNe~II from $\sim 10$--15 days and beyond \citep{2017ApJ...850...89G}. The sample was selected based on the absence of H$\alpha$ absorption; hence, none of the LSNe~II presented here show the typical hydrogen P~Cygni profile. Furthermore, the H$\alpha$ emission feature of all the studied LSNe~II show, at some point, an excess when fitted to a (skewed or normal) Gaussian model. We consider this excess to be produced by an additional component which is responsible for the observed broadening. In addition, the LSNe~II in the sample show hints of \ion{He}{I} even at late phases (see below). The most extreme case is SN~2017hxz, for which the \ion{He}{I} $\lambda$5876 profile becomes much stronger as the object evolves. Below we discuss the identification, characteristics, and implications of the presence/shape of both the H$\alpha$ and \ion{He}{I} features.

\subsubsection{Strong and wide H$\alpha$ emission}

The observed H$\alpha$ emission excess could be the reason for the emission feature becoming wider and developing a boxy profile. If this was the case, given the gradual evolution of the measured H$\alpha$ velocities and pEWs, the excess should emerge slowly and gradually. However, it is not clear whether the apparent gradual evolution is caused by a lack of spectral resolution, by the low-S/N spectra, or by the low cadence in the spectral observations. 

As can be seen in Fig.~\ref{fig:velpew}, the H$\alpha$ FWHM velocities of the LSNe~II in general are much higher than those of regular SNe~II at all of the considered epochs ($\sim 25$ to 100 days). \cite{2017ApJ...850...89G} studied a large sample of regular SNe~II and found that higher-velocity SNe~II have weaker spectral lines. The spectral evolution of the LSNe~II studied in this work is different from the typical spectral evolution observed in regular SNe~II. It can be seen in Fig.~\ref{fig:velpew} that the strength of the H$\alpha$ feature of LSNe~II is smaller than for regular SNe~II at early times, although they become much more prominent at late times (except maybe for SN~2018aql). We conclude that the excess emission and therefore the boxy profile arise from emission from higher-velocity material that is sustained throughout the LSN~II evolution, but we can not unequivocally associate this high-velocity material with larger explosion energies.

\subsubsection{Helium presence}
\label{sec:heid}

The spectral series of the studied LSNe~II show a clear emission (with weak/nonexistent absorption) feature at $\sim 5800$~\AA.
In regular SN~II early-time spectra, this feature is commonly identified as \ion{He}{I} $\lambda$5876, but at late times, when the temperature has decreased and is no longer able to excite the \ion{He}{I} ions, it could be identified as \ion{Na}{I}~D. \cite{1973ApJ...185..303K} argued in favour of such late-time feature identification based on the absence of other \ion{He}{I} lines and on the presumption that weak \ion{Na}{I}~D is visible together with strong \ion{Ca}{II} features that are usually seen in relatively late-time SN~II spectra.

We identify the feature as \ion{He}{I} $\lambda$5876 even at late times for several reasons. The \ion{Na}{I}~D lines typically appear together with other metal lines, but our LSNe~II do not develop strong (if any) metal lines in their spectra. In addition, the spectra remain blue for longer than regular SNe~II, implying that the temperature stays high or that high-energy photons produced by nonthermal excitation from CSM interaction exist, and possibly meaning that \ion{He}{I} ions can be excited for longer times. Finally, in most cases there is evidence of \ion{He}{I} $\lambda$7065 and \ion{He}{I} $\lambda$7281. 
Nevertheless, we do see notches on top of the emission part of the feature identified as \ion{He}{I} $\lambda$5876 starting at $\sim$ 50 days, the most noticeable case being SN~2018eph. Hence, we cannot rule out that this feature becomes a blend of \ion{He}{I} $\lambda$5876 and \ion{Na}{I}~D at late phases.

\subsection{Comparison with other LSNe~II}
\label{sec:comptolit}

In order to better interpret our data, we compare our sample with other LSNe~II in the literature. 
Our goal is to find events similar to those presented here. We considered only comparison events with publicly available $V$-band photometry and at least three publicly available good S/N ($> 5$) spectral observations at different phases. This is because we selected our sample based on the $V$ peak brightness, and because we want to be able to perform meaningful spectral comparisons. Moreover, since we are interested in understanding the observed features of our sample and not in gathering a sample of LSNe~II, the final requirement to consider an event as a comparison object is that it should be part of a study that provides some interpretation of the observed characteristics rather than a pure data release. Note that we do not put any constrain on the shape of the H$\alpha$ profile to consider an event as part as the comparison sample. We aim at discovering if any event that exists in the literature shows similar behaviour to those presented here.

To gather the comparison sample we inspected the Open Supernova Catalog \citep[OSC\footnote{Although the front end of the OSC is no longer accessible, the catalog is still available on GitHub containing all the transients uploaded through 8 April 2022. Note that there are no further updates after this date.};][]{2017ApJ...835...64G} and the SAO/NASA Astrophysics Data System (ADS\footnote{\url{https://ui.adsabs.harvard.edu}.}). We found six objects matching our criteria: SN~2013fc \citep{2016MNRAS.456..323K}, SN~2016ija \citep{2018ApJ...853...62T}, ASASSN-15nx \citep{2018ApJ...862..107B}, SN~2016gsd \citep{2020MNRAS.493.1761R}, SN~2016egz\footnote{The data for SN~2016egz were obtained upon request to \cite{2021ApJ...913...55H}.} \citep{2021ApJ...913...55H}, and SN~2018hfm \citep{2022MNRAS.509.2013Z}. The adopted explosion date, distance and extinction for each comparison event were obtained from the cited references. As far as we know, this comparison sample includes most of the LSNe~II with publicly available good-coverage observations (see above) present in the literature. We also include as comparison events SN~1979C and SN~1998S for the reasons mentioned in Section~\ref{sec:intro}, as well as other regular SNe~II to evaluate the differences between luminous and regular SN~II events. Below we describe the light curve and spectral comparisons, we find that out of the six comparison lumious events only two, SN~2016egz and SN~2018hfm, display spectral characteristics similar to those seen in our sample.

\subsubsection{Spectra}

%spectra
Here we inspect the spectral behaviour of our sample of LSNe~II against the above-mentioned comparison events (when spectra are available at similar phases). \cite{2017ApJ...850...89G} present spectral features observed for regular SNe~II at $\sim$ 10, 30, and 70 days. The spectral observations for our sample of LSNe~II started on average at $\sim$ 15 days. Thus, considering the mentioned work, we compare LSNe~II with SNe~II at $\sim$ 15, 30, and 70 days. The regular SNe~II selected for comparison are at $\sim$ 15 days for SN~2004et \citep{2014MNRAS.442..844F} because there is a spectrum taken at the considered phase, and at $\sim$ 30 and 70 days for SN~2003hn and SN~2003bn (respectively) given that they are the ones displayed in the plots of \cite{2017ApJ...850...89G}.

\begin{figure}
	\includegraphics[width=\columnwidth]{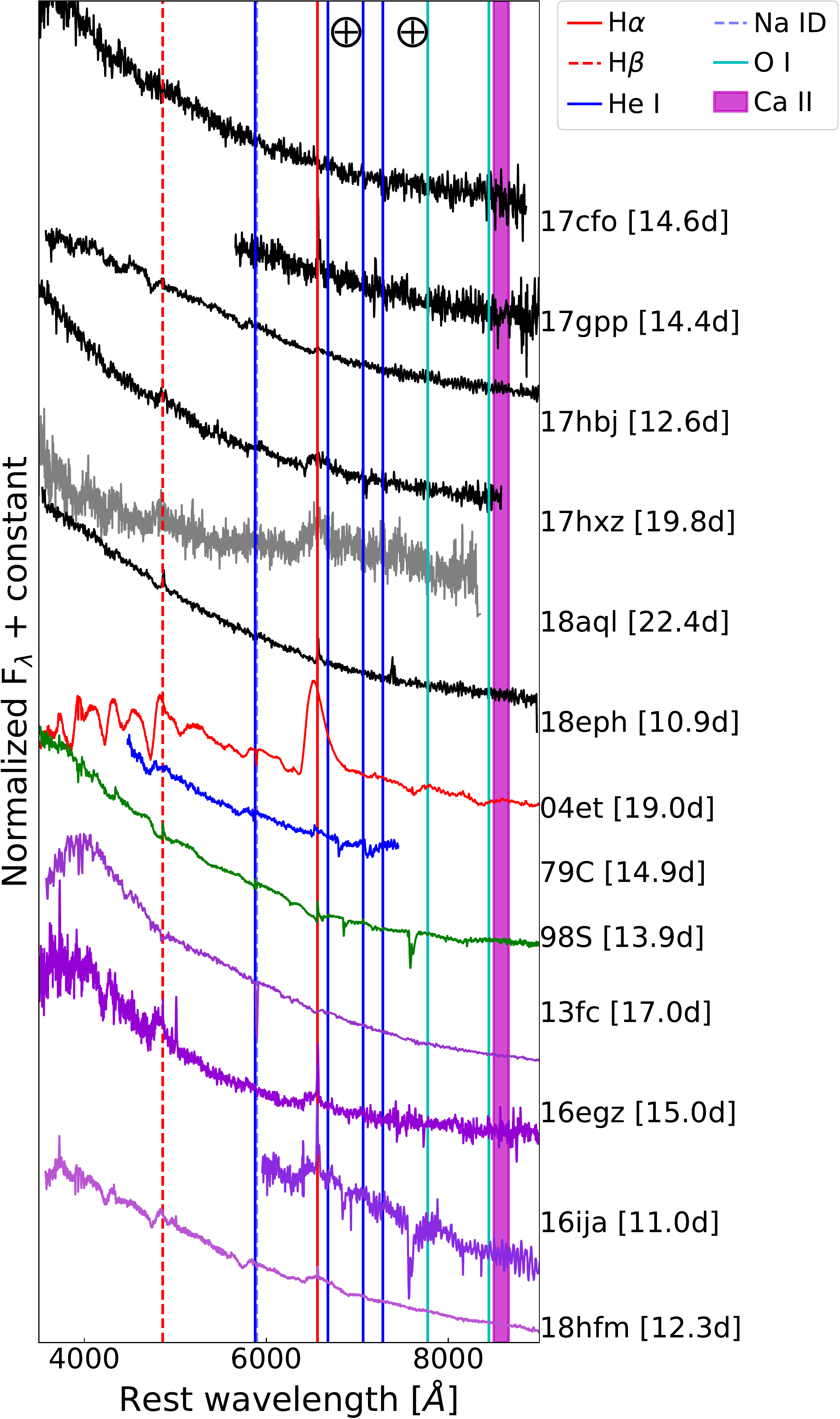}
    \caption{Normalised spectra of LSNe~II spectra at $\sim$ 15 days. In black and grey are the spectra of our LSN~II sample. In red the normal Type~II SN~2004et \protect\citep[from][]{2014MNRAS.442..844F}, in blue SN~1979C, in green SN~1998S, and in different shades of purple LSNe~II obtained from the literature. Some of the presented spectra have not been telluric corrected, the telluric regions are marked with $\oplus$ symbols.}
    \label{fig:speccomp10d}
\end{figure}

\begin{figure}
	\includegraphics[width=\columnwidth]{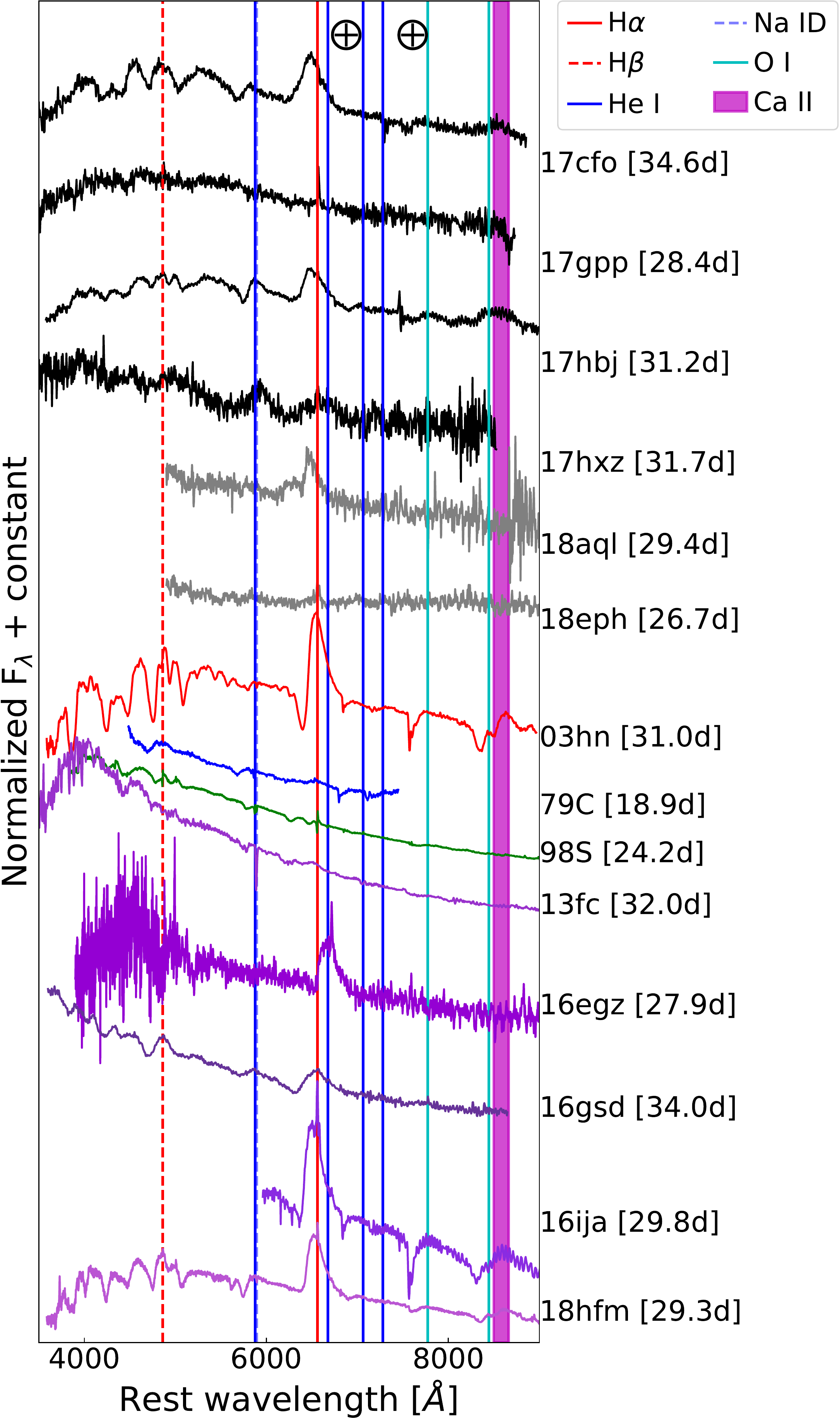}
    \caption{Normalised spectra of LSNe~II spectra at $\sim$ 30 days. In black and grey are the spectra of our LSN~II sample. In red the normal Type~II SN~2003hn \protect\citep[from][]{2017ApJ...850...89G}, in blue SN~1979C, in green SN~1998S, and in different shades of purple LSNe~II obtained from the literature. Some of the presented spectra have not been telluric corrected, the telluric regions are marked with $\oplus$ symbols.}
    \label{fig:speccomp30d}
\end{figure}

\begin{figure}
	\includegraphics[width=\columnwidth]{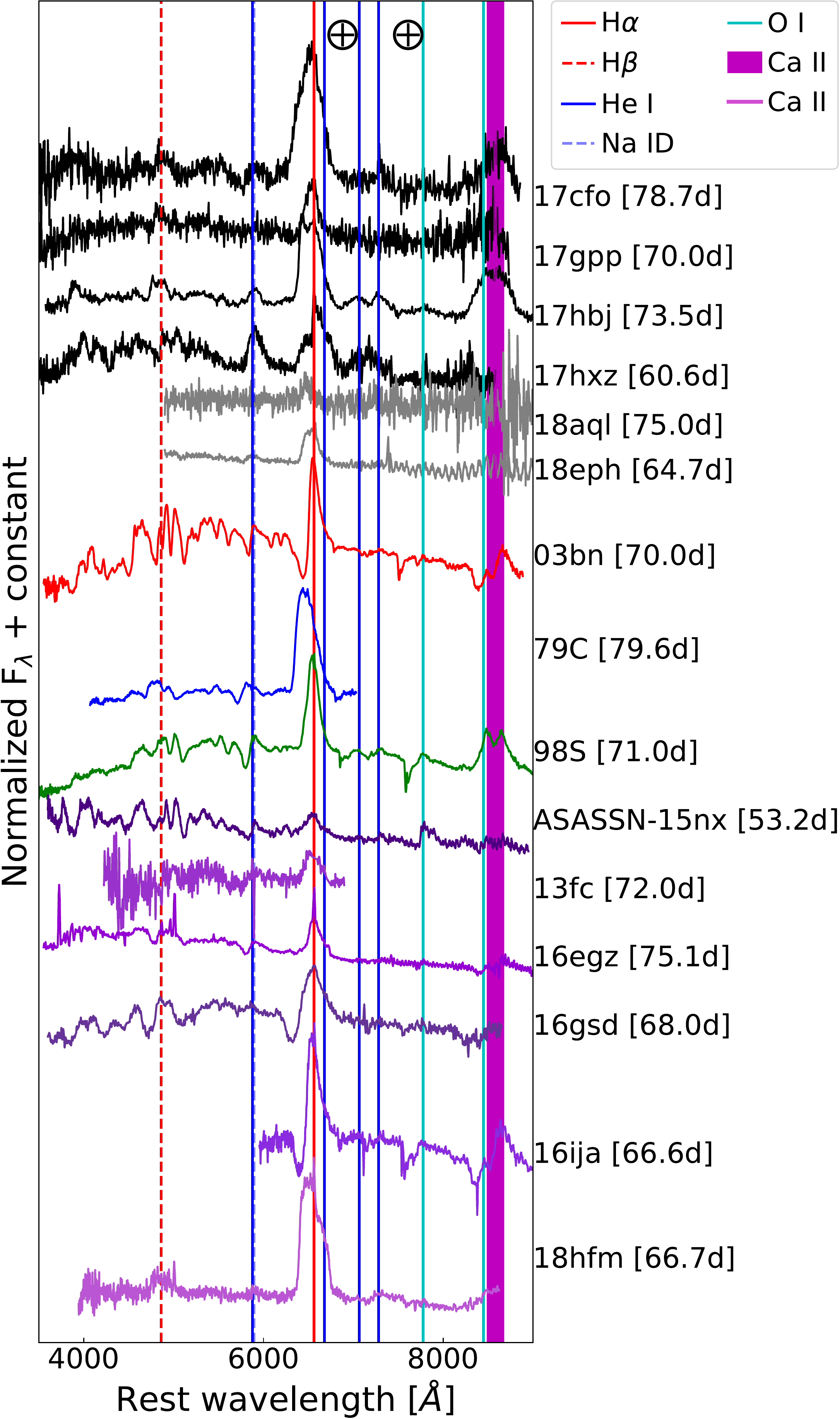}
    \caption{Normalised spectra of LSNe~II spectra at $\sim$ 70 days. In black and grey are the spectra of our LSN~II sample. In red the normal Type~II SN~2003bn \protect\citep[from][]{2017ApJ...850...89G}, in blue SN~1979C, in green SN~1998S, and in different shades of purple LSNe~II obtained from the literature. Some of the presented spectra have not been telluric corrected, the telluric regions are marked with $\oplus$ symbols.}
    \label{fig:speccomp70d}
\end{figure}

In Fig.~\ref{fig:speccomp10d} we see that at $\sim$ 15 days the H$\alpha$ emission and \ion{He}{I} $\lambda$5876 feature are much weaker in LSNe~II than in regular SNe~II. LSNe~II also lack many of the metal lines observed in regular events at this phase. It can be seen in Fig.~\ref{fig:speccomp30d} that at $\sim$ 30 days the H$\alpha$ emission is broader and the \ion{He}{I} $\lambda$5876 feature is stronger in LSNe~II than in regular SNe~II. The exceptions are SN~2017gpp and SN~2018eph where the respective features are still weak, similar to what is seen in the spectrum of SN~1979C. Also, the \ion{Ca}{II} NIR3, when observable, is wider in LSNe~II than in regular SNe~II. Finally, it can be observed in Fig.~\ref{fig:speccomp70d} that at $\sim$ 70 days the H$\alpha$ emission of LSNe~II is still broader than that of regular SNe~II. In addition, the H$\alpha$ emission of LSNe~II exhibits clear signatures of additional component(s), such as double peaks. The presence of multiple components is not obvious in SN~1979C and SN~1998S at this epoch. A broad \ion{He}{I} $\lambda$5876 feature is still present in LSNe~II, while normal events have a \ion{Na}{I}~D feature. \cite{2001MNRAS.325..907F} propose that SN~1998S shows a blend of \ion{He}{I} and \ion{Na}{I} at this epoch. The \ion{Ca}{II} NIR3 present in the spectra of our sample of LSNe~II remains broad, but it does not show multiple peaks as seen in SN~1998S. It is worth noting that we are trying to find LSNe~II similar to the ones presented in this work; hence, although considered as comparison events, SN~2016gsd and SN~2016ija would not have been selected as part of our sample because of the clear H$\alpha$ absorption features present at several epochs. Also, SN~2013fc would not have been selected because of the presence of narrow spectral features, although \cite{2016MNRAS.456..323K} note that these may well be dominated by host-galaxy lines. On the other hand, similarly to the LSNe~II in our sample, SN~2016egz and SN~2018hfm do not exhibit narrow lines or H$\alpha$ absorption features at any time in their evolution (see \citealt{2021ApJ...913...55H} and \citealt{2022MNRAS.509.2013Z}, respectively). SN~2016egz and SN~2018hfm also show signatures of \ion{He}{I} $\lambda$5876, although SN~2016egz develops an absorption component at $\sim$ 47 days that is not seen in our sample. In addition, they have broad \ion{Ca}{II} NIR3, which is seen to disappear in the available nebular spectra of SN~2016egz.

\subsubsection{Light curves}

\begin{figure}
	\includegraphics[width=\columnwidth]{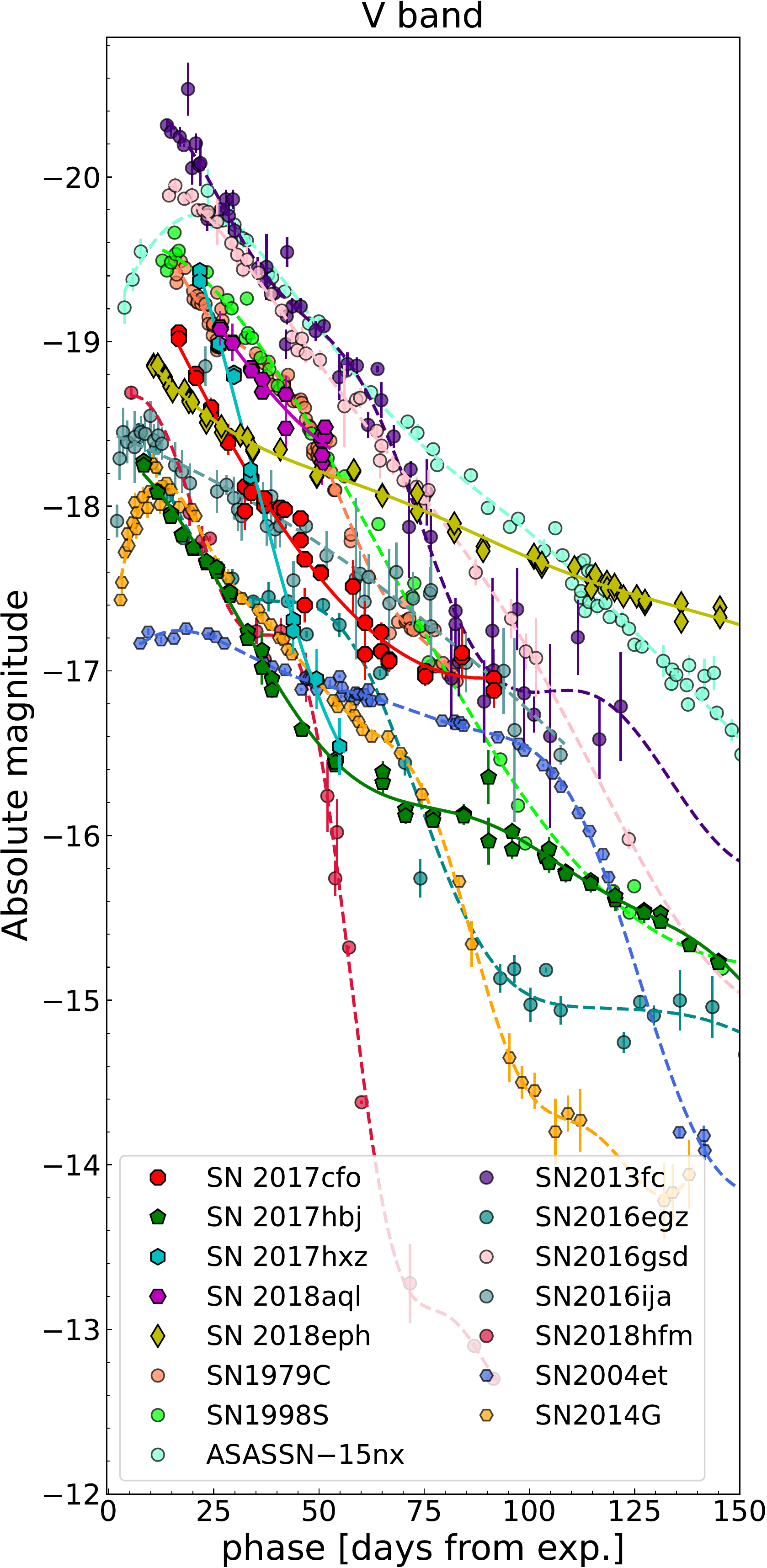}
    \caption{LSN~II light curves. In solid lines we present our sample of LSNe~II using a variety of markers. LSNe~II found in the literature, as well as SN~2004et and SN~2014G, are shown with dashed lines and circles and hexagons respectively.  Markers represent observed $V$-band photometry while lines represent GP interpolations.}
    \label{fig:lccomp}
\end{figure}

%light curves
Here we inspect the $V$-band light-curve behaviour of our sample against that of the comparison sample defined above. We include SN~2004et \citep{2010MNRAS.404..981M} and SN~2014G \citep{2016MNRAS.462..137T} as examples of events treated as regular SNe~II. The only criterion to select SN~2004et is that it has been extensively studied in the literature. SN~2014G was selected as it is  a well-studied fast decliner with a good dataset. All considered $V$ light curves can be seen in Fig.~\ref{fig:lccomp}. SN~2017cfo and SN~2017hbj present a decline rate similar to that of SN~2014G up to $\sim 75$ days, when the light curve of SN~2014G transitions to the radioactive tail. At this point both SN~2017cfo and SN~2017hbj show a subtle flattening that for the latter continues up to $\sim 100$ days, after which their light curves decline again. It is interesting to note that SN~2016egz exhibits a short plateau that starts around the same phase at which SN~2017cfo shows a break that leads to a subtle bump (see Section~\ref{sec:brightandfast}). Unfortunately, the quality of the light curve of SN~2017cfo prevents us from accurately identifying whether the bump is related to noise or if it could be an even shorter plateau. SN~2017hxz is the fastest decliner of the whole set (including the comparison LSNe~II); this is consistent with it being the most luminous in our sample, although several comparison events display brighter absolute magnitudes. The light curve of SN~2018aql seems to show a behaviour similar to that of SN~1979C, SN~1998S, ASASSN-15nx, and SN~2016gsd, although the observed range is too short to be certain. SN~2018eph is the slowest decliner of the sample, showing a continuous decline for $\sim$ 200 days and no sign of transition to the radioactive phase. 

\subsection{The blue, bright, and fast-declining SN~2017hxz}
\label{sec:17hxzstandsout}

SN~2017hxz has a number of properties that stand out from the rest of the sample. It displays the most luminous and fastest-declining $V$-band light curve of our sample of LSNe~II. Its spectra exhibit an extreme broadening of the H$\alpha$ feature (shown by the steep pEW evolution in Fig.~\ref{fig:velpew}). The feature we identified as \ion{He}{I} $\lambda$5876 presents an extreme evolution in comparison to the other LSNe~II in our sample. In addition, SN~2017hxz has the fastest and bluest  $B-V$ colour evolution among the LSN~II sample at all times (see Fig.~\ref{fig:colors}). 

\begin{figure}
	\includegraphics[width=\columnwidth]{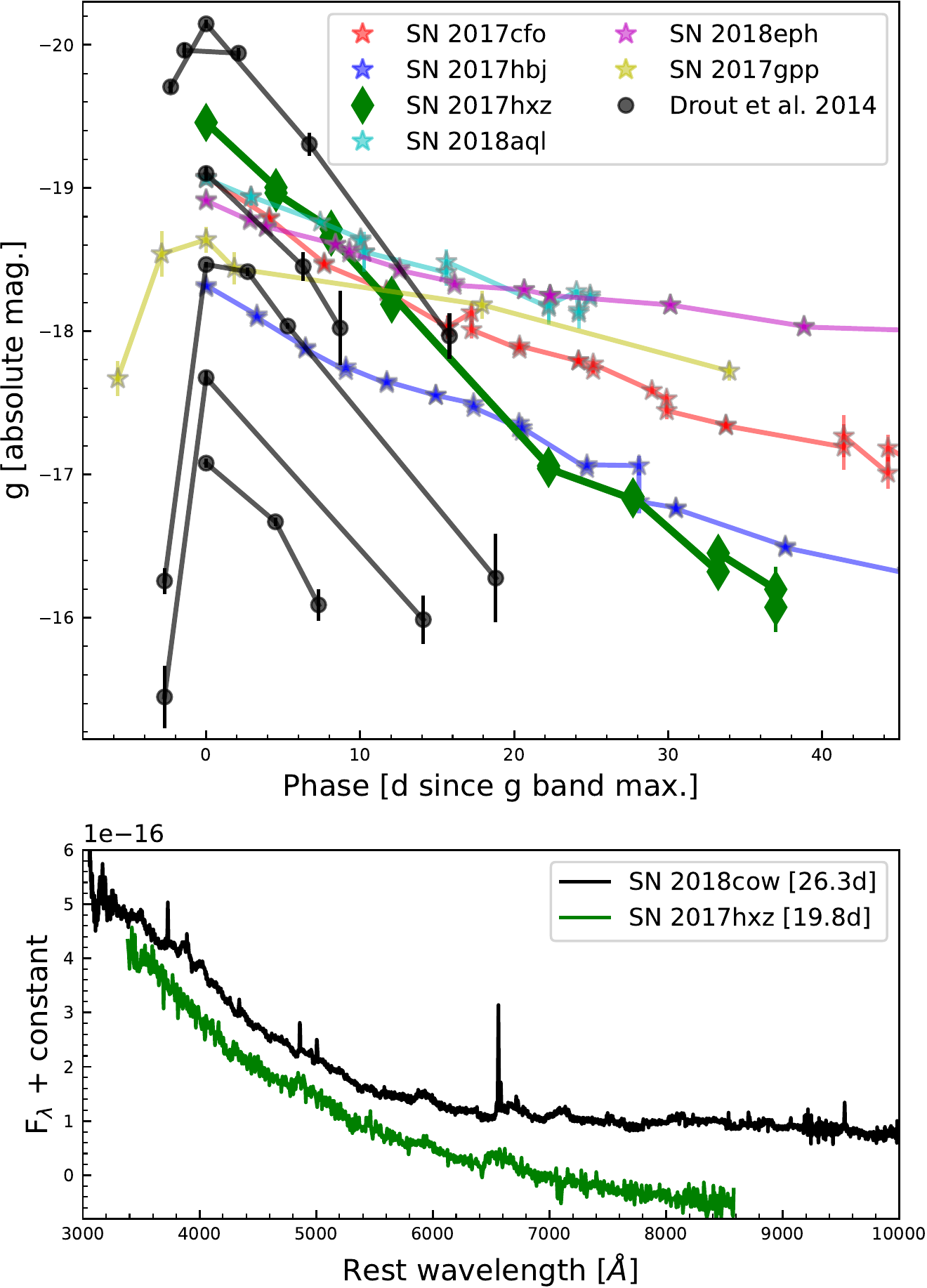}
    \caption{Comparison of SN~2017hxz with FBOTs. {\it Top panel:} $g$-band light curves of the FBOTs gold sample (black circles) of \protect\cite{2014ApJ...794...23D} versus the $g$ light curves of our sample of LSNe~II (coloured stars). The phase range is cropped for better visualisation. It can be seen that SN~2017hxz (green) shows similar behaviour to that of the FBOTs. {\it Bottom panel:} the spectrum of AT~2018cow (in black; observed by Christoffer Fremling and Yashvi Sharma and reduced by Christoffer Fremling) compared to the spectrum of SN~2017hxz (in green). The phase with respect to explosion is indicated in brackets (note that regular SNe~II do not show such blue spectra at these phases). The explosion epoch and redshift of AT~2018cow were obtained from \protect\cite{2018ApJ...865L...3P}.}
    \label{fig:fbots}
\end{figure}

Here we consider an interpretation of the nature of SN~2017hxz that could explain all these peculiarities. We compare SN~2017hxz with the so-called ``fast blue optical transients'' (FBOTs), which are characterised by blue colours and rapid evolution of their light curves. \cite{2014ApJ...794...23D} presented a sample of these events from Pan-STARRS1, but poor spectroscopic coverage prevented the authors from determining whether the events are hydrogen-rich. We compared the $gr$ photometry of the ``gold'' sample of \cite{2014ApJ...794...23D} to the $g$-band light curve of SN~2017hxz. The top panel of Fig.~\ref{fig:fbots} shows that SN~2017hxz falls well within this gold sample, displaying a similar evolution and light-curve decline rate. We also compare the first spectrum of SN~2017hxz with one of the latest publicly available spectra of AT~2018cow in WISeREP\footnote{\url{https://www.wiserep.org}} \citep{2012PASP..124..668Y}. Both spectra, shown in the bottom panel of Fig.~\ref{fig:fbots}, were observed at similar phases ($\sim$ 6 rest-frame days from explosion apart). We see a reasonable visual match. The presence of H$\alpha$ in the spectral evolution of AT~2018cow is unclear \citep[e.g.,][]{2018ApJ...865L...3P,2019MNRAS.488.3772F}. On the other hand, the presence of \ion{He}{I} is discussed by \cite{2018ApJ...865L...3P}; they mention that this challenges a magnetar or accretion scenario for AT~2018cow. However, \cite{2021arXiv210508811H} conclude that to explain the luminous millimeter, X-ray, and radio emission observed in events similar to AT~2018cow, an additional powering mechanism should be in place. These authors claim that the dominant powering mechanism of fast transients is interaction. We note that the most luminous transients in the sample of \cite{2021arXiv210508811H} are Type Ibn/IIn SNe, while the less luminous are Type IIb/Ib. Our sample only includes hydrogen-rich events with no persistent narrow spectral lines. The observed similarities of SN~2017hxz and the FBOTs family, together with the shared characteristics of SN~2017hxz and the presented sample of LSNe~II, suggest a link between other classes of fast transients and LSNe~II.

\subsection{Possible scenarios to explain the LSN~II observed features}
\label{sec:possiblescena}

So far we have presented the characteristics of a sample of six LSNe~II that stand out from a larger sample because of their spectral evolution (see Section~\ref{sec:summary} for a summary). 
We found not only that our sample shares characteristics with previously studied LSNe~II, but also that one of our LSNe~II (SN~2017hxz) shows characteristics similar to those observed in FBOTs (see Section~\ref{sec:17hxzstandsout}). Given that the aim of this work is to understand which type of progenitor conditions are such that they explode producing all the observed characteristics, we searched the literature for theoretical scenarios that propose explanations to the characteristics mentioned above.

There exist several models that explain the powering source that causes extra luminosity in SN light curves. The most popular explanations include large productions of $^{56}$Ni, a central engine (fallback accretion to a black hole or magnetar spindown), and/or CSM interaction (see Section~\ref{sec:intro}). Although we do not present any specific modelling here, we propose an interaction with a relatively low-density CSM as the powering mechanism of our LSN~II sample because of the following reasons.

\begin{figure}
	\includegraphics[width=\columnwidth]{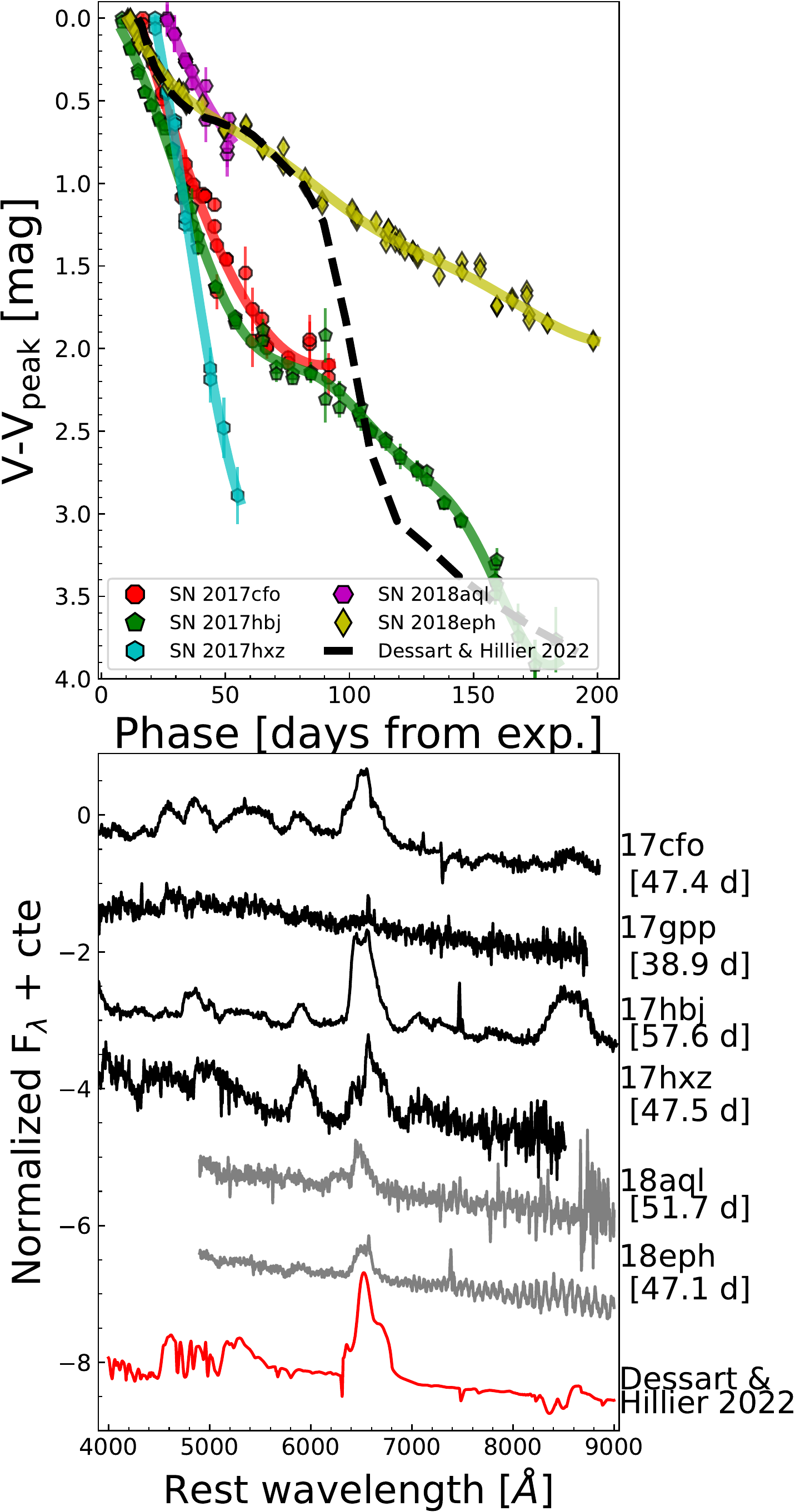}
    \caption{{\it Top panel:} light curve of the Pwr1e42 model presented by \protect\cite{2022A&A...660L...9D} (black dashed line) compared to our LSNe~II sample (observations presented in different markers and GP interpolation in colored solid lines) {\it Bottom panel:} spectrum at 50 days for the same Pwr1e42 model (red) compared to the spectra of our LSNe~II taken at a similar phase. The name of the SN and phase to which each spectrum corresponds is annotated to the right. The spectrum of SN~2018eph is affected by CCD fringing at the longest wavelengths.}
    \label{fig:compLuc}
\end{figure}

\begin{itemize}
    \item Many SN~II progenitors suffer mass loss just before explosion \citep{2016ApJ...818....3K,2021ApJ...912...46B}. It is thus logical to assume that this mass loss might be responsible for the generation of CSM that will later interact with the SN ejecta. In fact, there is evidence of early CSM interaction for most SNe~II \citep[e.g.,][]{2015MNRAS.451.2212G,2018NatAs...2..808F,2020ApJ...891L..32M}. The expansion of a shell produced by the reverse shock of the collision between the SN ejecta and CSM could explain the presence of broad, boxy emission profiles \citep{1995A&A...299..715P,2016MNRAS.456.1269B}. The presence of CSM could contribute to the SN continuum making it stronger, which would explain the lack of metal lines \citep{2000PASP..112..217B}. Moreover, additional thermal energy produced by interaction could explain the observed blue colours.
    \item One of the selection criteria for the presented LSN~II sample is the identification of an excess blueward of the H$\alpha$ rest wavelength that we interpret as the presence of an additional (or multiple additional) component contributing to the emission profile. \cite{2016MNRAS.456.3296B} find additional H$\alpha$ components in the late-time spectra of a fast-declining SN~II (SN~1996al) and attribute them to the interaction of the ejecta with an asymmetric CSM. They also argue that if the CSM is less asymmetric, the extra components might be seen at earlier times. In addition, we find \ion{He}{I} emission in the spectral series of our LSNe~II, and \cite{2016MNRAS.456.3296B} also find evidence of \ion{He}{I} throughout the evolution of the spectral series of SN~1996al that they attribute to either high-velocity $^{56}$Ni or interaction with CSM. Despite SN~1996al not being considered as a comparison event owing to its (slightly) fainter peak magnitude ($M_{V} = -$18.2~mag), the similar features can be considered to have a common origin.
    \item Although the $V$-band maximum has not been observed for our sample of LSNe~II, in average the first photometric point has been observed at 16.8 days after explosion, which means that the rise time should be at most 16.8 days, in average. This discards configurations that produce light curves with rise times of several weeks such as extremely massive or very dense progenitors (see Section~\ref{sec:intro}).
    \item We see several similarities between the characteristics of our sample of LSNe~II and those seen for SN~1979C and SN~1998S. CSM interaction has been invoked to explain the features of both events \citep[e.g.,][]{1993A&A...273..106B,2016MNRAS.458.2094D,2017hsn..book..403S}. As opposed to what is seen in the spectral sequence of SN~1998S, we do not detect long-lasting narrow lines in the spectral sequences of our LSNe~II; thus, the possibly existent CSM would not be as dense in our sample.
    \item We note similarities between the LSNe~II in our sample and other LSNe~II found in the literature for which CSM interaction has been claimed to explain their characteristics, which further supports the assumption that CSM might be producing the observed features of our LSN~II sample (see Figs. \ref{fig:speccomp10d}, \ref{fig:speccomp30d} and \ref{fig:speccomp70d}).
    \item Recently, \cite{2022MNRAS.516.1193K} examined a sample of SLSNe~II (some of which show similar spectral features to those observed in the LSNe~II presented here but are not included as comparison events because of the lack of $V$-band light curves) and favoured a CSM interaction powering mechanism based on the observed UV excess, although they do not discard a central engine for the brightest events which might even need both mechanisms. Unfortunately, we do not have UV data, but we do see similar blue, (almost) featureless early-time spectra with little metal-line evolution and broad H$\alpha$ absorption.    
    \item Finally, we compare the light curves and $\sim$ 50 days spectra of our LSNe~II to the light curve and spectral results of the models of \cite{2022A&A...660L...9D}. Model Pwr1e42 present the best spectral match although the peak associated light curve is dimmer ($-$17.8 mag in the $V$ band) than that observed in our events.  It is important to note that we are comparing our observations to models that were not produced to fit them but to study the diversity in the long-term radiative interaction signatures of the ejecta of a Type~II explosion produced by a 15~M$_{\sun}$ star that evolves at solar metalicity with a CSM produced by a mass-loss rate of up to 10$^{-3}$~M$_{\sun}$~yr$^{-1}$. Nevertheless, we can see in the top panel of Fig.~\ref{fig:compLuc} that, if we normalize the light curves with respect to their peak magnitude, the comparison model matches the LSNe~II light curves quite nicely, especially at early times. The match is remarkably good to SN~2018eph up to $\sim$ 90 days. The model shows a change of curvature followed by a subsequent drop at $\sim$ 30 days, this behaviour is comparable to that seen at $\sim$ 70 days in SN~2017hbj. In the bottom panel of Fig.~\ref{fig:compLuc} we can see that the model spectrum has an H$\alpha$ profile similar to those seen in our sample, suggesting that a scenario in which the ejecta interacts with a CSM that is not dense enough to be optically thick to electron scattering on large scales may indeed be the origin of our sample of LSNe~II. \cite{2022A&A...660L...9D} find that in such a scenario, the interaction power will trigger an ionisation wave that could weaken some metal lines. They also find that the spectra will develop broad and boxy features. This is consistent with the characteristics of our sample.
\end{itemize}

We propose that a typical red supergiant (RSG) SN~II progenitor that is surrounded by CSM that is not dense enough to be optically thick to electron scattering on large scales (yet denser than that present in regular SNe~II), produced by low wind mass-loss rates, may be able to account for all the observed characteristics in our LSN~II sample. Not only has the presence of CSM around SN progenitors already been studied by several authors \citep[e.g.,][]{2017ApJ...838...28M,2017MNRAS.469L.108M,2017NatPh..13..510Y,2018NatAs...2..808F}, but recently \cite{2022A&A...660L...9D} showed that an explosion of a progenitor similar as the one we propose will display high-velocity features and broad-boxy spectral profiles without persistent narrow lines. It has been proposed that a continuum in CSM density exists amid the progenitors of regular SNe~II and those of SNe~IIn \citep{2015MNRAS.449.1876S,2017hsn..book..403S}, so it is natural to speculate that our sample might be produced by progenitors with an intermediate CSM density. A caveat to our progenitor assumption is our lack of UV data. \cite{2022A&A...660L...9D} claim that, when considering their models, only early-time UV observations could effectively assess the presence of interaction \citep[as in][]{2022MNRAS.516.1193K}, while the presence of broad and boxy H$\alpha$ emission profiles only suggests an interaction scenario. Hence, we cannot completely discard other powering mechanisms. However, considering that the fast ejecta is located at large radii, a magnetar scenario is unlikely since magnetar power, which is injected in the inner ejecta, cannot cause broad, boxy line profiles at early times.

The comparison events that are the most similar to our LSNe~II are SN~2018hfm and SN~2016egz, in the sense that they also show broad, boxy H$\alpha$ profiles without absorption components. \cite{2021ApJ...913...55H} find that SN~2016egz could be explained by a $\sim$ 18--22~M$_{\sun}$ progenitor with small hydrogen envelope mass and enhanced mass loss that would produce the CSM with which the SN ejecta will interact give rise to the early luminous peak. The model proposed by \cite{2022MNRAS.509.2013Z} to explain SN~2018hfm considers the results presented by \cite{2018MNRAS.473.3863L} of a single-star progenitor of 27~M$_{\odot}$ with a large radius that retains only a small fraction of its hydrogen envelope before explosion. It should be noted that \cite{2022MNRAS.509.2013Z} consider the models of \cite{2018MNRAS.473.3863L} because of the low explosion energies that they infer from the modelling of the bolometric light curve. Indeed, the models of \cite{2018MNRAS.473.3863L} are not for luminous events but for low-luminosity SNe (that show fairly classical H$\alpha$ features). These scenarios are different from the one presented by \cite{2022A&A...660L...9D}, who propose the explosion of a 15~M$_{\odot}$ star interacting with a relatively low-density CSM. However, the three scenarios invoke CSM interaction to explain the observed features. Given that we do not observe plateaus in our sample (at least not one similar to that of SN~2016egz) and that we consider more investigation is needed to determine the accuracy of bolometric light-curve calculations in the presence of interaction, we prefer the models of \cite{2022A&A...660L...9D}. \cite{2022MNRAS.509.2013Z} also discuss a possible electron-capture explosion scenario based on the late-time spectral features of SN~2018hfm. Unfortunately, we do not have enough late-time spectra to study this possibility. 

\section{Conclusions}
\label{sec:conclusion}

In this work we have presented and characterised the optical light curves and spectral evolution of six LSNe~II: SN~2017cfo, SN~2017gpp, SN~2017hbj, SN~2017hxz, SN~2018aql, and SN~2018eph. They were selected from a larger sample because they share common photometric and spectroscopic evolution. Their optical light curves are luminous and rapidly declining, they exhibit blue colours, and they show blue early-time spectra, weak or nonexistent metal lines, and broad and boxy H$\alpha$ emission profiles. None of them develops an H$\alpha$ absorption component. Their H$\alpha$ lines show high-velocity and steep-pEW evolution, and also signatures of multiple components from a given phase onward. We note similarities in the characteristics of our LSNe~II and those observed in SN~1979C, SN~1998S, and other LSNe~II in the literature. 

We propose that our LSNe~II arise from RSG progenitors that are surrounded by CSM that is not dense enough to be optically thick to electron scattering on large scales, yet denser than that present in regular SNe~II, based on the above-mentioned similarities and on comparisons with the models presented by \cite{2022A&A...660L...9D}. Such models can provide ejecta-CSM interaction that accounts for the observed features of our sample without producing narrow emission lines typical of SNe~IIn. We note similarities between the decline rate and spectral features of SN~2017hxz and FBOTs, specifically those in the gold sample of \citet{2014ApJ...794...23D} and AT~2018cow. We speculate that these similarities may suggest a link between FBOTs and LSNe~II, although more events are needed to perform a thorough comparison of both families. We note that we present only a subset of a larger sample of LSNe~II; further study is needed to evaluate if our conclusions can be extrapolated to the entire sample.

\section*{Acknowledgements}

We thank the anonymous referee for the valuable revision.

\noindent P.J.P. thanks Joe Lyman for guidance with some of the used software. P.J.P. thanks the useful input of Tuomas Kangas. P.J.P. acknowledges funding support by the ESO Ph.D. studentship program. 

\noindent Based on observations collected at the European Organisation for Astronomical Research in the Southern Hemisphere, Chile, as part of ePESSTO+ (the advanced Public ESO Spectroscopic Survey for Transient Objects Survey). ePESSTO+ observations were obtained under ESO program IDs 1103.D-0328, 106.216C, 108.220C (PI: Inserra). 

\noindent This work makes use of observations from the Las Cumbres Observatory global telescope network (LCO). LCO data have been obtained via OPTICON. The OPTICON project has received funding from the European Union's Horizon 2020 research and innovation programme under grant no. 730890. The LCO team is supported by NSF grants AST-1911225 and AST-1911151. This work was funded by ANID, Millennium Science Initiative, ICN12\_009. 

\noindent L.M. acknowledges support from a CONICET fellowship and UNRN~PI2018~40B885 grant.

\noindent A.V.F. received generous financial support from the Christopher R. Redlich Fund and many individual donors. 

\noindent M.N. is supported by the European Research Council (ERC) under the European Union's Horizon 2020 research and innovation programme (grant agreement No.~948381) and by a Fellowship from the Alan Turing Institute. 

\noindent T.E.M.B. acknowledges financial support from the Spanish Ministerio de Ciencia e Innovaci\'on (MCIN), the Agencia Estatal de Investigaci\'on (AEI) 10.13039/501100011033, and the European Union Next Generation EU/PRTR funds under the 2021 Juan de la Cierva program FJC2021-047124-I and the PID2020-115253GA-I00 HOSTFLOWS project, from Centro Superior de Investigaciones Cient\'ificas (CSIC) under the PIE project 20215AT016, and the program Unidad de Excelencia Mar\'ia de Maeztu CEX2020-001058-M. 

\noindent Support for G.P. is provided by the Ministry of Economy, Development, and Tourism's Millennium Science Initiative through grant IC120009, awarded to The Millennium Institute of Astrophysics (MAS). P.C. acknowledges support via an Academy of Finland grant (340613; P.I. R. Kotak). 

\noindent M.G. is supported by the EU Horizon 2020 research and innovation programme under grant agreement No 101004719. 

\noindent L.G. acknowledges financial support from the Spanish Ministerio de Ciencia e Innovaci\'on (MCIN), the Agencia Estatal de Investigaci\'on (AEI) 10.13039/501100011033, and the European Social Fund (ESF) "Investing in your future" under the 2019 Ram\'on y Cajal program RYC2019-027683-I and the PID2020-115253GA-I00 HOSTFLOWS project, from Centro Superior de Investigaciones Cient\'ificas (CSIC) under the PIE project 20215AT016, and the program Unidad de Excelencia Mar\'ia de Maeztu CEX2020-001058-M. 

\noindent X.W. is supported by he National Natural Science Foundation of China (NSFC grants 12288102, 12033003, and 11633002), the Scholar Program of Beijing Academy of Science and Technology (DZ:BS202002), and the Tencent Xplorer Prize.

\noindent T.M.R. acknowledges the financial support of the Vilho, Yrjö and Kalle Väisälä Foundation of the Finnish academy of Science and Letters.

\noindent The identification of SN~2017gpp was done by volunteers of the Supernova Sighting project in the \url{Zooniverse.org} platform. 

\noindent We have used the NASA/IPAC Extragalactic Database (NED), which is operated by the Jet Propulsion Laboratory, California Institute of Technology, under contract with the National Aeronautics and Space Administration (NASA). We have also used the ``Aladin sky atlas'' developed at CDS, Strasbourg Observatory, France.

%%%%%%%%%%%%%%%%%%%%%%%%%%%%%%%%%%%%%%%%%%%%%%%%%%
\section*{Data Availability}

The photometric data in this paper are presented in Appendix \ref{app}, and the spectra can be found in the ESO archive (\url{https://archive.eso.org/scienceportal/home}) and WISeREP (\url{https://www.wiserep.org}).

%%%%%%%%%%%%%%%%%%%%%%%%%%%%%%%%%%%%%%%%%%%%%%%%%%
\section*{software}

{\sc numpy} \citep{harris2020array},
{\sc matplotlib} \citep{Hunter:2007}, 
{\sc pandas} \citep{mckinney-proc-scipy-2010},
{\sc scipy} \citep{2020SciPy-NMeth},
{\sc astropy} \citep{astropy:2018},
{\sc lmfit} \citep{2014zndo.....11813N},
{\sc PyAstronomy} \citep{pya},
{\sc sklearn} \citep{scikit-learn},
{\sc GPy} \citep{gpy2014}



\bibliographystyle{mnras}
\bibliography{luminous} % if your bibtex file is called example.bib

%\begin{thebibliography}{99}
%\end{thebibliography}

%%%%%%%%%%%%%%%%%%%%%%%%%%%%%%%%%%%%%%%%%%%%%%%%%%

%%%%%%%%%%%%%%%%% APPENDICES %%%%%%%%%%%%%%%%%%%%%
\newpage
\appendix
\begingroup
\let\clearpage\relax 
\onecolumn 
\section{Spectral and photometric observations log}
\label{app}

% If you want to present additional material which would interrupt the flow of the main paper,
% it can be placed in an Appendix which appears after the list of references.

%\begin{landscape}
\begin{ThreePartTable}
\begin{TableNotes}
      \small  
      \item The first column gives the SN name. Column~2 lists the spectral observation UT date. The third, fourth, and fifth columns respectively indicate the instrument, telescope, and observatory where the spectrum was obtained. Column~6 gives the phase of the spectrum with respect to the explosion. The seventh column lists the S/N of the spectrum. The last column indicates whether an excess is observed in the H$\alpha$ emission feature.
      \begin{itemize}
          \item[*] Publicly available on WISeREP
      \end{itemize}
\end{TableNotes}
\onecolumn
\begin{longtable}{lccclccc}
    \caption{Spectral observations log.}
    \label{tab:lum-specs}
    \endfirsthead
    \hline
    Object         & UT date    & Instrument & Telescope     & Observatory    & Phase   & S/N   & H$\alpha$ excess  \\
               &            &            &               &                & [d]    &        &                   \\
    \hline
SN~2017cfo & 2017-03-25  & EFOSC2         & NTT           & La Silla       & 15.8  & 20.3   & no    		  \\
           & 2017-03-27  & FLOYDS          & 2m0-02        & Siding Spring  & 17.0  & 14.1   & no    		  \\
           & 2017-03-31  & FLOYDS          & 2m0-02        & Siding Spring  & 20.9  & 5.4    &  $\cdots$    	  \\
           & 2017-04-04  & FLOYDS         & 2m0-01        & Haleakala      & 24.6  & 16.8   & no    		  \\
           & 2017-04-12  & FLOYDS          & 2m0-02        & Siding Spring  & 32.4  & 1.9    &  $\cdots$    	  \\
           & 2017-04-15  & EFOSC2         & NTT           & La Silla       & 35.9  & 23.8   & no    		  \\
           & 2017-04-16  & FLOYDS          & 2m0-02        & Siding Spring  & 36.2  & 7.1    & no    		  \\
           & 2017-04-27  & FLOYDS         & 2m0-01        & Haleakala      & 46.7  & 8.6    & no    		  \\
           & 2017-04-27  & EFOSC2         & NTT La        & Silla          & 47.4  & 41.8   &  yes  		  \\
           & 2017-05-06  & FLOYDS         & 2m0-01        & Haleakala      & 55.3  & 1.6    &  $\cdots$		  \\
           & 2017-05-31  & EFOSC2         & NTT           & La Silla       & 79.9  & 6.5    &  yes  		  \\
           & 2017-05-31  & EFOSC2         & NTT           & La Silla       & 79.9  & 4.7    &  $\cdots$		  \\
\hline
SN~2017gpp & 2017-09-10  & EFOSC2         & NTT           & La Silla       & 11.5  & 5.4   &  $\cdots$  	  \\
           & 2017-09-13  & EFOSC2         & NTT           & La Silla       & 14.3  & 5.4   &  $\cdots$  	  \\
           & 2017-09-13  & EFOSC2         & NTT           & La Silla       & 14.4  & 15.6   & no  		  \\
           & 2017-09-28  & EFOSC2         & NTT           & La Silla       & 28.4  & 22.6  & no  		  \\
           & 2017-10-09  & EFOSC2         & NTT           & La Silla       & 38.8  & 25.6  & no  		  \\
           & 2017-10-09  & EFOSC2         & NTT           & La Silla       & 38.9  & 15.2  & no  		  \\
           & 2017-10-27  & EFOSC2         & NTT           & La Silla       & 55.8  & 5.0   &  $\cdots$		  \\
           & 2017-11-11  & EFOSC2         & NTT           & La Silla       & 70.0  & 8.9   & no  		  \\
           & 2017-11-26  & EFOSC2         & NTT           & La Silla       & 84.2  & 8.4   & yes		  \\
\hline
SN~2017hbj & 2017-10-09  & EFOSC2         & NTT           & La Silla       & 12.7  & 65.6  & no  		  \\
           & 2017-10-10  & EFOSC2         & NTT           & La Silla       & 13.4  & 34.5  & no  		  \\
           & 2017-10-18  & FLOYDS         & 2m0-02        & Siding Spring  & 20.8  & 10.9  & no  		  \\
           & 2017-10-28  & EFOSC2         & NTT           & La Silla       & 31.2  & 61.9  &  yes		  \\
           & 2017-11-08  & EFOSC2         & NTT           & La Silla       & 42.1  & 35.8  &  yes		  \\
           & 2017-11-24  & EFOSC2         & NTT           & La Silla       & 57.6  & 47.2  &  yes		  \\
           & 2017-12-10  & EFOSC2         & NTT           & La Silla       & 73.5  & 31.5  &  yes		  \\
           & 2017-12-11  & EFOSC2         & NTT           & La Silla       & 74.3  & 19.1  &  yes		  \\
           & 2017-12-11  & EFOSC2         & NTT           & La Silla       & 74.4  & 25.2  &  yes		  \\
           & 2018-03-25  & EFOSC2         & NTT           & La Silla       & 176.4 & 3.9   &  $\cdots$		  \\
\hline
SN~2017hxz & 2017-11-11  & EFOSC2         & NTT           & La Silla       & 19.8  & 28.2  & no   		  \\
           & 2017-11-12  & EFOSC2         & NTT           & La Silla       & 20.6  & 39.8  & no   		  \\
           & 2017-11-12  & EFOSC2         & NTT           & La Silla       & 20.6  & 49.9  & no   		  \\
           & 2017-11-14  & FLOYDS         & 2m0-01        & Haleakala      & 21.8  & 34.0  & no   		  \\
           & 2017-11-24  & EFOSC2         & NTT           & La Silla       & 31.7  & 13.4  & no   		  \\
           & 2017-11-28  & EFOSC2         & NTT           & La Silla       & 35.5  & 13.2  &  yes 		  \\
           & 2017-12-11  & EFOSC2         & NTT           & La Silla       & 47.5  & 11.9  &  yes 		  \\
           & 2017-12-25  & EFOSC2         & NTT           & La Silla       & 60.6  & 9.0   &  yes		  \\
\hline
SN~2018aql & 2018-04-09  & YFOSC\tnote{*} & Lijiang 2.4~m & YNAO	   	   & 22.3  & 10.6  & no  		  \\
           & 2018-04-13  & FLOYDS         & 2m0-01        & Haleakala      & 26.6  & 29.9  & no  		  \\
           & 2018-05-10  & FLOYDS         & 2m0-01        & Haleakala      & 51.7  & 17.2  &  yes		  \\
           & 2018-05-15  & LRS2           & HET           & MCDONALD	   & 56.3  & 22.0  &  yes		  \\
           & 2018-05-23  & FLOYDS         & 2m0-01        & Haleakala      & 63.8  & 16.0  &  yes		  \\
           & 2018-06-04  & FLOYDS         & 2m0-01        & Haleakala      & 75.0  & 5.5   &  yes		  \\
\hline
SN~2018eph & 2018-08-04  & EFOSC2         & NTT           & La Silla       & 4.1   & 18.0  & no   		  \\
           & 2018-08-06  & FLOYDS         & 2m0-02        & Siding Spring  & 5.5   & 33.0  & no	 	  \\
           & 2018-08-09  & FLOYDS         & 2m0-02        & Siding Spring  & 8.4   & 23.3  & no	 	  \\
           & 2018-08-10  & FLOYDS         & 2m0-02        & Siding Spring  & 9.2   & 29.9  & no	 	  \\
           & 2018-08-11  & EFOSC2         & NTT           & La Silla       & 10.9  & 32.3  & no     	  \\
           & 2018-08-12  & FLOYDS         & 2m0-02        & Siding Spring  & 11.3  & 19.5  & no	 	  \\
           & 2018-08-14  & FLOYDS         & 2m0-02        & Siding Spring  & 13.1  & 20.6  & no	 	  \\
           & 2018-08-16  & FLOYDS         & 2m0-02        & Siding Spring  & 15.2  & 36.2  & no	 	  \\
           & 2018-08-16  & EFOSC2         & NTT           & La Silla       & 15.7  & 39.1  & no     	  \\
           & 2018-08-16  & EFOSC2         & NTT           & La Silla       & 15.7  & 49.0  & no     	  \\
           & 2018-08-19  & FLOYDS         & 2m0-02        & Siding Spring  & 18.1  & 19.3  & no	 	  \\
           & 2018-08-21  & FLOYDS         & 2m0-02        & Siding Spring  & 20.0  & 42.4  & no	 	  \\
           & 2018-08-24  & FLOYDS         & 2m0-02        & Siding Spring  & 22.9  & 17.8  & no	 	  \\
           & 2018-08-28  & FLOYDS         & 2m0-02        & Siding Spring  & 26.7  & 34.4  & no	 	  \\
           & 2018-09-07  & FLOYDS         & 2m0-02        & Siding Spring  & 36.4  & 18.8  & no	 	  \\
           & 2018-09-07  & EFOSC2         & NTT           & La Silla       & 37.1  & 80.6  & no     	  \\
           & 2018-09-15  & EFOSC2         & NTT           & La Silla       & 44.8  & 39.6  &  yes 		  \\
           & 2018-09-15  & EFOSC2         & NTT           & La Silla       & 44.8  & 24.8  & no     	  \\
           & 2018-09-17  & FLOYDS         & 2m0-02        & Siding Spring  & 46.1  & 38.9  &  yes 		  \\
           & 2018-09-18  & FLOYDS         & 2m0-02        & Siding Spring  & 47.1  & 43.9  &  yes 		  \\
           & 2018-10-06  & FLOYDS         & 2m0-02        & Siding Spring  & 64.7  & 42.4  &  yes 		  \\
           & 2018-10-18  & EFOSC2         & NTT           & La Silla       & 76.9  & 38.4  &  yes 		  \\
           & 2018-10-30  & FLOYDS         & 2m0-02        & Siding Spring  & 87.9  & 33.5  &  yes 		  \\
           & 2018-11-01  & EFOSC2         & NTT           & La Silla       & 90.5  & 24.4  &  yes 		  \\
           & 2018-11-14  & EFOSC2         & NTT           & La Silla       & 103.1 & 16.3  &  yes 		  \\
           & 2019-03-08  & EFOSC2         & NTT           & La Silla       & 213.7 & 5.7   & no	          \\
\hline
\insertTableNotes
\end{longtable}
\twocolumn
\end{ThreePartTable}
%\end{landscape}

%\captionsetup{width=20cm}
%\begin{landscape}
\begin{ThreePartTable}
 \begin{TableNotes}
       \small  
       \item The first column gives the SN name. Column~2 indicates the photometric observation UT date. The next five columns list the photometric magnitude in the $griBV$ bands, respectively. The uncertainty associated with each photometric magnitude is indicated in parentheses.
 \end{TableNotes}
\onecolumn
\begin{longtable}{lcccccc}
    \caption{LCOGT photometric observations log.}
    \label{tab:lum-photlcogt}
    \endfirsthead
    \hline
    Object           &  UT Date    &  $g$           &  $r$           &  $i$               &   $B$           &  $V$              \\
                   &             & [mag]          & [mag]          & [mag]              & [mag]           & [mag]\\
    \hline
SN~2017cfo     & 2017-03-26  & 17.26(0.01)  & 17.29(0.01)  & 17.30(0.02)      &  17.43(0.03)  & 17.26(0.02) 	  \\
	           & 2017-03-26  & 17.25(0.01)  & 17.29(0.01)  & 17.33(0.02)	  &  17.47(0.05)  & 17.30(0.03) 	  \\
	           & 2017-03-30  & 17.53(0.01)  & 17.46(0.02)  & 17.47(0.02)	  &  17.82(0.04)  & 17.51(0.02) 	  \\
	           & 2017-03-30  & 17.54(0.01)  & 17.43(0.02)  & 17.49(0.03)	  &  17.78(0.04)  & 17.54(0.02) 	  \\
	           & 2017-04-03  & 17.85(0.03)  & 17.66(0.02)  & 17.57(0.03)	  &  18.07(0.08)  & 17.75(0.06) 	  \\
	           & 2017-04-03  & 17.86(0.03)  & 17.63(0.03)  & 17.66(0.04)	  &  18.13(0.07)  & 17.72(0.06) 	  \\
	           & 2017-04-07  & 18.05(0.05)  & 17.79(0.06)  & 17.75(0.08)	  &  18.39(0.09)  & 17.89(0.07) 	  \\
	           & 2017-04-07  & 18.06(0.05)  & 17.79(0.06)  & 17.81(0.09)	  &  18.33(0.09)  & 17.93(0.08) 	  \\
	           & 2017-04-12  & 18.30(0.08)  & 17.87(0.05)  & 17.94(0.05)	  &  18.64(0.17)  & 18.20(0.11) 	  \\
	           & 2017-04-12  & 18.30(0.08)  & 17.91(0.05)  & 17.84(0.05)	  &  18.52(0.15)  & 18.35(0.12) 	  \\
	           & 2017-04-13  & 18.20(0.06)  & 17.83(0.09)  & 17.90(0.08)	  &  18.72(0.09)  & 18.23(0.07) 	  \\
	           & 2017-04-13  & 18.32(0.06)  & 17.96(0.07)  & 17.91(0.07)	  &  18.84(0.08)  & 18.14(0.09) 	  \\
	           & 2017-04-16  & 18.45(0.02)  & 17.97(0.02)  & 17.95(0.03)	  &  18.93(0.06)  & 18.31(0.04) 	  \\
	           & 2017-04-16  & 18.43(0.03)  & 18.01(0.02)  & 17.97(0.03)	  &  18.95(0.06)  & 18.27(0.04) 	  \\
	           & 2017-04-20  & 18.54(0.02)  & $\cdots$     & $\cdots$	      &  19.03(0.05)  & 18.34(0.03) 	  \\
	           & 2017-04-20  & 18.54(0.02)  & $\cdots$     & $\cdots$	      &  19.08(0.06)  & 18.32(0.03) 	  \\
	           & 2017-04-21  & 18.60(0.02)  & 18.06(0.02)  & 18.05(0.03)	  &  19.15(0.05)  & 18.34(0.03) 	  \\
	           & 2017-04-21  & 18.56(0.02)  & 18.04(0.02)  & 18.09(0.03)	  &  19.13(0.05)  & 18.34(0.03) 	  \\
	           & 2017-04-25  & 18.74(0.03)  & $\cdots$     & $\cdots$	      &  19.27(0.07)  & 18.52(0.05)  	  \\
	           & 2017-04-25  & $\cdots$     & $\cdots$     & $\cdots$	      &  19.27(0.08)  & 18.39(0.04) 	  \\
	           & 2017-04-26  & 18.80(0.03)  & 18.22(0.04)  & 18.15(0.03)	  &  $\cdots$     & 18.92(0.10) 	  \\
	           & 2017-04-26  & 18.88(0.06)  & 18.16(0.02)  & 18.16(0.05)	  &  19.40(0.19)  & 18.64(0.04) 	  \\
	           & 2017-04-30  & 18.98(0.02)  & 18.33(0.02)  & 18.26(0.04)	  &  19.47(0.05)  & 18.72(0.04) 	  \\
	           & 2017-04-30  & 18.99(0.03)  & 18.33(0.02)  & 18.26(0.04)	  &  19.42(0.06)  & 18.72(0.04) 	  \\
	           & 2017-05-08  & 19.14(0.16)  & 18.65(0.11)  & $\cdots$	      &  $\cdots$     & 18.80(0.16) 	  \\
	           & 2017-05-08  & 19.06(0.15)  & 18.47(0.09)  & $\cdots$	      &  $\cdots$     & $\cdots$	  \\
	           & 2017-05-11  & 19.32(0.11)  & 18.70(0.07)  & 18.69(0.08)	  &  $\cdots$     & 19.21(0.16) 	  \\
	           & 2017-05-11  & 19.15(0.09)  & 18.77(0.07)  & 18.56(0.08)	  &  19.58(0.18)  & 19.02(0.12) 	  \\
	           & 2017-05-15  & 19.38(0.04)  & $\cdots$     & 18.83(0.11)	  &  19.77(0.08)  & 19.08(0.06) 	  \\
	           & 2017-05-15  & 19.31(0.04)  & $\cdots$     & 18.58(0.08)	  &  19.82(0.09)  & 19.20(0.06) 	  \\
	           & 2017-05-17  & 19.33(0.03)  & 18.73(0.03)  & 18.66(0.07)	  &  19.94(0.08)  & 19.25(0.06) 	  \\
	           & 2017-05-17  & 19.39(0.04)  & 18.68(0.03)  & 18.79(0.07)	  &  19.90(0.08)  & 19.26(0.07) 	  \\
	           & 2017-05-26  & 19.46(0.03)  & 18.83(0.03)  & 18.97(0.06)	  &  19.99(0.06)  & 19.32(0.05) 	  \\
	           & 2017-05-26  & 19.48(0.11)  & 18.85(0.03)  & 18.85(0.05)	  &  19.95(0.07   & 19.35(0.06) 	  \\
 	           & 2017-06-04  & $\cdots$     & $\cdots$     & $\cdots$	      &  19.95(0.20)  & 19.23(0.14) 	  \\
 	           & 2017-06-04  & $\cdots$     & $\cdots$     & $\cdots$	      &  19.95(0.20)  & 19.21(0.15) 	  \\
 	           & 2017-06-12  & $\cdots$     & $\cdots$     & $\cdots$	      &  19.88(0.13)  & 19.36(0.07) 	  \\
 	           & 2017-06-12  & $\cdots$     & $\cdots$     & $\cdots$	      &  20.13(0.15)  & 19.43(0.10)	  \\
\hline
SN~2017gpp     & $\cdots$    & $\cdots$       & $\cdots$       & $\cdots$           &  $\cdots$       & $\cdots$	  \\
\hline
SN~2017hbj     & 2017-10-05  & 16.12(0.01)  & 16.25(0.01)  & 16.36(0.01)      &  16.37(0.03)  & 16.21(0.02)	  \\
 	           & 2017-10-05  & 16.18(0.01)  & 16.27(0.01)  & 16.41(0.02)	  &  16.39(0.02)  & 16.24(0.02)	  \\
	           & 2017-10-08  & 16.39(0.01)  & 16.39(0.01)  & 16.44(0.02)	  &  16.65(0.02)  & 16.39(0.02)	  \\
	           & 2017-10-08  & 16.41(0.01)  & 16.42(0.01)  & 16.44(0.01)	  &  16.66(0.02)  & 16.40(0.02)	  \\
	           & 2017-10-12  & 16.62(0.01)  & 16.51(0.01)  & 16.50(0.01)	  &  16.88(0.04)  & 16.53(0.02)	  \\
	           & 2017-10-12  & 16.63(0.01)  & 16.48(0.01)  & 16.50(0.01)	  &  16.88(0.04)  & 16.55(0.02)	  \\
	           & 2017-10-14  & 16.78(0.01)  & 16.60(0.01)  & 16.54(0.01)	  &  17.08(0.02)  & 16.66(0.01)	  \\
	           & 2017-10-14  & 16.75(0.01)  & 16.60(0.01)  & 16.56(0.01)	  &  17.09(0.02)  & 16.66(0.01)	  \\
	           & 2017-10-17  & 16.87(0.01)  & 16.65(0.01)  & 16.61(0.01)	  &  17.20(0.02)  & 16.74(0.01)	  \\
	           & 2017-10-17  & 16.85(0.01)  & 16.66(0.01)  & 16.61(0.01)	  &  17.20(0.03)  & 16.74(0.01)	  \\
	           & 2017-10-20  & 16.95(0.01)  & 16.71(0.01)  & 16.66(0.01)	  &  17.34(0.02)  & 16.83(0.02)	  \\
	           & 2017-10-20  & 16.95(0.01)  & 16.71(0.01)  & 16.66(0.01)	  &  17.31(0.02)  & 16.82(0.02)	  \\
	           & 2017-10-23  & 17.01(0.01)  & 16.75(0.01)  & 16.69(0.01)	  &  17.39(0.04)  & 16.88(0.02)	  \\
	           & 2017-10-23  & 17.04(0.01)  & 16.72(0.01)  & 16.70(0.01)	  &  17.39(0.04)  & 16.86(0.02)	  \\
	           & 2017-10-26  & 17.18(0.02)  & $\cdots$     & $\cdots$     	  &  17.56(0.03)  & 17.01(0.03)	  \\
	           & 2017-10-26  & 17.15(0.02)  & $\cdots$     & $\cdots$     	  &  17.56(0.03)  & 17.00(0.03)	  \\
	           & 2017-10-26  & 17.20(0.01)  & 16.84(0.01)  & 16.82(0.02)	  &  17.67(0.03)  & 17.03(0.02)	  \\
	           & 2017-10-26  & 17.18(0.01)  & 16.87(0.01)  & 16.80(0.02)	  &  17.58(0.03)  & 17.01(0.02)	  \\
	           & 2017-10-30  & 17.46(0.01)  & 17.04(0.01)  & 17.02(0.02)	  &  17.87(0.03)  & 17.25(0.02)	  \\
     	       & 2017-10-30  & 17.44(0.01)  & 17.04(0.01)  & 17.03(0.02)	  &  17.89(0.03)  & 17.29(0.02)	  \\
	           & 2017-11-03  & 17.44(0.07)  & 17.19(0.06)  & 17.15(0.06)	  &  18.01(0.20)  & 17.36(0.12)	  \\
	           & 2017-11-03  & 17.69(0.08)  & 17.17(0.07)  & 17.23(0.06)	  &  17.77(0.13)  & 17.46(0.10)	  \\
	           & 2017-11-05  & 17.74(0.03)  & 17.25(0.03)  & 17.23(0.04)	  &  18.11(0.05)  & 17.53(0.05)	  \\
    	       & 2017-11-05  & 17.74(0.04)  & 17.31(0.03)  & 17.32(0.04)	  &  18.15(0.05)  & 17.60(0.04)	  \\
	           & 2017-11-12  & 18.00(0.01)  & 17.42(0.01)  & 17.47(0.01)	  &  18.39(0.05)  & 17.84(0.02)	  \\
	           & 2017-11-12  & 18.02(0.01)  & 17.42(0.01)  & 17.47(0.01)	  &  18.38(0.05)  & 17.84(0.02)	  \\
	           & 2017-11-20  & 18.20(0.01)  & 17.54(0.01)  & $\cdots$     	  &  18.55(0.05)  & 18.03(0.02)	  \\
	           & 2017-11-20  & 18.18(0.01)  & 17.54(0.01)  & $\cdots$     	  &  18.61(0.05)  & 18.06(0.02)	  \\
    	       & 2017-11-21  & 18.20(0.02)  & 17.58(0.02)  & 17.67(0.04)	  &  18.63(0.05)  & 18.02(0.04)	  \\
    	       & 2017-11-21  & 18.20(0.02)  & 17.57(0.02)  & 17.65(0.04)	  &  18.61(0.05)  & 18.04(0.04)	  \\
    	       & 2017-11-26  & 18.10(0.05)  & $\cdots$     & $\cdots$     	  &  18.62(0.18)  & $\cdots$     	  \\
    	       & 2017-11-26  & 18.28(0.06)  & 17.76(0.15)  & 17.78(0.17)	  &  $\cdots$     & $\cdots$     	  \\
    	       & 2017-12-02  & 18.34(0.05)  & 17.62(0.04)  & 17.84(0.05)	  &  18.86(0.09)  & 18.16(0.08)	  \\
    	       & 2017-12-02  & 18.40(0.06)  & 17.72(0.04)  & 17.81(0.04)	  &  18.62(0.07)  & 18.10(0.07)	  \\
    	       & 2017-12-07  & 18.43(0.03)  & 17.72(0.03)  & 17.99(0.05)	  &  18.78(0.05)  & 18.33(0.05)	  \\
    	       & 2017-12-07  & 18.40(0.03)  & 17.74(0.03)  & 17.94(0.05)	  &  18.85(0.05)  & 18.36(0.05)	  \\
    	       & 2017-12-14  & 18.48(0.02)  & 17.74(0.02)  & 18.00(0.03)	  &  18.86(0.03)  & 18.36(0.03)	  \\
    	       & 2017-12-14  & 18.47(0.02)  & 17.75(0.02)  & 18.01(0.03)	  &  18.73(0.03)  & 18.39(0.04)	  \\
    	       & 2017-12-20  & $\cdots$     & $\cdots$     & $\cdots$	      &  19.07(0.09)  & $\cdots$     	  \\
    	       & 2017-12-22  & 18.58(0.04)  & 17.81(0.03)  & 18.20(0.05)	  &  18.91(0.10)  & 18.35(0.06)	  \\
    	       & 2017-12-22  & 18.53(0.03)  & 17.79(0.03)  & 18.26(0.06)	  &  18.93(0.09)  & 18.37(0.05)	  \\
    	       & 2017-12-28  & 18.63(0.11)  & 17.91(0.08)  & $\cdots$     	  &  $\cdots$     & 18.13(0.16)	  \\
    	       & 2017-12-28  & 18.50(0.10)  & 17.90(0.07)  & 17.98(0.10)	  &  $\cdots$     & 18.52(0.14)	  \\
    	       & 2018-01-02  & 18.61(0.04)  & 17.91(0.03)  & 18.22(0.05)	  &  18.75(0.09)  & 18.46(0.05)	  \\
    	       & 2018-01-02  & 18.59(0.05)  & 17.84(0.03)  & 18.19(0.05)	  &  18.88(0.10)  & 18.57(0.06)	  \\
    	       & 2018-01-08  & $\cdots$     & $\cdots$     & $\cdots$	      &  18.96(0.05)  & $\cdots$	  \\
    	       & 2018-01-10  & 18.67(0.02)  & 17.92(0.02)  & 18.27(0.03)	  &  19.00(0.07)  & 18.60(0.03)	  \\
    	       & 2018-01-10  & 18.64(0.02)  & 17.88(0.02)  & 18.36(0.03)	  &  18.88(0.06)  & 18.62(0.03)	  \\
    	       & 2018-01-11  & 18.50(0.04)  & 17.98(0.16)  & $\cdots$     	  &  19.43(0.16)  & 18.57(0.07)	  \\
    	       & 2018-01-11  & 18.50(0.07)  & 17.90(0.19)  & 18.27(0.15)	  &  18.92(0.07)  & 18.65(0.06)	  \\
    	       & 2018-01-15  & 18.65(0.02)  & 18.00(0.03)  & 18.48(0.06)	  &  19.03(0.04)  & 18.71(0.05)	  \\
     	       & 2018-01-15  & 18.67(0.02)  & 17.97(0.03)  & 18.48(0.06)	  &  19.07(0.03)  & 18.72(0.05)	  \\
               & 2018-01-21  & 18.65(0.02)  & 17.95(0.03)  & 18.39(0.05)	  &  19.11(0.04)  & 18.76(0.04)	  \\
	           & 2018-01-21  & 18.67(0.02)  & 18.01(0.03)  & 18.52(0.07)	  &  19.13(0.04)  & 18.78(0.05)	  \\
    	       & 2018-01-27  & 18.81(0.04)  & 18.03(0.03)  & 18.53(0.06)	  &  19.05(0.08)  & 18.88(0.06)	  \\
    	       & 2018-01-27  & 18.84(0.04)  & 18.04(0.03)  & 18.59(0.06)	  &  19.04(0.09)  & 18.85(0.06)	  \\
    	       & 2018-01-31  & $\cdots$     & $\cdots$     & $\cdots$	      &  $\cdots$     & $\cdots$	  \\
    	       & 2018-02-02  & $\cdots$     & $\cdots$     & $\cdots$	      &  19.18(0.05)  & $\cdots$	  \\
     	       & 2018-02-03  & 18.90(0.03)  & 18.23(0.04)  & 18.73(0.07)	  &  19.22(0.04)  & 18.94(0.05)	  \\
	           & 2018-02-03  & 18.89(0.03)  & 18.23(0.04)  & 18.68(0.08)	  &  19.27(0.05)  & 18.95(0.06)	  \\
    	       & 2018-02-07  & 18.96(0.02)  & 18.24(0.03)  & 18.77(0.06)	  &  19.32(0.04)  & 18.96(0.04)	  \\
    	       & 2018-02-07  & 18.93(0.02)  & 18.27(0.03)  & 18.85(0.06)	  &  19.31(0.04)  & 19.01(0.04)	  \\
    	       & 2018-02-14  & 19.05(0.02)  & 18.36(0.03)  & 18.92(0.07)	  &  19.38(0.04)  & 19.14(0.05)	  \\
    	       & 2018-02-14  & 19.05(0.02)  & 18.40(0.03)  & 19.01(0.08)	  &  19.39(0.04)  & 19.15(0.04)	  \\
    	       & 2018-02-21  & 19.29(0.03)  & 18.45(0.02)  & 18.91(0.04)	  &  19.49(0.08)  & 19.25(0.04)	  \\
    	       & 2018-02-21  & 19.22(0.03)  & 18.46(0.02)  & 19.00(0.04)	  &  19.52(0.07)  & 19.26(0.04)	  \\
    	       & 2018-03-01  & $\cdots$     & 18.79(0.14)  & $\cdots$     	  &  $\cdots$     & $\cdots$     	  \\
    	       & 2018-03-01  & $\cdots$     & 18.44(0.10)  & $\cdots$     	  &  $\cdots$     & $\cdots$	  \\
    	       & 2018-03-07  & 19.61(0.03)  & 18.76(0.03)  & 19.38(0.07)	  &  19.73(0.07)  & 19.52(0.05)	  \\
    	       & 2018-03-07  & 19.58(0.03)  & 18.77(0.03)  & 19.52(0.07)	  &  19.85(0.08)  & 19.70(0.06)	  \\
    	       & 2018-03-07  & $\cdots$     & $\cdots$     & $\cdots$	      &  19.88(0.08)  & 19.62(0.05)	  \\
    	       & 2018-03-07  & $\cdots$     & $\cdots$     & $\cdots$	      &  19.72(0.07)  & 19.62(0.05)	  \\
    	       & 2018-03-08  & 19.50(0.04)  & 18.79(0.04)  & 19.25(0.08)	  &  20.09(0.15)  & 19.64(0.08)	  \\
    	       & 2018-03-08  & 19.59(0.04)  & 18.79(0.04)  & 19.40(0.09)	  &  19.85(0.09)  & 19.49(0.07)	  \\
    	       & 2018-03-17  & 19.92(0.05)  & 19.02(0.04)  & 19.52(0.10)	  &  20.05(0.07)  & 19.96(0.13)	  \\
    	       & 2018-03-17  & 19.78(0.04)  & 18.99(0.04)  & 19.69(0.12)	  &  20.15(0.08)  & 19.85(0.09)	  \\
    	       & 2018-03-24  & 19.98(0.09)  & 19.24(0.07)  & 19.95(0.19)      &  20.25(0.13)  & 20.33(0.19)	  \\
    	       & 2018-03-24  & 20.00(0.09)  & 19.27(0.07)  & $\cdots$     	  &  20.13(0.12)  & 20.13(0.15)	  \\
    	       & 2018-04-01  & 20.03(0.14)  & 19.47(0.12)  & $\cdots$     	  &  20.52(0.19)  & $\cdots$	  \\
    	       & 2018-04-01  & 19.82(0.11)  & 19.61(0.15)  & $\cdots$     	  &  $\cdots$     & 19.97(0.20)	  \\
\hline
SN~2017hxz     & 2017-11-13  & 18.30(0.01)  & 18.35(0.02)  & 18.33(0.03)      &  18.41(0.02)  & 18.30(0.02) 	  \\
    	       & 2017-11-13  & 18.30(0.01)  & 18.38(0.02)  & 18.44(0.04)	  &  18.44(0.02)  & 18.36(0.02) 	  \\
    	       & 2017-11-18  & 18.75(0.01)  & 18.75(0.02)  & 18.66(0.02)	  &  18.87(0.03)  & 18.75(0.02) 	  \\
    	       & 2017-11-18  & 18.79(0.01)  & 18.74(0.02)  & 18.65(0.03)	  &  18.87(0.03)  & 18.74(0.02) 	  \\
    	       & 2017-11-22  & 19.04(0.02)  & 19.12(0.04)  & 19.02(0.05)	  &  19.08(0.07)  & 18.92(0.05) 	  \\
    	       & 2017-11-22  & 19.10(0.02)  & 19.05(0.03)  & 19.00(0.06)	  &  18.93(0.06)  & 18.94(0.04) 	  \\
    	       & 2017-11-26  & 19.57(0.03)  & 19.72(0.06)  & 19.41(0.09)	  &  19.78(0.04)  & 19.55(0.05) 	  \\
    	       & 2017-11-26  & 19.50(0.02)  & 19.47(0.04)  & 19.22(0.07)	  &  19.82(0.06)  & 19.51(0.05) 	  \\
    	       & 2017-12-07  & 20.70(0.11)  & 20.52(0.11)  & 19.94(0.10)	  &  $\cdots$     & 20.42(0.12) 	  \\
    	       & 2017-12-07  & 20.71(0.11)  & 20.43(0.11)  & 20.07(0.13)	  &  $\cdots$     & 20.48(0.14) 	  \\
    	       & 2017-12-13  & 20.93(0.08)  & 21.06(0.18)  & $\cdots$	      &  21.47(0.14)  & $\cdots$	  \\
    	       & 2017-12-13  & 20.91(0.08)  & 21.10(0.19)  & 20.25(0.17)	  &  21.26(0.13)  & 20.78(0.18) 	  \\
    	       & 2017-12-19  & 21.43(0.11)  & 21.10(0.13)  & 20.74(0.15)	  &  21.41(0.19)  & 21.19(0.17) 	  \\
    	       & 2017-12-19  & 21.30(0.11)  & 21.19(0.15)  & $\cdots$	      &  $\cdots$     & $\cdots$	  \\
    	       & 2017-12-23  & 21.56(0.16)  & $\cdots$     & $\cdots$	      &  $\cdots$	  & $\cdots$	  \\
    	       & 2017-12-23  & 21.68(0.17)  & $\cdots$     & $\cdots$         &  $\cdots$     & $\cdots$	  \\
\hline
SN~2018aql     & 2018-04-12  & 18.53(0.04)  & 18.44(0.12)  & 18.21(0.17)      &  18.61(0.11)  & 18.50(0.10)	  \\
    	       & 2018-04-12  & 18.54(0.05)  & 18.41(0.07)  & $\cdots$     	  &  18.61(0.10)  & 18.52(0.10)	  \\
    	       & 2018-04-15  & 18.67(0.06)  & 18.52(0.13)  & $\cdots$     	  &  18.78(0.10)  & 18.60(0.11)	  \\
    	       & 2018-04-15  & 18.66(0.06)  & 18.51(0.13)  & $\cdots$     	  &  18.79(0.10)  & 18.60(0.11)	  \\
    	       & 2018-04-20  & 18.84(0.01)  & $\cdots$     & $\cdots$     	  &  19.10(0.02)  & 18.75(0.02)	  \\
    	       & 2018-04-20  & $\cdots$	    & $\cdots$     & $\cdots$         &  19.12(0.02)  & 18.77(0.02)	  \\
    	       & 2018-04-20  & $\cdots$	    & $\cdots$     & $\cdots$         &  19.14(0.02)  & 18.75(0.02)	  \\
    	       & 2018-04-20  & $\cdots$	    & $\cdots$     & $\cdots$         &  19.11(0.02)  & 18.77(0.02)	  \\
    	       & 2018-04-23  & 18.96(0.03)  & 18.66(0.03)  & 18.44(0.03)	  &  19.19(0.05)  & 18.90(0.05)	  \\
    	       & 2018-04-23  & 18.97(0.03)  & 18.70(0.03)  & 18.44(0.03)	  &  19.39(0.06)  & 18.82(0.04)	  \\
    	       & 2018-04-24  & 19.05(0.11)  & 18.76(0.15)  & $\cdots$    	  &  $\cdots$     & $\cdots$	  \\
    	       & 2018-04-24  & 19.05(0.14)  & 18.79(0.16)  & $\cdots$     	  &  $\cdots$     & $\cdots$	  \\
    	       & 2018-04-29  & 19.19(0.09)  & 18.98(0.10)  & 18.42(0.08)	  &  19.44(0.14)  & 19.12(0.13)	  \\
    	       & 2018-04-29  & 19.12(0.09)  & 18.85(0.09)  & 18.54(0.08)	  &  19.51(0.17)  & 18.91(0.11)	  \\
    	       & 2018-05-07  & 19.43(0.12)  & $\cdots$     & $\cdots$     	  &  $\cdots$     & $\cdots$	  \\
    	       & 2018-05-07  & 19.42(0.12)  & 19.06(0.19)  & $\cdots$     	  &  $\cdots$     & $\cdots$	  \\
    	       & 2018-05-08  & 19.32(0.02)  & 18.94(0.02)  & 18.62(0.03)	  &  19.71(0.03)  & 19.19(0.03)	  \\
    	       & 2018-05-08  & 19.45(0.07)  & 19.08(0.12)  & $\cdots$     	  &  19.70(0.04)  & 19.16(0.03)	  \\
    	       & 2018-05-08  & 19.47(0.11)  & 19.10(0.13)  & $\cdots$     	  &  19.69(0.15)  & 19.33(0.13)	  \\
    	       & 2018-05-08  & $\cdots$     & $\cdots$     & $\cdots$	      &  19.79(0.13)  & 19.28(0.12)	  \\
    	       & 2018-05-09  & 19.35(0.02)  & 18.96(0.02)  & 18.67(0.03)	  &  19.78(0.04)  & 19.17(0.03)	  \\
    	       & 2018-05-09  & 19.37(0.02)  & 18.91(0.02)  & 18.65(0.02)	  &  20.19(0.06)  & 19.11(0.03)	  \\
\hline
SN~2018eph     & 2018-08-08  & $\cdots$     & $\cdots$       & $\cdots$         &  16.65(0.01)  & $\cdots$	  \\ 
    	       & 2018-08-11  & 16.60(0.01)  & $\cdots$       & 16.67(0.01)	    &  16.81(0.02)  & 16.65(0.01)     \\
    	       & 2018-08-11  & 16.61(0.01)  & $\cdots$       & 16.70(0.01)	    &  16.83(0.02)  & 16.66(0.02)     \\
    	       & 2018-08-12  & $\cdots$     & 16.657(0.016)  & 16.71(0.02)	    &  16.83(0.03)  & 16.65(0.02)     \\
    	       & 2018-08-12  & $\cdots$     & 16.659(0.016)  & 16.72(0.02)	    &  16.84(0.03)  & 16.65(0.02)     \\
    	       & 2018-08-14  & 16.75(0.01)  & 16.705(0.013)  & 16.72(0.01)	    &  16.99(0.02)  & 16.72(0.02)     \\
    	       & 2018-08-14  & 16.73(0.01)  & 16.699(0.012)  & 16.74(0.01)	    &  16.99(0.02)  & 16.73(0.02)     \\
    	       & 2018-08-15  & 16.80(0.01)  & 16.770(0.015)  & 16.78(0.01)	    &  17.08(0.02)  & 16.78(0.02)     \\
    	       & 2018-08-15  & 16.79(0.01)  & 16.774(0.015)  & 16.79(0.02)	    &  17.05(0.02)  & 16.78(0.02)     \\
    	       & 2018-08-16  & $\cdots$     & $\cdots$       & $\cdots$  	    &  17.06(0.03)  & 16.81(0.01)     \\
    	       & 2018-08-16  & $\cdots$     & $\cdots$       & $\cdots$  	    &  17.07(0.02)  & 16.81(0.01)     \\
    	       & 2018-08-19  & $\cdots$     & $\cdots$       & $\cdots$  	    &  17.08(0.03)  & 16.80(0.02)     \\
    	       & 2018-08-19  & $\cdots$     & $\cdots$       & $\cdots$  	    &  17.13(0.03)  & 16.80(0.02)     \\
    	       & 2018-08-20  & 16.91(0.01)  & $\cdots$       & 16.77(0.01)	    &  17.21(0.02)  & 16.86(0.01)     \\
    	       & 2018-08-20  & 16.92(0.01)  & $\cdots$       & 16.79(0.01)	    &  17.21(0.02)  & 16.87(0.01)     \\
    	       & 2018-08-21  & 16.95(0.01)  & $\cdots$       & 16.81(0.01)	    &  17.26(0.02)  & 16.87(0.02)     \\
    	       & 2018-08-21  & 16.97(0.01)  & $\cdots$       & 16.79(0.01)	    &  17.23(0.03)  & 16.89(0.02)     \\
    	       & 2018-08-22  & $\cdots$     & $\cdots$       & 16.79(0.01)	    &  17.28(0.03)  & 16.92(0.02)     \\
    	       & 2018-08-22  & $\cdots$     & $\cdots$       & 16.80(0.01)	    &  17.29(0.03)  & 16.90(0.01)     \\
    	       & 2018-08-24  & 17.08(0.02)  & 16.916(0.026)  & 16.83(0.02)	    &  17.32(0.04)  & 17.00(0.03)     \\
    	       & 2018-08-24  & 17.09(0.02)  & 16.909(0.025)  & 16.86(0.02)	    &  17.39(0.04)  & 16.96(0.03)     \\
    	       & 2018-08-28  & 17.18(0.01)  & 16.975(0.014)  & 16.87(0.01)	    &  17.50(0.03)  & 17.06(0.02)     \\
    	       & 2018-08-28  & 17.20(0.01)  & 16.971(0.014)  & 16.89(0.01)	    &  17.51(0.03)  & 17.02(0.02)     \\
    	       & 2018-09-01  & 17.23(0.01)  & 17.011(0.017)  & 16.91(0.02)	    &  17.58(0.03)  & 17.07(0.02)     \\
    	       & 2018-09-01  & 17.23(0.01)  & 16.982(0.016)  & 16.96(0.02)	    &  17.61(0.03)  & 17.07(0.02)     \\
    	       & 2018-09-03  & 17.27(0.02)  & 16.990(0.020)  & 16.88(0.02)	    &  17.68(0.03)  & 17.09(0.03)     \\
    	       & 2018-09-03  & 17.27(0.02)  & 16.991(0.020)  & 16.92(0.02)	    &  17.62(0.03)  & 17.09(0.03)     \\
    	       & 2018-09-03  & 17.27(0.02)  & 16.993(0.020)  & 16.91(0.02)	    &  17.63(0.03)  & 17.10(0.02)     \\
    	       & 2018-09-03  & 17.27(0.02)  & 16.996(0.020)  & 16.91(0.02)	    &  $\cdots$     & $\cdots$	  \\
    	       & 2018-09-04  & $\cdots$     & $\cdots$       & 16.91(0.01)	    &  17.68(0.03)  & 17.17(0.02)     \\
    	       & 2018-09-04  & $\cdots$     & $\cdots$       & 16.92(0.01)	    &  17.70(0.02)  & 17.16(0.01)     \\
    	       & 2018-09-11  & 17.34(0.01)  & 17.020(0.013)  & 16.88(0.01)	    &  17.75(0.03)  & 17.16(0.02)     \\
    	       & 2018-09-11  & 17.34(0.01)  & 16.982(0.012)  & 16.93(0.02)	    &  17.75(0.12)  & 17.16(0.02)     \\
    	       & 2018-09-20  & 17.49(0.01)  & 17.110(0.016)  & 17.03(0.02)	    &  17.91(0.03)  & 17.33(0.02)     \\
    	       & 2018-09-20  & 17.49(0.01)  & 17.108(0.015)  & 17.03(0.02)	    &  17.93(0.03)  & 17.32(0.02)     \\
    	       & 2018-09-29  & 17.52(0.02)  & 17.078(0.017)  & 17.03(0.02)	    &  17.92(0.03)  & 17.30(0.02)     \\
    	       & 2018-09-29  & 17.50(0.02)  & 17.089(0.017)  & 17.01(0.02)	    &  17.89(0.03)  & 17.29(0.02)     \\
    	       & 2018-10-06  & $\cdots$     & 17.151(0.013)  & 17.13(0.02)	    &  18.05(0.03)  & 17.44(0.02)     \\
    	       & 2018-10-06  & $\cdots$     & 17.160(0.013)  & 17.11(0.02)	    &  18.07(0.03)  & 17.44(0.02)     \\
    	       & 2018-10-15  & 17.73(0.03)  & 17.137(0.022)  & 17.15(0.02)	    &  18.08(0.05)  & 17.54(0.03)     \\
    	       & 2018-10-15  & $\cdots$     & $\cdots$       & $\cdots$         &  18.11(0.04)  & 17.43(0.03)     \\
    	       & 2018-10-24  & 17.76(0.02)  & 17.354(0.020)  & 17.37(0.02)	    &  18.28(0.05)  & 17.66(0.03)     \\
    	       & 2018-10-24  & 17.75(0.02)  & 17.336(0.020)  & 17.31(0.02)	    &  18.29(0.04)  & 17.61(0.03)     \\
    	       & 2018-10-31  & $\cdots$     & $\cdots$       & 17.37(0.02)	    &  18.28(0.03)  & 17.75(0.02)     \\
    	       & 2018-10-31  & $\cdots$     & $\cdots$       & 17.35(0.02)	    &  18.30(0.03)  & 17.78(0.02)     \\
    	       & 2018-11-12  & $\cdots$     & $\cdots$       & 17.53(0.01)	    &  18.40(0.04)  & 17.82(0.02)     \\
    	       & 2018-11-12  & $\cdots$     & $\cdots$       & 17.49(0.01)	    &  18.43(0.03)  & 17.80(0.02)     \\
     	       & 2018-11-14  & $\cdots$     & $\cdots$       & $\cdots$	        &  18.44(0.03)  & 17.87(0.02)     \\
	           & 2018-11-14  & $\cdots$     & $\cdots$       & $\cdots$	        &  18.43(0.03)  & 17.85(0.02)     \\
	           & 2018-11-21  & 17.60(0.02)  & $\cdots$       & $\cdots$     	&  $\cdots$     & $\cdots$	  \\
    	       & 2018-11-22  & 17.30(0.09)  & $\cdots$       & $\cdots$     	&  18.49(0.05)  & 17.88(0.04)     \\
    	       & 2018-11-22  & 17.34(0.03)  & $\cdots$       & $\cdots$     	&  $\cdots$     & $\cdots$	  \\
    	       & 2018-11-24  & 16.48(0.06)  & $\cdots$       & 17.61(0.02)	    &  $\cdots$     & $\cdots$	  \\
    	       & 2018-11-24  & 16.47(0.06)  & $\cdots$       & 17.60(0.02)	    &  $\cdots$     & $\cdots$	  \\
    	       & 2018-11-26  & 18.08(0.01)  & 17.549(0.012)  & 17.63(0.01)	    &  18.58(0.03)  & 17.96(0.02)     \\
    	       & 2018-11-26  & 18.11(0.01)  & 17.585(0.012)  & 17.62(0.01)	    &  18.51(0.03)  & 18.01(0.02)     \\
    	       & 2018-11-27  & $\cdots$     & $\cdots$       & $\cdots$	        &  18.53(0.03)  & 17.92(0.03)     \\
    	       & 2018-11-27  & $\cdots$     & $\cdots$       & $\cdots$	        &  18.68(0.08)  & 17.92(0.04)     \\
    	       & 2018-11-28  & 18.07(0.02)  & $\cdots$       & 17.61(0.02)	    &  $\cdots$     & $\cdots$	  \\
    	       & 2018-11-30  & 18.27(0.03)  & 17.513(0.017)  & 17.66(0.01)	    &  18.60(0.03)  & 17.96(0.02)     \\
    	       & 2018-11-30  & 18.11(0.02)  & 17.514(0.017)  & 17.66(0.01)	    &  18.60(0.03)  & 18.00(0.02)     \\
    	       & 2018-11-30  & $\cdots$     & $\cdots$       & 17.70(0.02)	    &  $\cdots$     & $\cdots$	  \\
    	       & 2018-12-02  & 17.84(0.01)  & 17.548(0.014)  & 17.67(0.02)	    &  18.57(0.03)  & 18.02(0.02)     \\
    	       & 2018-12-02  & $\cdots$     & $\cdots$       & $\cdots$	        &  18.59(0.02)  & 17.98(0.02)     \\
    	       & 2018-12-04  & 18.15(0.02)  & 17.565(0.017)  & 17.73(0.02)	    &  18.62(0.02)  & 18.05(0.02)     \\
    	       & 2018-12-04  & 18.15(0.02)  & 17.567(0.016)  & 17.61(0.02)	    &  $\cdots$     & $\cdots$	  \\
    	       & 2018-12-07  & 18.20(0.03)  & $\cdots$       & $\cdots$     	&  18.66(0.03)  & 18.05(0.03)     \\
    	       & 2018-12-07  & $\cdots$     & $\cdots$       & $\cdots$	        &  18.69(0.04)  & 18.05(0.03)     \\
    	       & 2018-12-09  & $\cdots$     & $\cdots$       & $\cdots$	        &  18.69(0.02)  & 18.10(0.02)     \\
    	       & 2018-12-09  & $\cdots$     & $\cdots$       & $\cdots$	        &  18.72(0.02)  & 18.08(0.02)     \\
    	       & 2018-12-18  & 18.31(0.02)  & 17.674(0.015)  & 17.89(0.02)	    &  18.80(0.03)  & 18.09(0.02)     \\
    	       & 2018-12-18  & 18.32(0.02)  & 17.670(0.015)  & 17.87(0.02)	    &  18.75(0.03)  & 18.21(0.02)     \\
    	       & 2018-12-28  & $\cdots$     & $\cdots$       & $\cdots$	        &  18.83(0.03)  & 18.12(0.03)     \\
    	       & 2018-12-28  & $\cdots$     & $\cdots$       & $\cdots$	        &  18.75(0.05)  & 18.18(0.03)     \\
    	       & 2019-01-04  & $\cdots$     & $\cdots$       & $\cdots$	        &  18.76(0.03)  & 18.13(0.03)     \\
    	       & 2019-01-04  & $\cdots$     & $\cdots$       & $\cdots$	        &  18.80(0.03)  & 18.16(0.03)     \\
    	       & 2019-01-11  & $\cdots$     & $\cdots$       & $\cdots$	        &  18.95(0.03)  & 18.39(0.02)     \\
    	       & 2019-01-11  & $\cdots$     & $\cdots$       & $\cdots$	        &  18.98(0.03)  & 18.39(0.02)     \\
    	       & 2019-01-11  & $\cdots$     & $\cdots$       & $\cdots$	        &  19.05(0.02)  & 18.39(0.02)     \\
    	       & 2019-01-11  & $\cdots$     & $\cdots$       & $\cdots$	        &  19.04(0.03)  & 18.38(0.02)     \\
    	       & 2019-01-18  & 18.57(0.03)  & 17.773(0.029)  & 18.01(0.09)	    &  19.24(0.13)  & 18.34(0.05)     \\
    	       & 2019-01-18  & 18.50(0.03)  & 17.858(0.035)  & 18.00(0.08)	    &  19.05(0.07)  & 18.36(0.05)     \\
    	       & 2019-01-24  & 18.47(0.03)  & $\cdots$       & $\cdots$         &  19.08(0.07)  & 18.29(0.04)     \\ 
    	       & 2019-01-24  & 18.47(0.03)  & $\cdots$       & $\cdots$         &  19.00(0.06)  & 18.33(0.04)     \\
    	       & 2019-01-25  & $\cdots$     & $\cdots$       & $\cdots$	        &  19.13(0.04)  & 18.44(0.02)     \\
    	       & 2019-01-25  & $\cdots$     & $\cdots$       & $\cdots$	        &  19.12(0.04)  & 18.48(0.03)     \\
    	       & 2019-02-01  & $\cdots$     & $\cdots$       & $\cdots$	        &  19.19(0.02)  & 18.49(0.02)     \\
               & 2019-02-01  & $\cdots$     & $\cdots$       & $\cdots$	        &  19.15(0.02)  & 18.49(0.02)     \\
    	       & 2019-02-20  & 18.81(0.03)  & 18.069(0.020)  & 17.97(0.02)	    &  19.27(0.05)  & 18.59(0.03)     \\
    	       & 2019-02-20  & 18.69(0.03)  & 18.095(0.021)  & 17.96(0.03)	    &  19.25(0.05)  & 18.60(0.03)     \\
    	       & 2019-03-08  & $\cdots$     & $\cdots$       & $\cdots$	        &  19.33(0.03)  & $\cdots$	  \\
    	       & 2019-03-08  & $\cdots$     & $\cdots$       & $\cdots$	        &  19.30(0.03)  & $\cdots$          \\
\hline
\insertTableNotes
\end{longtable}
%\twocolumn
\end{ThreePartTable}
%\end{landscape}

%\begin{landscape}
\begin{ThreePartTable}
 \begin{TableNotes}
      \small  
      \item The first column gives the SN name. Column~2 indicates the photometric observation UT date. Then the $o$ photometric magnitude is listed. The uncertainty associated with each photometric magnitude is indicated in parentheses. The scheme is repeated from left to right for SN~2017cfo, SN~2017hxz, and SN~2018aql, respectively.
%       \begin{itemize}
%           \item
%       \end{itemize}
 \end{TableNotes}
%\setlength\tabcolsep{2.5pt} % default value: 6pt
%\onecolumn
\begin{longtable}{lcc||lcc||lcc}
    \caption{ATLAS photometry.}
    \label{tab:lum-photatlas}
    \endfirsthead
    \hline
    Object             &  UT Date    &  $o$  & SN             & UT Date     &  $o$    & SN             & UT Date     &  $o$\\
                   &             & [mag] &                &             & [mag]   &                &             & [mag]\\
    \hline
SN~2017cfo & 2017-03-15 & 18.33(0.13)   & SN~2017hxz  & 2017-10-27 & 19.58(0.07) & SN~2018aql     & 2018-03-29 & 18.11(0.05)\\
           & 2017-03-17 & 17.62(0.03)   &             & 2017-11-06 & 18.68(0.29) &                & 2018-03-31 & 18.20(0.19)\\
           & 2017-03-19 & 17.40(0.04)   &             & 2017-11-08 & 18.08(0.14) &                & 2018-04-06 & 18.48(0.07)\\
           & 2017-03-23 & 17.36(0.03)   &             & 2017-11-10 & 18.13(0.11) &                & 2018-04-12 & 18.87(0.08)\\
           & 2017-03-27 & 17.47(0.04)   &             & 2017-11-14 & 18.08(0.04) &                & 2018-04-24 & 19.62(0.13)\\
           & 2017-03-28 & 17.52(0.05)   &             & 2017-11-18 & 18.28(0.14) &                & 2018-04-26 & 19.79(0.19)\\
           & 2017-04-01 & 17.69(0.04)   &             & 2017-11-22 & 18.65(0.05) &                & 2018-05-10 & 20.20(0.26)\\
           & 2017-04-04 & 17.95(0.06)   &             & 2017-12-04 & 18.90(0.34) &                & 2018-05-16 & 20.32(0.32)\\
           & 2017-04-08 & 17.87(0.18)   &             & 2017-12-08 & 19.90(0.29) &                &            &            \\ 
           & 2017-04-09 & 18.23(0.22)   &             & 2017-12-10 & 19.71(0.27) &                &            &            \\ 
           & 2017-04-12 & 18.24(0.08)   &             & 2017-12-12 & 20.39(0.37) &                &            &            \\ 
           & 2017-04-16 & 18.30(0.05)   &             & 2017-12-24 & 20.41(0.36) &                &            &            \\
           & 2017-04-20 & 18.38(0.12)   &             &                          &                &            &            \\
           & 2017-04-21 & 18.30(0.10)   &	          &				             &		          &	           &	    \\
           & 2017-04-25 & 18.70(0.30)   &	          &				             &		          &	           &	    \\
           & 2017-05-07 & 19.26(0.13)   &	          &				             &		          &	           &	    \\
           & 2017-05-13 & 19.45(0.14)   &	          &				             &		          &	           &	    \\
           & 2017-05-15 & 19.23(0.18)   &	          &				             &		          &	           &	    \\
           & 2017-06-13 & 19.60(0.20)   &	          &				             &		          &	           &	    \\
           & 2017-06-21 & 19.76(0.13)   &	          &				             &		          &	           &	    \\
           & 2017-06-24 & 19.91(0.27)   &	          &				             &		          &	           &	    \\
           & 2017-06-28 & 19.80(0.27)   &	          &				             &		          &	           &	    \\

\hline
\insertTableNotes
\end{longtable}
%\twocolumn
\end{ThreePartTable}
%\end{landscape}

%
%

%\begin{landscape}
\begin{ThreePartTable}
 \begin{TableNotes}
      \small  
      \item The first column gives the SN name. Column~2 indicates the photometric observation UT date. The next three columns list the $gri$ photometric magnitudes, respectively. The uncertainty associated with each photometric magnitude is indicated in parentheses.
 \end{TableNotes}
\begin{longtable}{lcccc}
    \caption{SkyMapper photometry.}
    \label{tab:lum-photskymap}
    \endfirsthead
    \hline
    Object            & UT Date    &  $g$           &  $r$           &  $i$   \\
                  &            & [mag]          & [mag]          & [mag]  \\
    \hline
SN~2017gpp   & 2017-08-31  & 19.36(0.12)  & 19.46(0.27)  & $\cdots$     \\
             & 2017-09-03  & 18.49(0.16)  & 18.42(0.15)  & $\cdots$     \\
             & 2017-09-06  & 18.39(0.09)  & 18.54(0.09)  & $\cdots$     \\
             & 2017-09-08  & 18.59(0.11)  & $\cdots$     & $\cdots$     \\
             & 2017-09-09  & $\cdots$     & $\cdots$     & 18.70(0.08)  \\
             & 2017-09-14  & $\cdots$     & $\cdots$     & 18.59(0.11)  \\
             & 2017-09-17  & $\cdots$     & $\cdots$     & 18.45(0.08)  \\
             & 2017-09-17  & $\cdots$     & $\cdots$     & 18.36(0.08)  \\
             & 2017-09-22  & $\cdots$     & $\cdots$     & 18.36(0.10)  \\
             & 2017-09-25  & 18.85(0.10)  & 18.79(0.10)  & $\cdots$     \\
             & 2017-10-01  & $\cdots$     & $\cdots$     & 18.58(0.08)  \\
             & 2017-10-09  & $\cdots$     & $\cdots$     & 18.84(0.10)  \\
             & 2017-10-09  & $\cdots$     & $\cdots$     & 18.64(0.10)  \\
             & 2017-10-12  & 19.31(0.07)  & 18.95(0.11)  & $\cdots$     \\
\hline
\insertTableNotes
\end{longtable}
\twocolumn
\end{ThreePartTable}
%\end{landscape}

\endgroup 
%%%%%%%%%%%%%%%%%%%%%%%%%%%%%%%%%%%%%%%%%%%%%%%%%%

% Don't change these lines
\bsp	% typesetting comment
\label{lastpage}
\end{document}